%% file: EXO-24-016_temp.tex
\pdfoutput=1
\documentclass[11pt,twoside,a4paper,cmspaper,final,collab]{cms-tdr}
\def\svnVersion{5abf61e}\def\svnDate{2026/11/05}\def\cmsCernNoTag{CERN-EP-2026-132}\def\cmsCernDate{\today}\def\cmsMessage{Published in the Journal of High Energy Physics as \href{http://dx.doi.org/10.1007/JHEP11(2026)004}{\doi{10.1007/JHEP11(2026)004}.}}
\begin{document}\cmsNoteHeader{EXO-24-016}

\newcommand{\lumivalscouting}{62.4\fbinv}
\newcommand{\lumivalthree}{27.8\fbinv}
\newcommand{\mmumu}{\ensuremath{m_{\PGm\PGm}}\xspace}
\newcommand{\ptmu}{\ensuremath{\pt^{\PGm}}\xspace}
\newcommand{\etamu}{\ensuremath{\eta^{\PGm}}\xspace}
\newcommand{\ptmuone}{\ensuremath{\pt^{\PGm_{1}}}\xspace}
\newcommand{\ptmutwo}{\ensuremath{\pt^{\PGm_{2}}}\xspace}
\newcommand{\ptmumu}{\ensuremath{\pt^{\PGm\PGm}}\xspace}
\newcommand{\ptmumuvec}{\ensuremath{\vec{p}_{\, \mathrm{T}}^{\PGm\PGm}}\xspace}
\newcommand{\drmumu}{\ensuremath{\Delta R(\PGm_{1},\PGm_{2})}\xspace}
\newcommand{\dphimumu}{\ensuremath{\Delta\phi(\PGm_{1},\PGm_{2})}\xspace}
\newcommand{\dphimudv}{\ensuremath{\Delta\phi(\ptmumuvec,\overrightarrow{\mathrm{SV}})}\xspace}
\newcommand{\vecdv}{\ensuremath{\overrightarrow{\mathrm{SV}}}\xspace}
\newcommand{\detamumu}{\ensuremath{\Delta\eta(\PGm_{1},\PGm_{2})}\xspace}
\newcommand{\logabsetaphi}{\ensuremath{\log_{10}(\abs{\Delta\eta_{\PGm\PGm}}/\abs{\Delta\phi_{\PGm\PGm}})}\xspace}
\newcommand{\lxy}{\ensuremath{l_\mathrm{xy}}\xspace}
\newcommand{\dxy}{\ensuremath{d_\mathrm{xy}}\xspace}
\newcommand{\sigmadxy}{\ensuremath{\sigma_{\dxy}}\xspace}
\newcommand{\scaleddxy}{\ensuremath{\abs{\dxy}/(\lxy\mmumu/\ptmumu)}\xspace}
\providecommand{\PSGp}{\HepSusyParticle{\PGp}{}{}\xspace}
\providecommand{\PSGh}{\HepSusyParticle{\PGh}{}{}\xspace}
\providecommand{\PGY}{\HepParticle{\Uppsi}{}{}\xspace}
\providecommand{\PAGY}{\HepAntiParticle{\Uppsi}{}{}\xspace}
\providecommand{\PApr}{\HepParticle{A}{}{\prime}\xspace}

\cmsNoteHeader{EXO-24-016}
\title{Search for long-lived particles decaying into muons in proton-proton collisions at \texorpdfstring{$\sqrt{s} = 13.6\TeV$}{sqrt(s) = 13.6 TeV} using data scouting}

\date{\today}

\abstract{
A search for long-lived particles decaying into muons is performed using proton-proton collisions at $\sqrt{s}=13.6\TeV$, collected by the CMS experiment at the LHC in 2022 and 2023, corresponding to an integrated luminosity of 62.4\fbinv. The data were collected using dedicated dimuon triggers with low transverse momentum thresholds, recorded with a high-rate data scouting trigger stream. This data stream retains a reduced amount of information at the high-level trigger, to explore otherwise inaccessible phase space at low multimuon invariant mass and nonzero displacement from the primary interaction vertex. No significant excess of events above the standard model prediction is found. Upper limits on branching fractions at 95\% confidence level are set for a wide range of mass and lifetime hypotheses in several beyond the standard model frameworks, where the Higgs boson decays into long-lived dark photons or into dark partons that produce showers containing long-lived particles, or where a long-lived scalar resonance is produced from the decay of a \PQb hadron. The resulting constraints improve and extend existing ones in large regions of the parameter space.
}

\hypersetup{
pdfauthor={CMS Collaboration},
pdftitle={Search for long-lived particles decaying into muons in proton-proton collisions at sqrt(s) = 13.6 TeV using the CMS scouting data sets},
pdfsubject={CMS},
pdfkeywords={CMS, scouting, long-lived particles, muons, resonance, dark sector}}

\maketitle

\section{Introduction}
\label{sec:introduction}

The origin of dark matter remains one of the outstanding puzzles in particle physics and cosmology, in spite of the cosmological evidence of its existence~\cite{DM1,DM2,DM3,DM4}.
Dark matter is expected to interact very weakly, if at all, with standard model (SM) particles, except via gravity.
This introduces the possibility of a hidden (dark) sector of matter~\cite{DM5,hahm_2015,CMS:2024zqs}.
Particles in the dark sector would interact with the SM particles only via weakly interacting mediators.

\begin{figure}[ht]
  \centering 
    \vspace{0.01\textwidth}
    \includegraphics[width=0.48\textwidth]{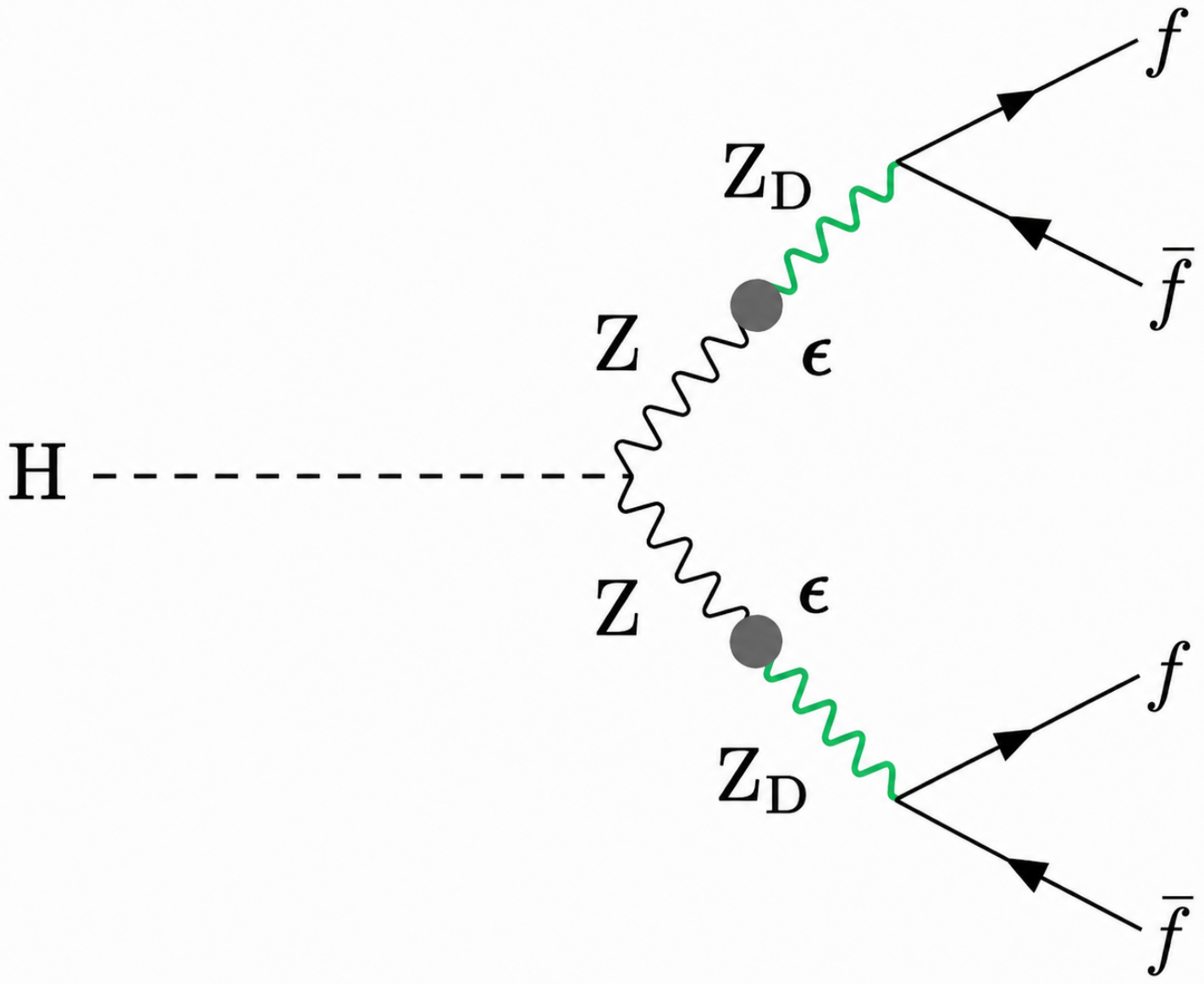}
    \includegraphics[width=0.48\textwidth]{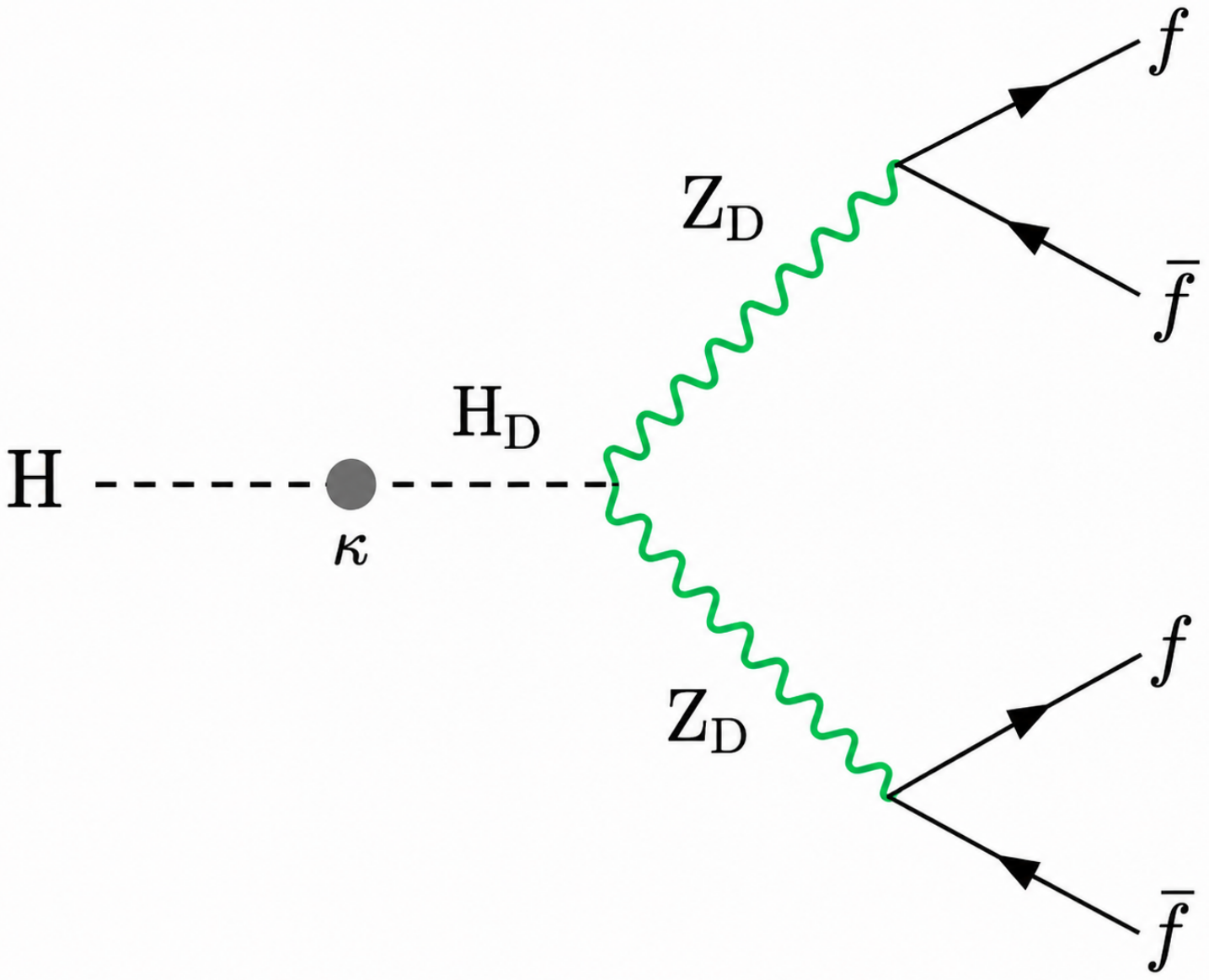}
    \vspace{0.01\textwidth}
    \caption{Diagrams illustrating an SM-like Higgs boson (\PH) decay to four fermions ($f$) 
      via two intermediate dark photons, $\PZ_{\mathrm{D}}$~\cite{hahm_2015}:
      through the hypercharge portal (left) and the Higgs portal (right) via a dark Higgs boson ($\PH_{\mathrm{D}}$).
      The long-lived particles are highlighted in green.
      }
    \label{fig:HZdZd}
\end{figure}

One dark-matter scenario, referred to as the hidden Abelian Higgs model (HAHM),
involves a spontaneously broken dark $U(1)_{\mathrm{D}}$ gauge symmetry, 
mediated by a dark photon, $\PZ_{\mathrm{D}}$~\cite{hahm_2015}.
In this scenario, the only renormalizable interaction with SM particles occurs via the kinetic mixing of a dark photon with a hypercharge gauge boson.
In addition, if a dark Higgs mechanism is responsible for the spontaneous breaking of the $U(1)_{\mathrm{D}}$ gauge symmetry, 
then the dark Higgs boson ($\PH_{\mathrm{D}}$) has a renormalizable coupling to the 125\GeV SM-like Higgs boson (\PH), 
resulting in mixing between the two physical scalar states.
Thus, the hidden sector may interact with the SM sector either through the hypercharge portal
via the kinetic mixing parameter $\epsilon$,
or through the Higgs portal via the Higgs mixing parameter $\kappa$.
The dark photon $\PZ_{\mathrm{D}}$ may also mix with the SM photon (\PGg) and the \PZ boson through the hypercharge portal.
If there are no hidden-sector states below the $\PZ_{\mathrm{D}}$ boson mass, the $\PZ_{\mathrm{D}}$ decays exclusively into SM particles,
with a sizable branching fraction to leptons, where the coupling of the SM fermions to the $\PZ_{\mathrm{D}}$ boson is proportional to $\epsilon$.
The $\PZ_{\mathrm{D}}$ boson is expected to be long-lived if $\epsilon\lesssim10^{-4}$,
with a mean proper decay length $c\tau_{0}^{\PZ_{\mathrm{D}}} \gtrsim 1\unit{\mum}$,
proportional to $\epsilon^{-2}$.
In this regime, the decays $\PH\to\PZ\PZ_{\mathrm{D}}$ and $\PH\to\PZ_{\mathrm{D}}\PZ_{\mathrm{D}}$ 
through the hypercharge portal have negligible branching fractions, hence the decays through the Higgs portal are expected to dominate.
The diagrams in Fig.~\ref{fig:HZdZd} illustrate the production of a pair of $\PZ_{\mathrm{D}}$ bosons in a Higgs boson decay.

\begin{figure}[th]
  \centering 
    \vspace{0.01\textwidth}
    \includegraphics[width=0.48\textwidth]{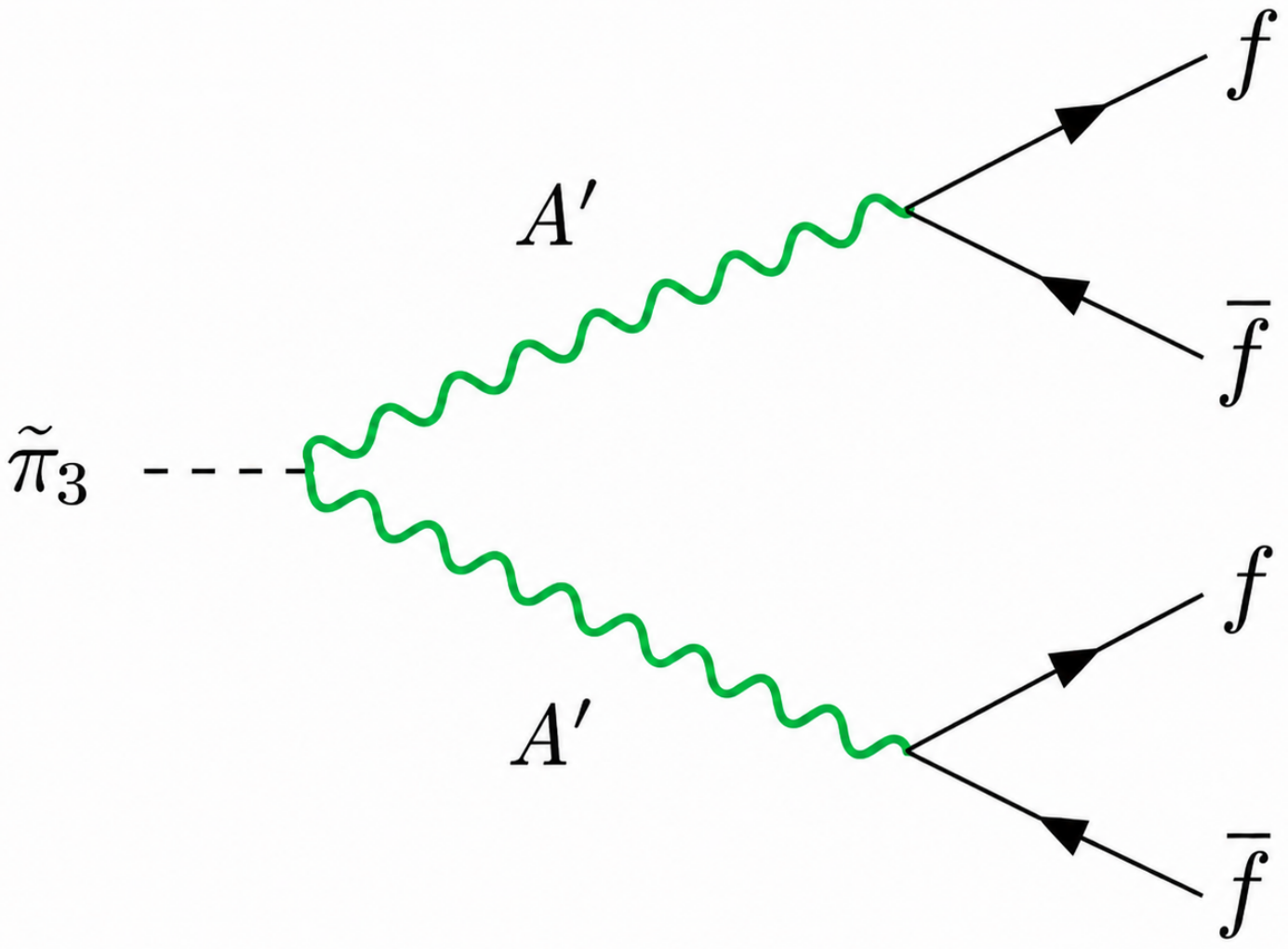}
    \includegraphics[width=0.48\textwidth]{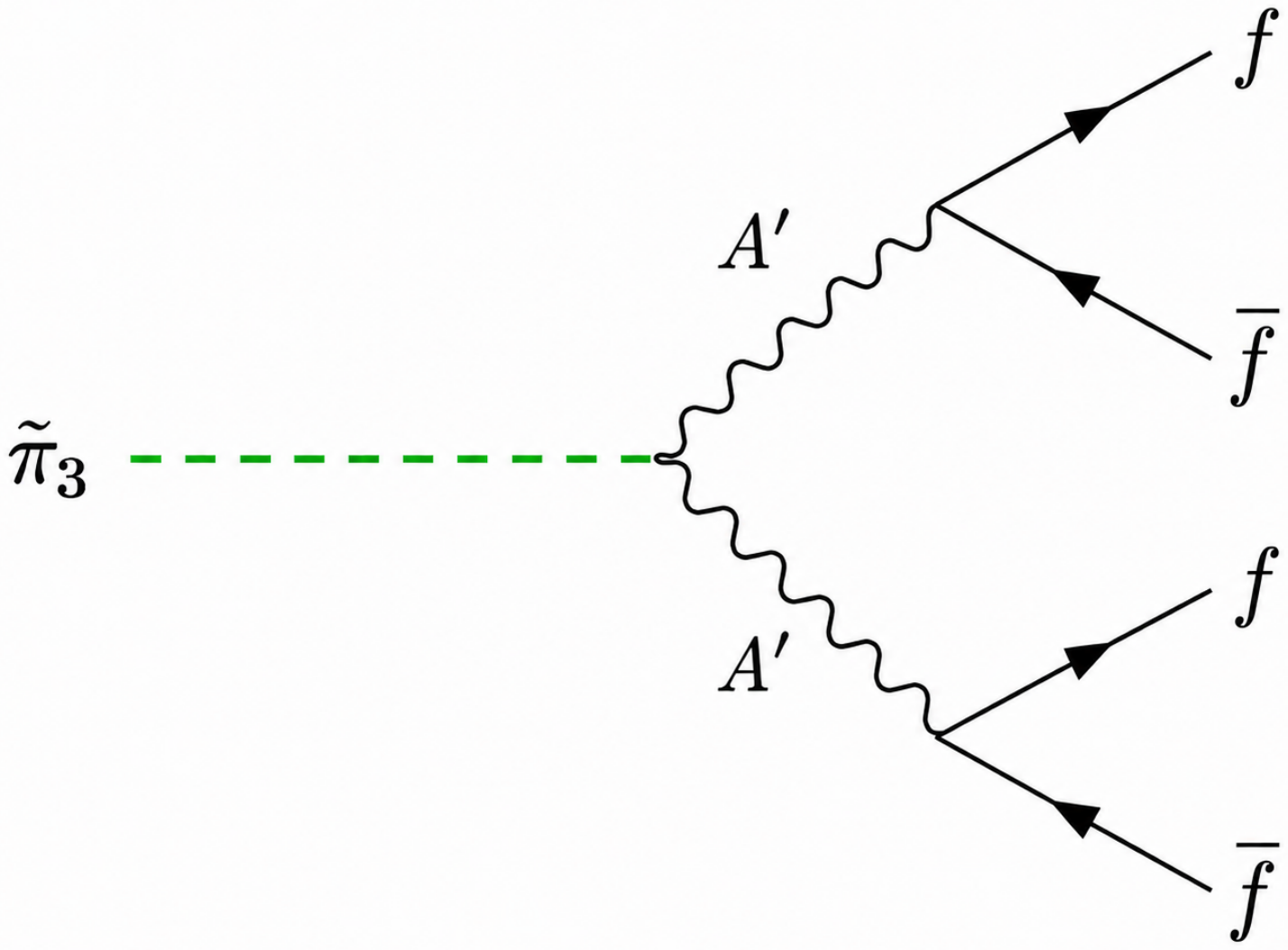}
    \vspace{0.01\textwidth}
    \caption{
      Diagrams illustrating two dark-shower scenarios~\cite{darkqcd},
      where a dark meson $\PSGh$ is allowed to decay into dark pions ($\PSGp_3$),
      which in turn decay into pairs of dark photons ($\PApr$).
      Scenario~A (left) assumes prompt $\PSGp_3$ decays into a pair of long-lived dark photons, $\PApr$, each decaying into fermions.
      Scenario~B1 (right) assumes long-lived $\PSGp_3$ decays into a pair of promptly decaying dark photons ($\PApr$).
      The long-lived particles are highlighted in green.
    }
    \label{fig:dqcd}
\end{figure}

In other hidden-sector models~\cite{darkqcd}, 
a new, confining dark gauge group is introduced, which can produce events with multiple nonisolated dark particles, referred to as ``dark showers'', 
analogous to phenomena in the quantum chromodynamics sector of the SM.
A spontaneously broken $U(1)$ gauge symmetry is involved, mediated by a dark photon $\PApr$ that couples to the SM photon.
The confining dark gauge group dynamically breaks the hidden-sector $SU(2) \otimes SU(2)$ flavor symmetry,
resulting in the addition of a set of three dark pions ($\PSGp_1$, $\PSGp_2$, and $\PSGp_3$)
and a heavier dark meson, $\PSGh$.
In these models, the SM Higgs boson can decay into a pair of dark partons $\PGY\PAGY$,
which then undergo showering and hadronization within the dark sector.
Different scenarios are considered: a model where the dark photon $\PApr$ is long-lived,
referred to as \textit{Scenario~A}~\cite{darkqcd},
and a model where the dark pion $\PSGp_3$ is long-lived instead,
referred to as \textit{Scenario~B1}~\cite{darkqcd}.
In both scenarios, it is assumed that the three
dark pions $\PSGp_1$, $\PSGp_2$, and $\PSGp_3$ have nearly degenerate masses $m_{\PSGp}$ and arise from the decay of $\PSGh$,
$\PSGh \to 3 \PSGp$, with the $\PSGh$ meson mass $m_{\PSGh} > 3 m_{\PSGp}$.
On average, five dark mesons per event are produced.
The dark pion $\PSGp_3$ is assumed to decay, in turn, into a pair of dark photons $\PApr$,
while $\PSGp_1$ and $\PSGp_2$ are assumed to be stable.
Thus, Scenario~A yields final states where multiple fermion pairs are produced in as many distinct vertices,
each corresponding to the decay of a dark photon $\PApr$.
Conversely, Scenario~B1 yields final states where four fermions emerge from the same displaced vertex.
Diagrams illustrating Scenarios~A and~B1 are shown in Fig.~\ref{fig:dqcd}.

Constraints have been placed on visible dark-photon decays by beam dump~\cite{ZDLimits-1},
fixed-target~\cite{ZDLimits-2}, $\Pep\Pem$ collider~\cite{ZDLimits-3}, and rare-decay experiments~\cite{ZDLimits-4}, 
as well as by the LHCb~\cite{ZDLimits-5a,ZDLimits-5b,ZDLimits-5c}, ATLAS~\cite{ATLAS:2022izj,ATLAS:2311.18298,ATLAS:2503.20445},
and CMS~\cite{EXO-19-018,EXO-20-014,EXO-21-005,EXO-23-014,CMS-EXO-24-008} experiments at the CERN LHC.

\begin{figure}[th]
  \centering 
    \vspace{0.01\textwidth}
    \includegraphics[height=0.25\textheight]{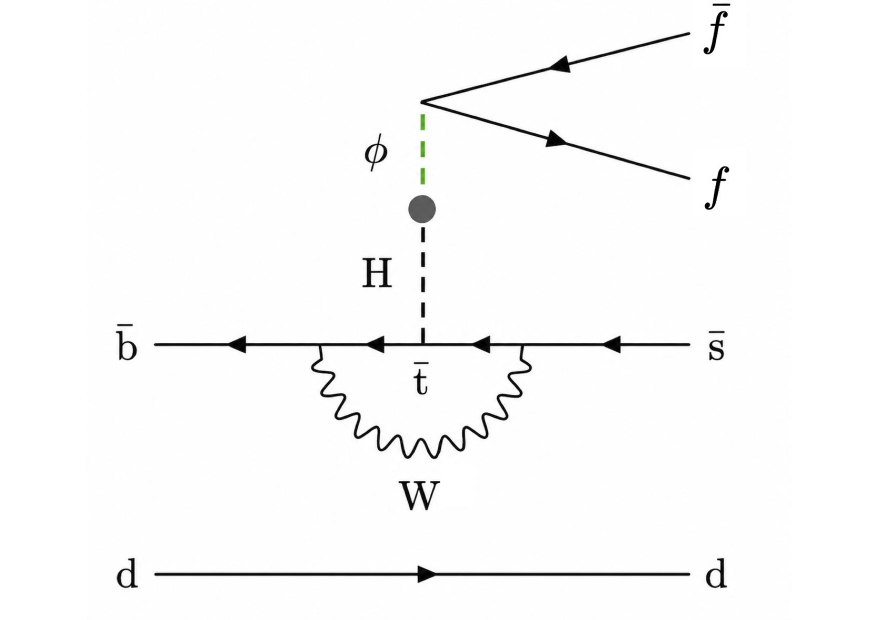}
    \vspace{0.01\textwidth}
    \caption{
      Diagram illustrating the production of a scalar resonance $\PGf$ in a \PQb hadron decay,
      through the mixing with an SM-like Higgs boson (\PH).
      The long-lived particle $\PGf$ is highlighted in green.
    }
    \label{fig:btophi}
\end{figure}

Other scenarios of physics beyond the SM (BSM) are also considered that may produce a long-lived, low-mass resonance decaying into a muon pair.
In one of the minimal extensions of the SM a singlet scalar field $\PGf$ is added,
which mixes with the Higgs boson with a mixing angle $\theta$ and thereby couples to all SM fermions~\cite{BPhi-inflaton,BPhi}.
The $\PGf$ boson is expected to be long-lived under the hypothesis of weak couplings to SM fermions,
corresponding to $\theta \lesssim 10^{-3}$.
Such a scalar resonance can be produced in decays of \PQb hadrons, $\Ph_\PQb \to \PGf X$.
This process is depicted in Fig.~\ref{fig:btophi}.
Constraints on this model have been previously set by the CHARM~\cite{BtoPhiLimits-0}, LHCb~\cite{BtoPhiLimits-1,BtoPhiLimits-2}, and CMS~\cite{EXO-20-014} experiments.

This paper presents a search for narrow, long-lived multimuon resonances, based on data collected with the CMS experiment
in proton-proton ($\Pp\Pp$) collisions at $\sqrt{s}=13.6\TeV$ in 2022 and 2023.
A dedicated dimuon trigger stream (\textit{scouting}~\cite{EXO-23-007}) with low transverse momentum (\pt) thresholds,
recorded at high rate by retaining a reduced amount of information at the high-level trigger (HLT) is used.
The rate of scouting triggers is higher than that 
of the standard triggers owing to less stringent requirements 
that enable multimuon resonance searches across mass and lifetime ranges 
that are otherwise inaccessible.
The selected data correspond to a total integrated luminosity of \lumivalscouting ($34.6$ and \lumivalthree collected in 2022 and 2023, respectively).

The search targets narrow, low-mass, long-lived resonances decaying into pairs of oppositely charged muons, 
where the lifetime of the long-lived particle is such that the transverse displacement (\lxy) of its decay vertex is within roughly $70\cm$
of the primary vertex (PV) of the hard scattering, \ie, sufficiently within the volume of the tracker detector to ensure accurate measurements.
The signal is expected to appear as a narrow peak on top of a smooth multimuon invariant mass distribution, 
with a resonance width that is assumed to be smaller than the experimental mass resolution.
The search is performed by looking for localized excesses in the multimuon invariant mass spectrum
across different ranges of \lxy.
We probe, for the first time in CMS, multimuon final states with large displacements and masses below 10\GeV,
which have not been explored so far owing to the limitations of the standard trigger streams and of the dimuon scouting triggers used by CMS before 2022. 
In particular, the improvements in the dimuon scouting triggers allow for the reconstruction of displaced muon pairs beyond 11\cm in \lxy,
which was a limitation of the analysis presented in Ref.~\cite{EXO-20-014},
where data collected with the scouting triggers available in 2017 and 2018 were used.
Moreover, the extended event content available in the scouting data collected since 2022~\cite{EXO-23-007} is exploited
to further improve the sensitivity of the search.

This paper is organized as follows.
The CMS detector, and the data sets and triggers used in the analysis are described in
Sections~\ref{sec:cmsdetector} and~\ref{sec:triggers}, respectively.
The signal Monte Carlo (MC) samples for each interpretation are presented in Section~\ref{sec:mc}.
The event selection and categorization strategies are discussed in Section~\ref{sec:selection},
while Section~\ref{sec:fit} outlines the modeling of the signal and the background.
Finally, results are presented in Section~\ref{sec:results}, and the paper is summarized in Section~\ref{sec:summary}.
Tabulated results are available in the HEPData record for this analysis~\cite{hepdata}.

\section{The CMS detector}
\label{sec:cmsdetector}

The central feature of the CMS apparatus is a superconducting solenoid of 6\unit{m} internal diameter
providing a magnetic field of 3.8\unit{T}. Within the solenoid volume are a silicon pixel and strip tracker, a
lead tungstate crystal electromagnetic calorimeter, and a brass and scintillator hadron calorimeter, each composed of
a barrel and two endcap sections. Forward calorimeters extend the pseudorapidity ($\eta$) coverage provided by the
barrel and endcap detectors. Muons are reconstructed using gas-ionization detectors interleaved with the layers of the steel flux-return yoke outside
the solenoid.
A global ``particle-flow'' (PF) algorithm~\cite{Sirunyan:2017ulk} aims to reconstruct all individual particles in an event,
combining information provided by the all-silicon tracker and by the calorimeters with data from the gas-ionization muon detectors.
The pixel tracker consists of four concentric barrel layers at radii of 2.9, 6.8, 10.9, and 16.0\cm, 
and three disks on each end at distances of 29.1, 39.6, and 51.6\cm from the center of the detector~\cite{phase1tracker}.
The region between 20 and 116\cm is occupied by the silicon strip tracker. 
Four barrel layers, supplemented by three disk in the endcaps,
compose the tracker inner barrel and disks,
which extends the radius up to 55\cm. Beyond this range, 
the tracker outer barrel and the tracker endcaps are situated,
adding six layers and nine additional disks at each end.
A more detailed description of the CMS detector, together with a definition of the coordinate system used and the relevant
kinematic variables, can be found in Refs.~\cite{CMS:2008xjf,CMS:2023gfb}.

A two-level trigger system is used to select events of potential physics interest.
The first level of the CMS trigger system (L1)~\cite{CMSL1}, composed of custom hardware processors, uses information from the
calorimeters and muon detectors to select events of interest in a fixed time interval of about 4\unit{\mus}.
The HLT processor farm further decreases the event rate from around 100\unit{kHz} to a few\unit{kHz}, before data storage.
A more detailed description of the CMS trigger system can be found in Refs.~\cite{CMStrigger,CMS:2024aqx}.

In addition to the standard HLT selection streams, 
a dedicated set of triggers exists that allows the exploration of additional phase space. 
This approach is referred to as \textit{data scouting}~\cite{EXO-23-007}.
The scouting trigger algorithms used in this search
select events containing muon pairs with a dimuon invariant mass greater than the muon mass ($\mmumu\gtrsim 2 m_{\PGm}$) at a rate of a few\unit{kHz}.
In order to allow the large rate within an acceptable bandwidth,
the event size is reduced by up to a factor of 1000
by only retaining limited information, as reconstructed at the HLT.

\section{Data sets and triggers}
\label{sec:triggers}

The standard trigger streams typically select events containing at least one particle (\eg, a lepton) or jet with large \pt.
Thus, such triggers are not necessarily ideal in a generic search for light, narrow resonances decaying into two muons with $\pt \lesssim 10\GeV$.
Muons are reconstructed at the HLT
by combining the information from the tracker and the muon detectors.
The retained limited information includes their four-momenta,
the number of hits left in the tracker and in the muon detectors, and detailed information on the muon tracks.
These muon tracks are used in pairs to form dimuon vertices, considering all possible pairs.
In the following, these vertices are called secondary vertices (SVs),
and they may be either displaced from the PV or not.
Candidate $\Pp\Pp$ interaction vertices are formed
using tracks reconstructed from hits only in the pixel detector,
and the candidate vertex with the largest value of summed track $\pt^{2}$ is taken to be the PV of the $\Pp\Pp$ interaction.
Events in the scouting data sets contain information on  the normalized $\chi^2$ of all reconstructed vertices, 
together with their position with respect to the nominal interaction point.

Between 2018 and 2022, the CMS muon scouting triggers~\cite{EXO-23-016} were updated,
broadening the accessible phase space and increasing the amount of information retained for each event.
The main improvement arises from the removal of the pixel hit multiplicity requirement,
which previously prevented any muon with $<$2~pixel hits from being accepted by the scouting triggers.
As a result, the muons with $\lxy > 11\cm$ are now accessible for the first time in a search using scouting data.
Additionally, more detailed information on tracks and PF candidates is retained
and used to improve the background rejection, as described in Section~\ref{sec:evtsel}.

Events are preselected at the L1 trigger
with at least two muons passing requirements on the \pt, $\eta$ and charge of each muon
and on their angular separation,
defined as $\drmumu = \sqrt{\smash[b]{[\dphimumu]^2 + [\detamumu]^2}}$,
where $\phi$ is the azimuthal angle and $\PGm_{1}$ ($\PGm_{2}$) is the leading (subleading) muon.

In 2022, events were required to pass at least one of three criteria:
\begin{itemize}
\item the presence of at least two opposite-sign (OS) L1 muons with
$\ptmu>4\GeV$ and $\drmumu<1.2$;
\item the presence of at least two OS L1 muons with $\ptmu>4.5\GeV$ and $\abs{\etamu}<2.0$; or
\item the presence of at least two L1 muons with $\ptmuone>15\GeV$ and $\ptmutwo>7\GeV$.
\end{itemize}
In 2023, two additional L1 triggers were added, requiring:
\begin{itemize}
\item the presence of at least two OS L1 muons with $\abs{\etamu}<1.4$ and $\drmumu<1.4$; or
\item the presence of at least two L1 muons with $\ptmu>8\GeV$.
\end{itemize}
The measurement of the muon momentum at the L1 trigger relies only on the information from the muon detectors.
After the L1 trigger preselections, only events containing two OS muons 
with $\ptmu>3\GeV$ and $\abs{\etamu}<2.4$ are retained at the HLT,
where the inner tracker information improves the muon \pt measurement and spatial resolution.

The efficiency of the scouting triggers is measured in data,
as a function  of the dimuon \lxy and of the minimum \ptmu of the pair,
using events with at least two muons selected with independent, orthogonal triggers based on calorimeter energy deposits.
The measured efficiency and its uncertainty are accounted for in the normalization of simulated events, as discussed in Section~\ref{sec:mc}.

Because only a reduced amount of information is available in the scouting data,
no additional calibration is performed for the muons used in this search.
The resulting dimuon mass resolution was measured to differ by less than 1\% from that obtained
using muons that undergo the full offline reconstruction~\cite{EXO-23-007},
which has a negligible effect on the results presented in this paper.

\section{Monte Carlo simulation}
\label{sec:mc}

The SM background is estimated directly from data in the dimuon (\mmumu) or four-muon ($m_{4\PGm}$) invariant mass spectrum,
parameterized by smooth analytical functions.
Thus, the search does not rely on MC simulation in order to estimate the background.

On the other hand, MC simulations of BSM signals are used to interpret the results of the search.
In particular, a model of Higgs boson production via the dominant mode of gluon-gluon fusion is simulated, with $\PH\to\PZ_{\mathrm{D}}\PZ_{\mathrm{D}}$,
where $\PZ_{\mathrm{D}}$ is a long-lived resonance and at least one $\PZ_{\mathrm{D}}$ decays into a pair of OS muons.
The phenomenology of this model is defined by four parameters:
\begin{itemize}
    \item the kinetic mixing $\epsilon$;
    \item the mass of the dark photon, $m_{\PZ_{\mathrm{D}}}$;
    \item the Higgs portal coupling, $\kappa$; and
    \item the mass of the dark Higgs boson, $m_{\PH_{\mathrm{D}}}$.
\end{itemize}
In the simulation used in this search, $m_{\PH_{\mathrm{D}}}$ and $\kappa$ are fixed to 400\GeV and 0.01, respectively,
in a regime where $\PH \to \PH_{\mathrm{D}}\PH_{\mathrm{D}}$ decays are kinematically forbidden.
Instead, the Higgs boson is set to always decay into a pair of $\PZ_{\mathrm{D}}$ bosons,
which in turn decay into SM fermions with branching fractions that depend on their mass~\cite{hahm_2015}.
The branching fraction for $\PZ_{\mathrm{D}}$ decays into muon pairs varies from about 50\% at a mass of 0.5\GeV to about 9\% at a mass of 50\GeV.
The lifetime of the $\PZ_{\mathrm{D}}$ boson for a given mass hypothesis is determined by $\epsilon$.
In this paper, dark photons are simulated with $c\tau_{0}^{\PZ_{\mathrm{D}}}=0.1$, 1, 10, and $100\cm$, except for masses below 1\GeV,
where only $c\tau_{0}^{\PZ_{\mathrm{D}}}=0.1$ and $1\cm$ scenarios are simulated to account for the variation in acceptance
due to the large Lorentz boost of the particles, which results in longer decay lengths for low-mass resonances.
A lifetime-based reweighting procedure is additionally employed to probe intermediate lifetime values.
The production of dark photons is modeled at leading order (LO) with \MGvATNLO~2.9.9~\cite{Alwall:2014hca}.
While only the dominant gluon-gluon fusion production mechanism is considered for the generation,
the Higgs boson production cross section is normalized to the most recent theoretical prediction for the sum of
all production modes for the mass of Higgs boson $m_\PH = 125\GeV$ at $\sqrt{s} = 13.6\TeV$,
which is 59.8\unit{pb}~\cite{Cepeda:2019klc, LHCHWG136TeV}.
Other Higgs boson production mechanisms are not considered in the generation for simplicity,
because they have negligible effects on the analysis acceptance.
The decays of dark photons are simulated using \PYTHIA~8.306~\cite{Bierlich:2022pfr}.
These samples are used to set upper limits on the branching fraction 
$\mathcal{B}(\PH\to\PZ_{\mathrm{D}}\PZ_{\mathrm{D}})$,
using values of $\mathcal{B}(\PZ_{\mathrm{D}}\to\PGm\PGm)$ from the model of Ref.~\cite{hahm_2015}.

Dark-shower signal models are simulated at LO with \PYTHIA~8.309 for selected mass combinations of Scenarios A and B1,
considering two dark flavors and three dark colors.
In these models, the SM Higgs boson is set to decay into a pair of dark partons $\PGY\PAGY$,
which then undergo showering and hadronization within the dark sector.
The Higgs boson production cross section is normalized to $59.8\unit{pb}$~\cite{Cepeda:2019klc, LHCHWG136TeV}.
The mass scale is set by assuming $\tilde{\Lambda} = m_{\PSGh} = 4 m_{\PSGp}$,
where $\tilde{\Lambda}$ is the dark-sector confinement scale.
The coupling of the dark photon to dark-sector quarks controls the isospin violation through the mixing angle $\theta$.
We consider $\sin\theta = 0.1$ to ensure that isospin breaking remains relatively small,
resulting in the dark pions having nearly degenerate masses.
The branching fraction $\mathcal{B}(\PSGp_3 \to \PApr\PApr)$ is set to unity.
Samples are generated with different mass hypotheses for the $\PSGp_3$ and the $\PApr$ particles,
and with mean proper decay lengths ranging from $0.01$ to $100\cm$ for $\PApr$ ($\PSGp_3$) in Scenario A (B1).
These samples are used to set upper limits on the branching fraction $\mathcal{B}(\PH\to\PGY\PAGY)$.

For models where a long-lived scalar resonance \PGf arises from the decay of a \PQb hadron,
the production of \PQb hadrons is initially generated at LO using \PYTHIA~8.306,
where the \PGf boson decays into a pair of OS muons with 100\% probability.
Subsequently, a fixed-order calculation at next-to-leading logarithmic (FONLL) precision in the strong coupling constant $\alpha_\mathrm{S}$
is used to normalize the signal and to reweight the \pt spectrum of \PQb hadrons.
These samples are used to set upper limits on the branching fraction product $\mathcal{B}(\Ph_\PQb \to \PGf X)\mathcal{B}(\PGf\to\PGm\PGm)$.

The generators use the CP5 tune~\cite{Sirunyan:2019dfx} and 
the NNPDF3.1 sets of parton distribution functions (PDFs)~\cite{Ball:2017nwa} at next-to-next-to-LO.
The interactions of particles with the CMS detector are simulated using \GEANTfour~\cite{Agostinelli:2002hh}.
Simulated minimum bias events are superimposed on the hard $\Pp\Pp$ interaction to describe
the effect of additional inelastic interactions within the same or neighboring bunch crossings (pileup).
The simulation is reweighted to match the distribution of the number of interactions per bunch crossing observed in data.
All simulated events are reconstructed with the same algorithms used for data.

\section{Event selection and categorization}
\label{sec:selection}

\subsection{Event selection}
\label{sec:evtsel}

The events collected using the scouting triggers (Section~\ref{sec:triggers})
are further selected to contain at least one pair of OS muons associated with an SV
that satisfies a number of quality criteria:
\begin{itemize}
\item $\lxy < 70\cm$;
\item $x$ coordinate uncertainty $<$0.05\cm;
\item $y$ coordinate uncertainty $<$0.05\cm;
\item $z$ coordinate uncertainty $<$0.10\cm;
\item The SV fit $\chi^{2}$ normalized to the number of degrees of freedom (dof) $\left(\chi^{2}/\mathrm{dof}\right)_\mathrm{SV} < 3$.
\end{itemize}
The $\left(\chi^{2}/\mathrm{dof}\right)_\mathrm{track}$ of each muon track is also required to be less than 3.
Furthermore, any SV located within a tracker module is not considered in order to suppress the contribution from material interaction vertices,
\ie, vertices from particles produced in the interactions with the detector material.

A large fraction of events contain only one OS muon pair and an associated SV.  
Among the events with more than one pair of OS muons, 
the pairs are ordered according to increasing $(\chi^{2}/\mathrm{dof})_\mathrm{SV}$
and only the first or the first two pairs are used in this search.
If two muon pairs are reconstructed, the positions of the corresponding SVs may be spatially resolved or may overlap within uncertainties,
thus appearing as a single four-muon ($4\PGm$) vertex.
In this case, the average position of the two overlapping SVs is used as the position of the $4\PGm$ vertex,
and the selections are adapted accordingly, as detailed in the following.

For decays of BSM particles into a single muon pair, we expect the transverse momentum of the dimuon system, \ptmumuvec, 
to be collinear with the SV vector, which is defined as the vector \vecdv connecting the PV to the SV. 
This does not apply to a muon pair where the two muons originate from different vertices
and erroneously form a common SV because their trajectories accidentally cross.
In order to suppress this background, we require both the azimuthal angle ($\abs{\dphimudv}$)
and the three-dimensional angle between the two vectors to be less than $\pi/2$.
This requirement is used in events with one or two resolved SVs.
In events where four muons are selected from two SVs that overlap in space within uncertainties, hereafter referred to as \textit{overlapping}~SVs,
we require the four-muon transverse momentum, $\vec{p}_\mathrm{T}^{4\PGm}$, to be collinear with
the vector connecting the PV with the overlapping SVs.
The same requirements on the azimuthal and three-dimensional angles are applied.

Similarly, the opening angle between two muons erroneously forming a common SV tends to be very large, independently of the boost of the dimuon system.
Hence, in order to further reduce the background from events with an erroneous muon SV association, we also require both the azimuthal and three-dimensional angles between the muons to be $<$2.8.
The main goal of such selections is to suppress background where SVs are formed by cosmic ray muons, muons from pileup interactions,
and muons from back-to-back quantum chromodynamics dijet events or Drell--Yan events. 

Muons from pileup vertices can overlap in the $r$-$\phi$ plane, despite being separated in $r$-$z$ at their origin, and,
as a consequence, they can be wrongly associated with a common SV.
These muon pairs are characterized by a large \detamumu and a small \dphimumu.
We exploit this feature to suppress this background contribution with the 
requirement $\logabsetaphi<1.25$.
This requirement is not applied in cases with two selected muon pairs.

Additional background may arise from ``prompt'' (\ie, nondisplaced) muons.
Muons from truly displaced vertices do not leave hits in the tracker layers between the interaction point and the SV.
Therefore, this background can be suppressed by applying a selection on the number of hits associated with each muon between the PV and the SV,
referred to as $N_\text{excess}^{\PGm\PGm}$:
\begin{itemize}
\item $N_\text{excess}^{\PGm\PGm} = 0$ for $\lxy < 11\cm$;
\item $N_\text{excess}^{\PGm\PGm} \leq 1$ for $\lxy \in [11, 16]\cm$;
\item $N_\text{excess}^{\PGm\PGm} \leq 2$ for $\lxy \in [16, 70]\cm$.
\end{itemize}
The requirement is loosened for higher displacements, because at large values of $\lxy$ it is more likely that random hits are wrongly associated with displaced muon tracks,
and is fully relaxed in events where four muons from two overlapping SVs are selected.

In order to identify SVs of interest,
the muon track transverse impact parameter, \dxy, is also used.
Specifically, we select events based on the \dxy significance ($\abs{\dxy}/\sigmadxy$) and the lifetime-scaled \dxy (\scaleddxy).
The first expression normalizes \dxy to its measurement uncertainty (\sigmadxy),
while the second normalizes \dxy to the lifetime corresponding to the observed SV transverse displacement. 
This allows the same selection requirements to be used for all lifetime hypotheses.
We require each muon in dimuon events to satisfy: 
\begin{itemize}
\item $\abs{\dxy}/\sigmadxy > 2$;
\item $\scaleddxy > 0.1$.
\end{itemize}
In events with two muon pairs, where there are few background events, the selected muons are only required to have $\abs{\dxy}/\sigmadxy > 2$.

Finally, in events with two muon pairs from spatially resolved SVs,
the two dimuon invariant masses are further required to differ by less than $5\%$ of their average.

The muon isolation is measured as the scalar \pt sum of the additional PF candidates
within a cone of $\Delta R = \sqrt{\smash[b]{(\Delta\eta)^2+(\Delta\phi)^2}} = 0.4$ around its direction,
including corrections for the contributions from pileup.
A muon is considered isolated if its isolation is less than 8\GeV and less than 20\% of its \pt.
These isolation criteria are only used to categorize events with only one selected muon pair, as described in Section~\ref{sec:strategy}.

\subsection{Event categorization}
\label{sec:strategy}

Once the objects have been selected, three types of events are considered in the search:
\begin{itemize}
  \item events with one muon pair;
  \item events with two muon pairs from two SVs whose positions are spatially resolved; and
  \item events with two muon pairs from two SVs whose positions overlap within uncertainties,
    consistent with a single $4\PGm$ vertex.
\end{itemize}

Because $4\PGm$ events are expected to provide sensitivity to a range of BSM scenarios and are depleted in background, we first select events with two muon pairs.
If an event contains four muons from two resolved SVs,
we then assign the event to the so-called \textit{multivertex} category.
If instead we identify four muons from two overlapping SVs, we assign the event to the \textit{overlapping vertex} category.
In both cases, the signal would appear as a narrow peak in the $m_{4\PGm}$ spectrum.

If only one muon pair is found, the event is categorized as a dimuon event.
For dimuon events, we first define two orthogonal regions in $\abs{\dphimudv}$,
distinguishing \textit{pointing} and \textit{nonpointing} topologies,
with $\abs{\dphimudv} < 0.02$ and $0.02 < \abs{\dphimudv} < \pi/2$, respectively.

Dimuon events in both the pointing and nonpointing categories are then further split according to $\lxy$,
to maximize the sensitivity to different lifetime hypotheses.
We define eight bins in $\lxy$, following the detector geometry:
\begin{itemize}
      \item within the beam pipe: [0, 0.2]\cm; [0.2, 1]\cm; [1, 2.4]\cm;
      \item between the beam pipe and the first pixel tracker layer: [2.4, 3.1]\cm;
      \item between the first and second pixel tracker layers: [3.1, 7]\cm;
      \item between the second and third pixel tracker layers: [7, 11]\cm;
      \item between the third and fourth pixel tracker layers: [11, 16]\cm; and
      \item outside the pixel tracker: [16, 70]\cm.
\end{itemize}
The last two \lxy intervals probe phase space that was inaccessible to previous results using scouting data~\cite{EXO-20-014}.

\begin{figure}[htb]
  \centering
  \includegraphics[width=.49\textwidth]{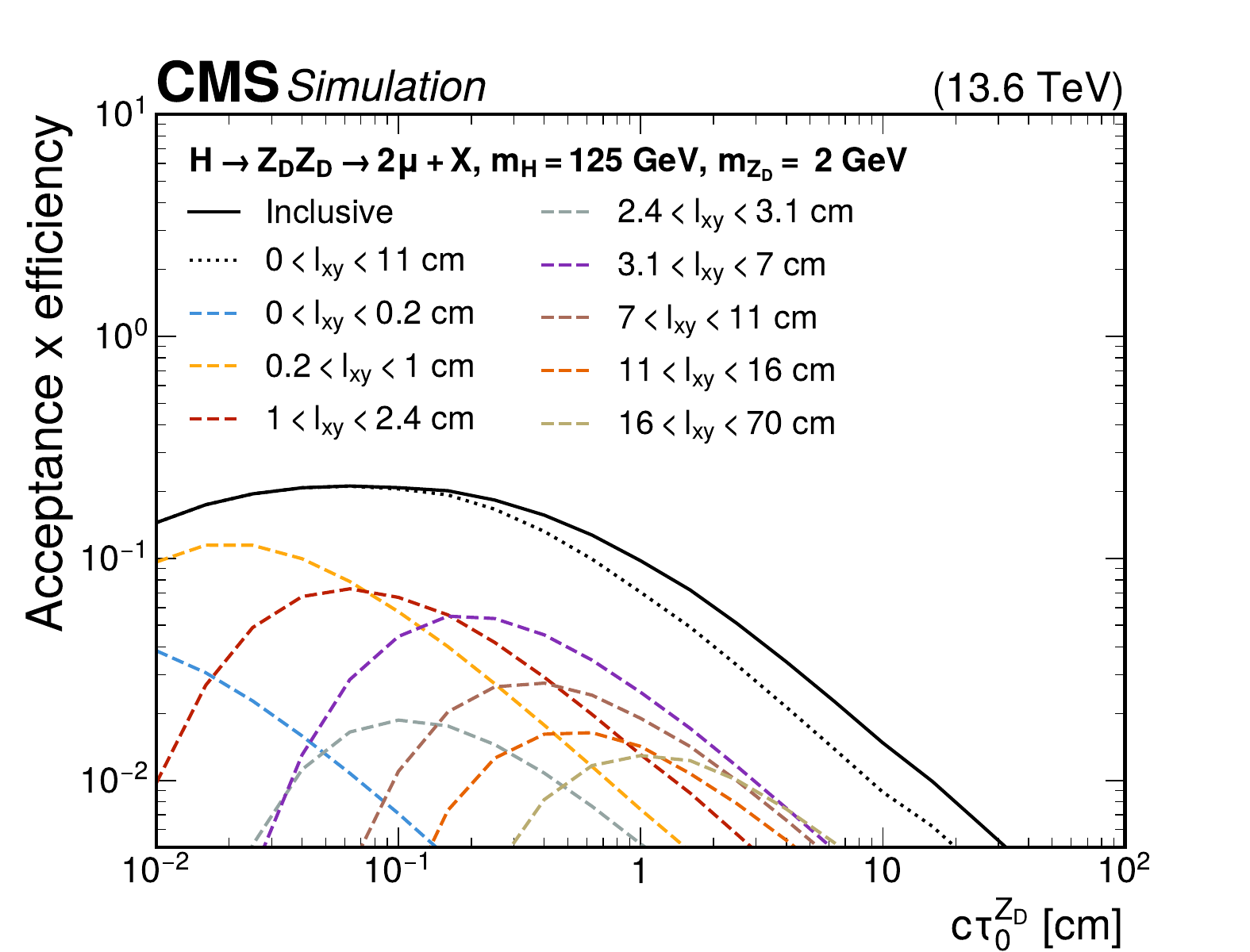}
  \includegraphics[width=.49\textwidth]{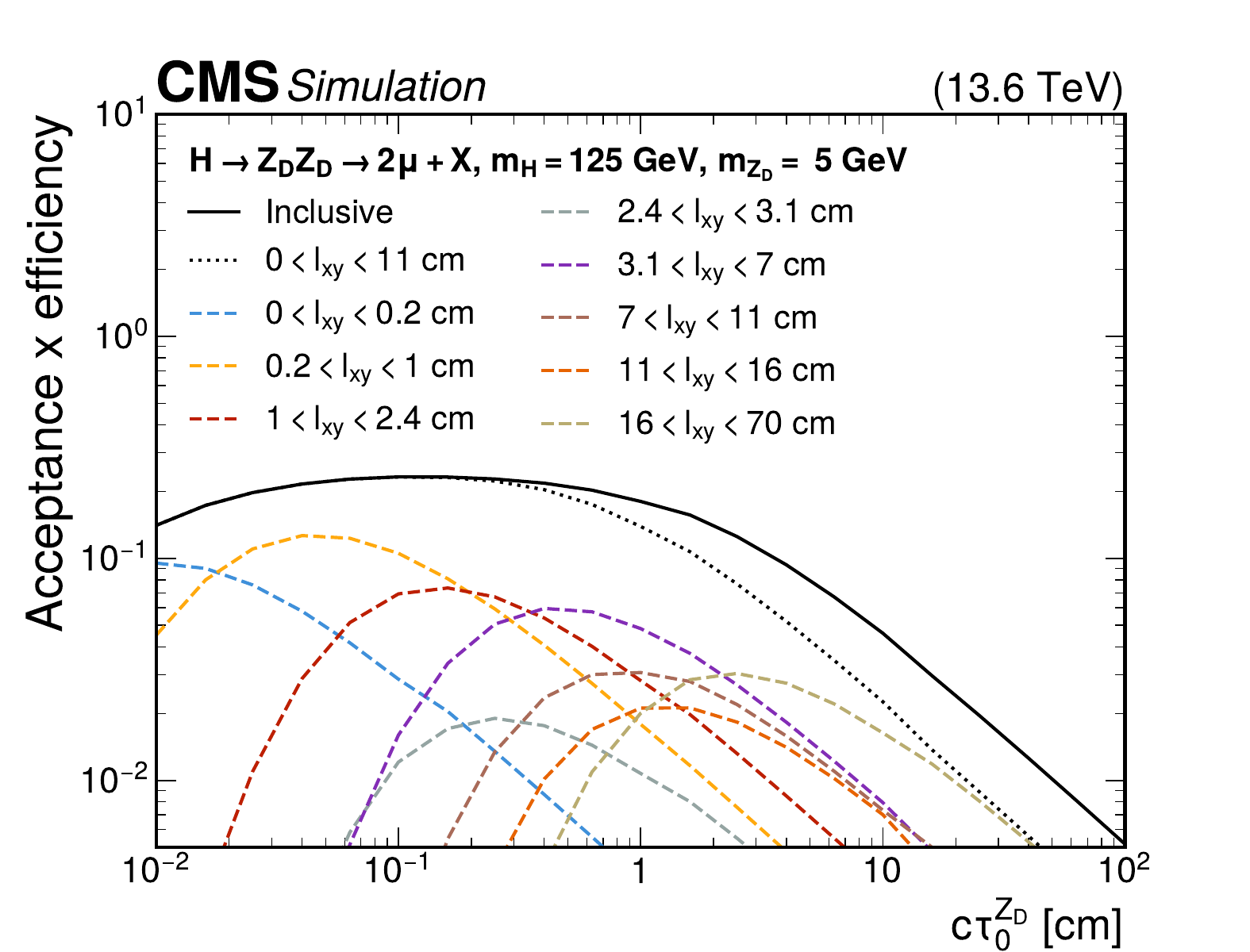}
  \includegraphics[width=.49\textwidth]{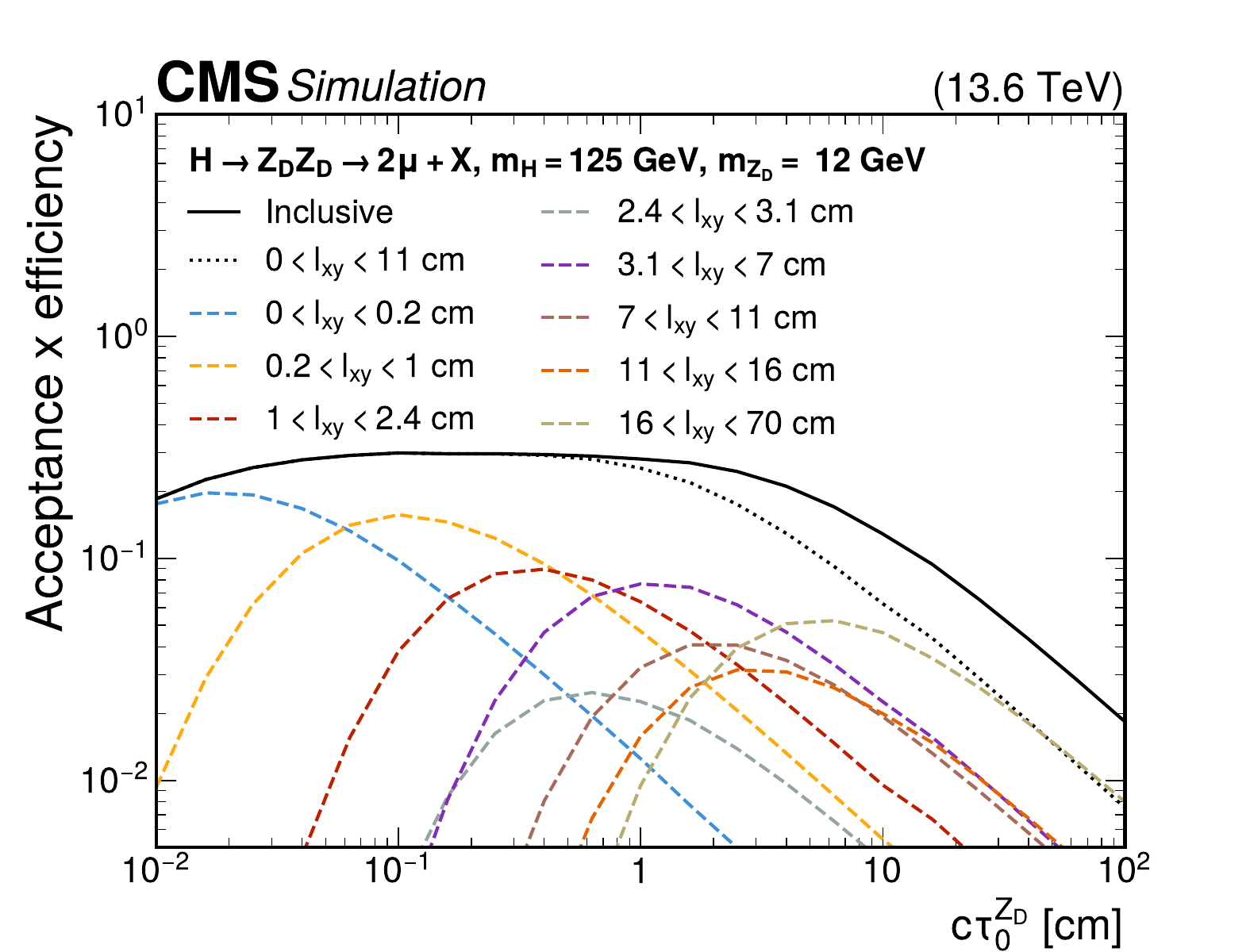}
  \includegraphics[width=.49\textwidth]{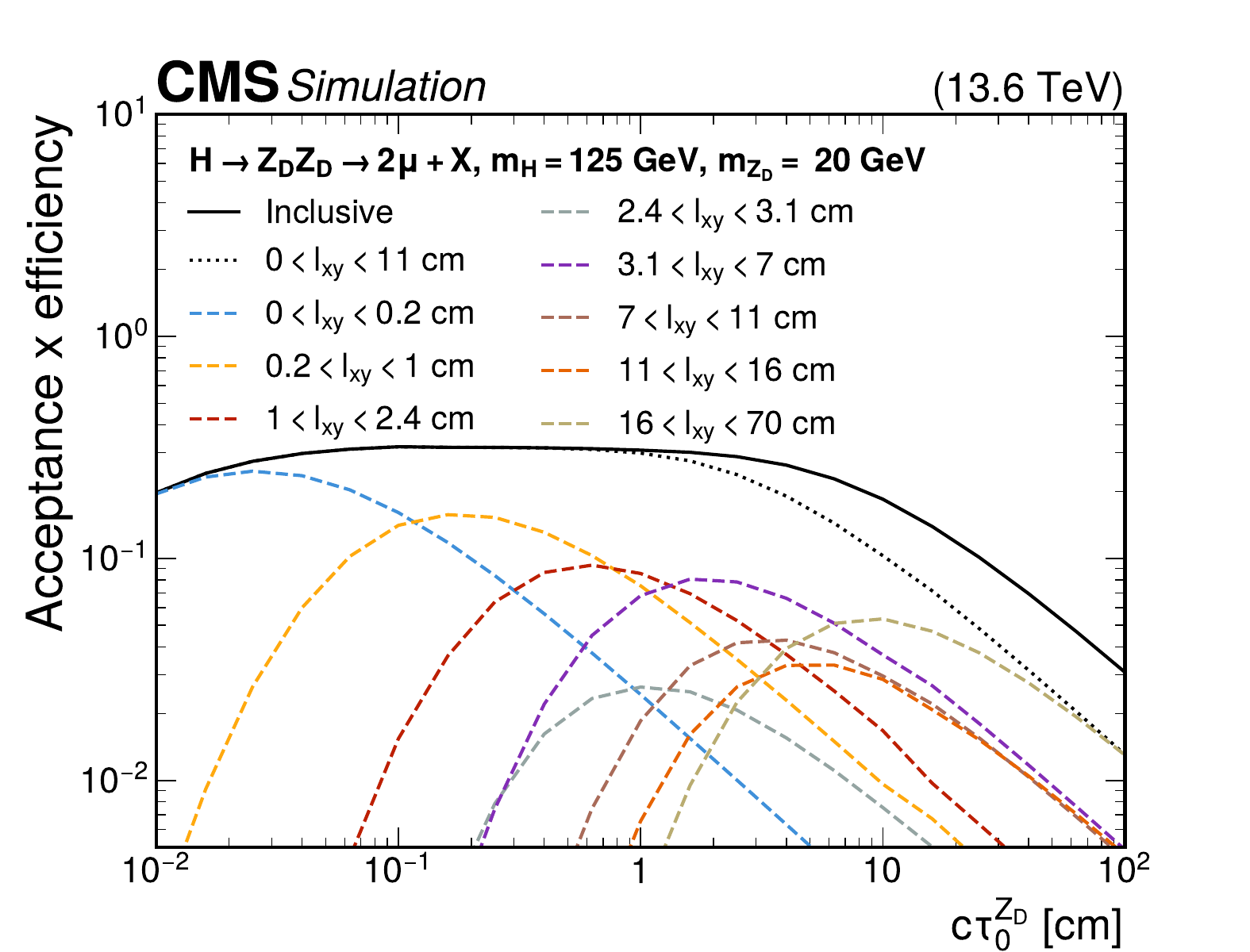}
  \caption{
    Overall signal acceptance times efficiency ($\mathcal{A}\varepsilon$) as measured for the HAHM signal model with $m_{\PZ_{\mathrm{D}}} = 2, 5, 12$, and $20\GeV$,
    as a function of the mean proper decay length of the dark photon, $c\tau_{0}^{\PZ_{\mathrm{D}}}$.
    The signal $\mathcal{A}\varepsilon$ is defined as the fraction of events with at least one dark photon decaying into a muon pair,
    reconstructed as a dimuon candidate within
    the \lxy bins of $[0, 0.2, 1, 2.4, 3.1, 7, 11, 16, 70]\cm$.
    The average $\mathcal{A}\varepsilon$ over the simulated samples corresponding to the 2022 and 2023 data-taking periods is shown.
  }
  \label{fig:signalefficiency}
\end{figure}

The overall signal acceptance times efficiency ($\mathcal{A}\varepsilon$) in the different \lxy categories is shown in Fig.~\ref{fig:signalefficiency}
for representative HAHM dark photon mass hypotheses, $m_{\PZ_{\mathrm{D}}} = 2, 5, 12$ and $20\GeV$,
as a function of the dark photon mean proper decay length, $c\tau_{0}^{\PZ_{\mathrm{D}}}$.
The signal $\mathcal{A}\varepsilon$ is defined as the fraction of events with at least one dark photon decaying into a muon pair,
reconstructed as a single dimuon candidate within the specified displacement range.
It is a luminosity-weighted average over the 2022 and 2023 data-taking periods.
The increase in acceptance resulting from the newly accessible $11 < \lxy < 70\cm$ interval is apparent for all $m_{\PZ_{\mathrm{D}}}$ hypotheses,
extending the analysis sensitivity up to $c\tau_{0}^{\PZ_{\mathrm{D}}} \gtrsim 1\unit{m}$,
and enhancing it for $c\tau_{0}^{\PZ_{\mathrm{D}}} \gtrsim 10\unit{mm}$ at small $m_{\PZ_{\mathrm{D}}}$.

Within each displacement interval,
the events belonging to the pointing category are also categorized according to muon isolation,
distinguishing between isolated and nonisolated topologies, and $\ptmumu$.
Isolated topologies consist of events where both muons are isolated;
nonisolated topologies consist of events where at least one of the two muons fails the isolation criteria defined in Section~\ref{sec:evtsel}.
The splitting in isolation aims at covering scenarios where the long-lived particle is created with hadronic activity in its vicinity, 
such as scalars produced in \PQb hadron decays, and those where the displaced muons are produced alone, such as in $\PZ_{\mathrm{D}}\to\PGm\PGm$ decays.
Two categories are defined in $\ptmumu$:
\begin{itemize}
    \item $\ptmumu < 25\GeV$;
    \item $\ptmumu > 25\GeV$.
\end{itemize}
The categorization in $\ptmumu$ aims at covering a wide spectrum of possible production scenarios,
depending on the particles involved in the process and on their multiplicity.
For example, dimuon pairs produced from a decay of a light particle, e.g., a \PQb hadron,
have lower $\ptmumu$ than others originating from a much heavier particle, such as a Higgs boson.

A total of 40 dimuon categories are defined, which are considered together with two four-muon categories.
In each event category, we define mass windows that slide along the multimuon invariant mass spectrum (as described in Section~\ref{sec:fit})
and perform a search for a resonant peak within each mass window.
Mass regions corresponding to known SM resonances decaying either into muon pairs or in a pair of charged hadrons are not considered, \ie, they are ``masked''.
The masked resonances are listed in Table~\ref{tab:maskedregions}.
For events with four selected muons, if any of the muon pairs is found within a masked region, this pair is rejected and the event is treated as a dimuon event.

\begin{table}[htbp!]
\caption{
  List of known SM resonances and corresponding masked mass windows, equal to $\pm5\sigma_{\text{mass}}$ around the mean mass,
  where mean and resolution ($\sigma_{\text{mass}}$) are determined from a fit to data.
  \label{tab:maskedregions}
}
\centering
\newcolumntype{.}{D{.}{.}{2.6}}
\newcolumntype{x}{D{.}{ }{2.4}}
\begin{tabular}{l.xrr@{ }l}
Resonance & \multicolumn{1}{c}{Mean mass [\GeVns{}]} & \multicolumn{1}{c}{$\sigma_{\text{mass}}$ [\MeVns{}]} &
\multicolumn{3}{c}{Window [\GeVns{}]} \\
& & &
\multicolumn{3}{c}{(mean $\pm$ $5\sigma_{\text{mass}}$)} \\
\hline
$\PKzS$                   & 0.46  & 9.  & & [0.41, & 0.50] \\
$\PGh$                    & 0.55  & 8.  & & [0.51, & 0.59] \\
$\PGrzP{770}/\PGoP{782}$ & 0.78  & 11. & & [0.73, & 0.83] \\
$\PGf(1020)$              & 1.02  & 12. & & [0.96, & 1.08] \\
$\PJGy$                   & 3.09  & 35. & & [2.91, & 3.27] \\
$\PGyP{2S}$               & 3.68  & 42. & & [3.47, & 3.89] \\
$\PGUP{1S}$               & 9.45  & 93. & & [8.99, & 9.91] \\
$\PGUP{2S}$               & 10.10 & 91. & & [9.64, & 10.56] \\
$\PGUP{3S}$               & 10.34 & 87. & & [9.90, & 10.78] \\
\end{tabular}
\end{table}

\begin{figure}[tbp]
  \centering
  \includegraphics[width=.49\textwidth]{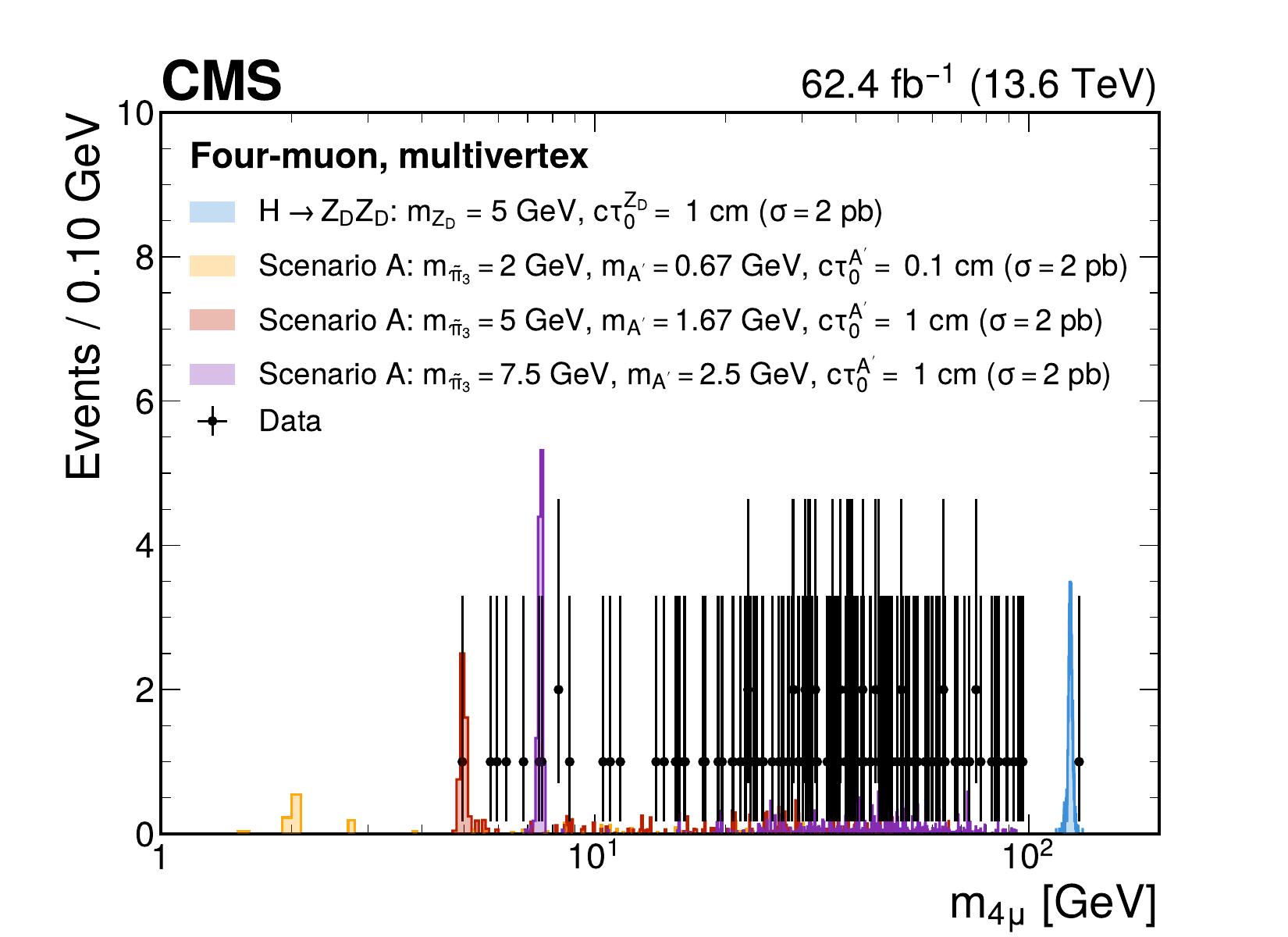}
  \includegraphics[width=.49\textwidth]{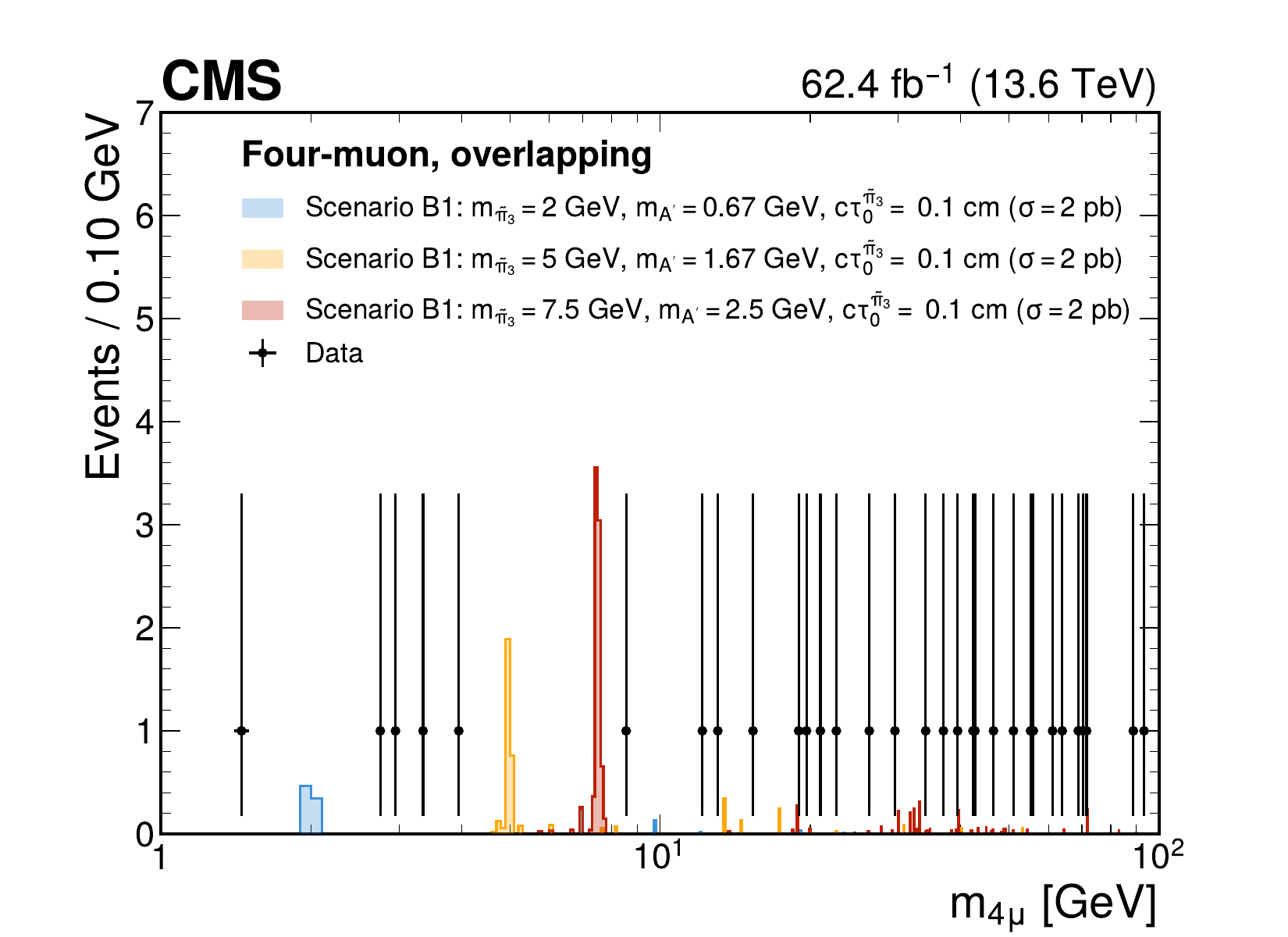}
  \caption{
    Four-muon invariant mass distributions of selected events in the multivertex (left)
    and overlapping vertex (right) categories,
    for data (black points with uncertainties) and representative signal models (colored histograms),
    assuming a cross section $\sigma = 2\unit{pb}$ for all models, for visualization purposes.
  }
  \label{fig:4mu}
\end{figure}

\begin{figure}[tbp]
  \centering
  \includegraphics[width=.49\textwidth]{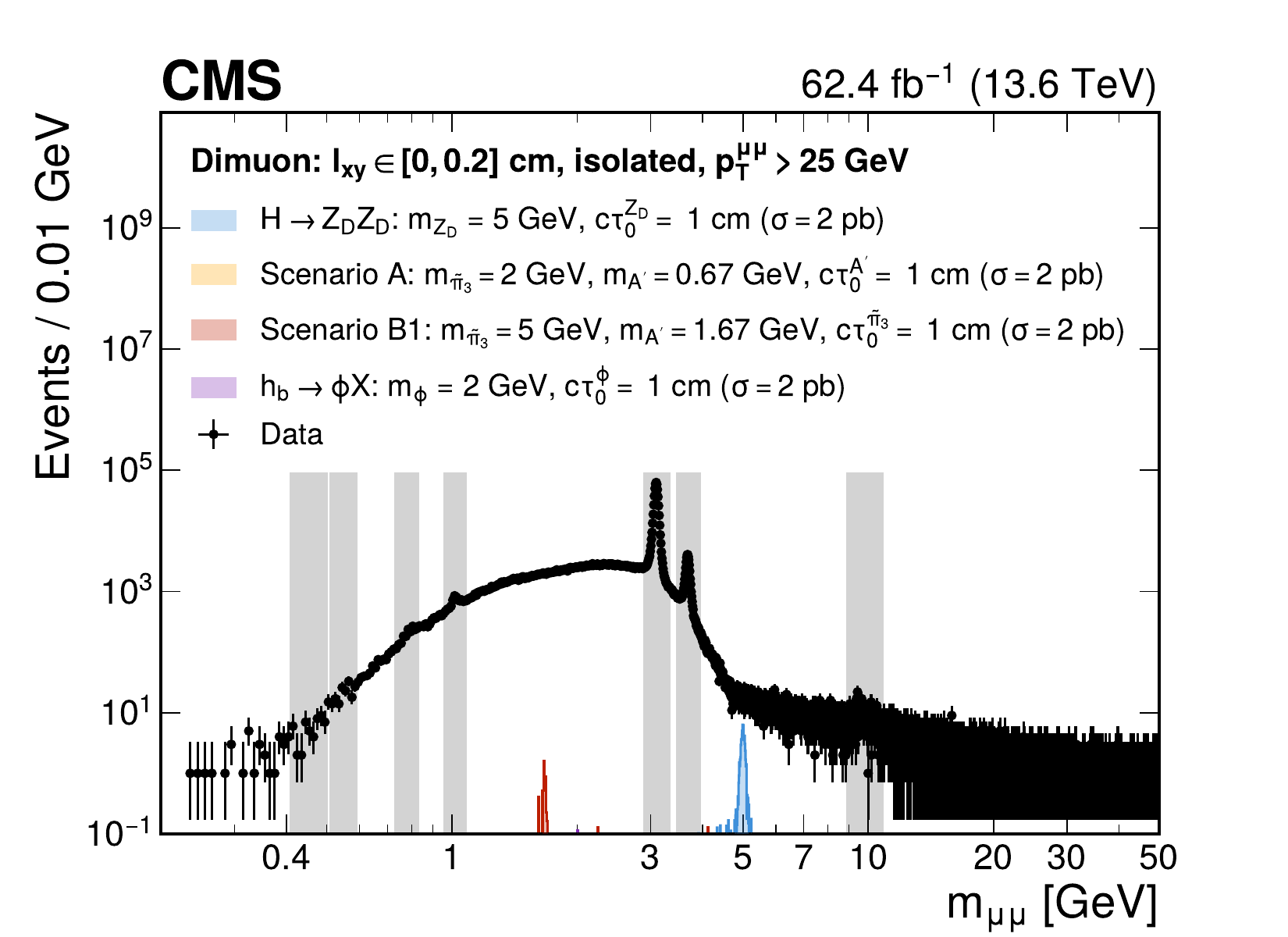}
  \includegraphics[width=.49\textwidth]{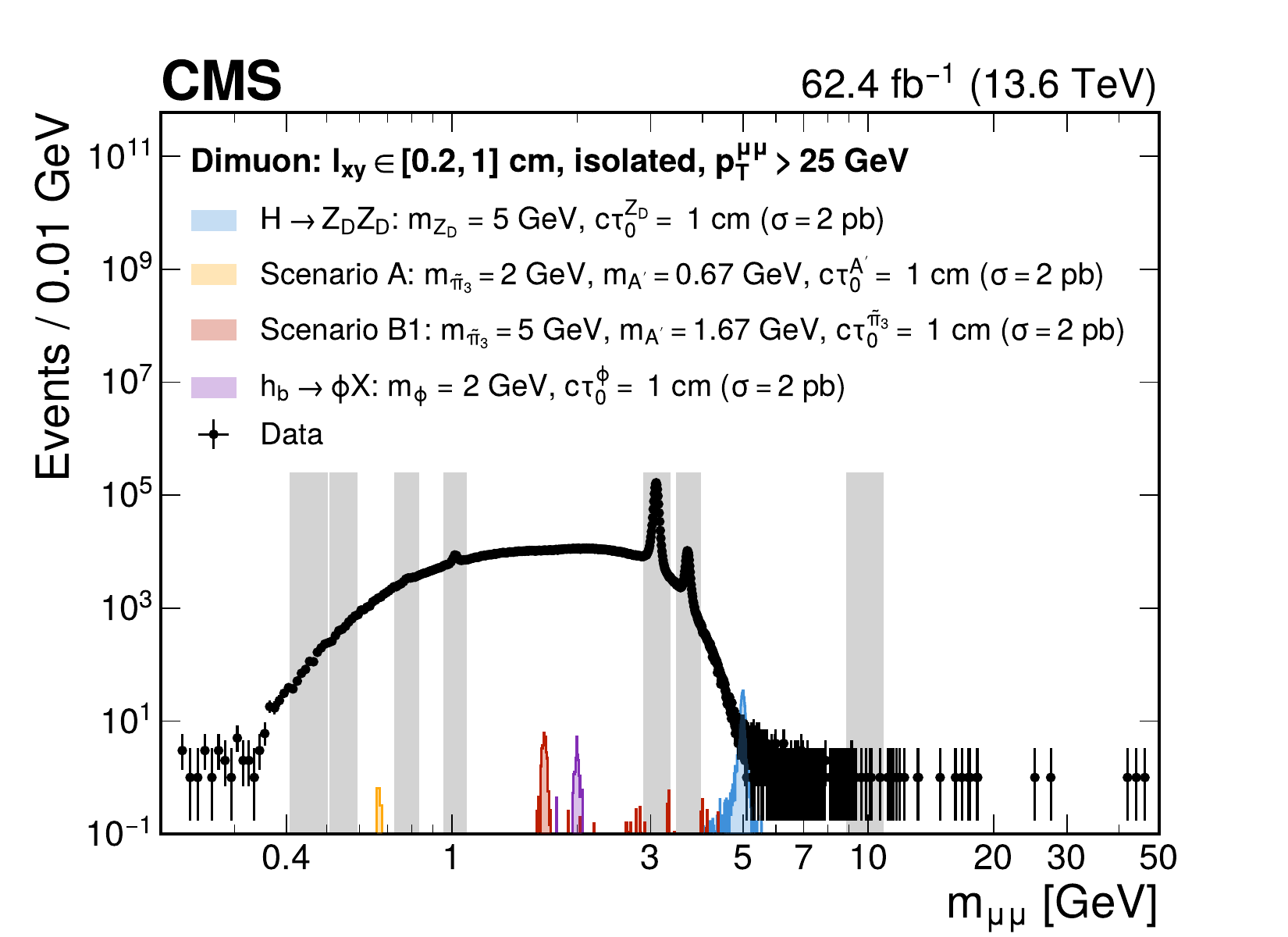}
  \includegraphics[width=.49\textwidth]{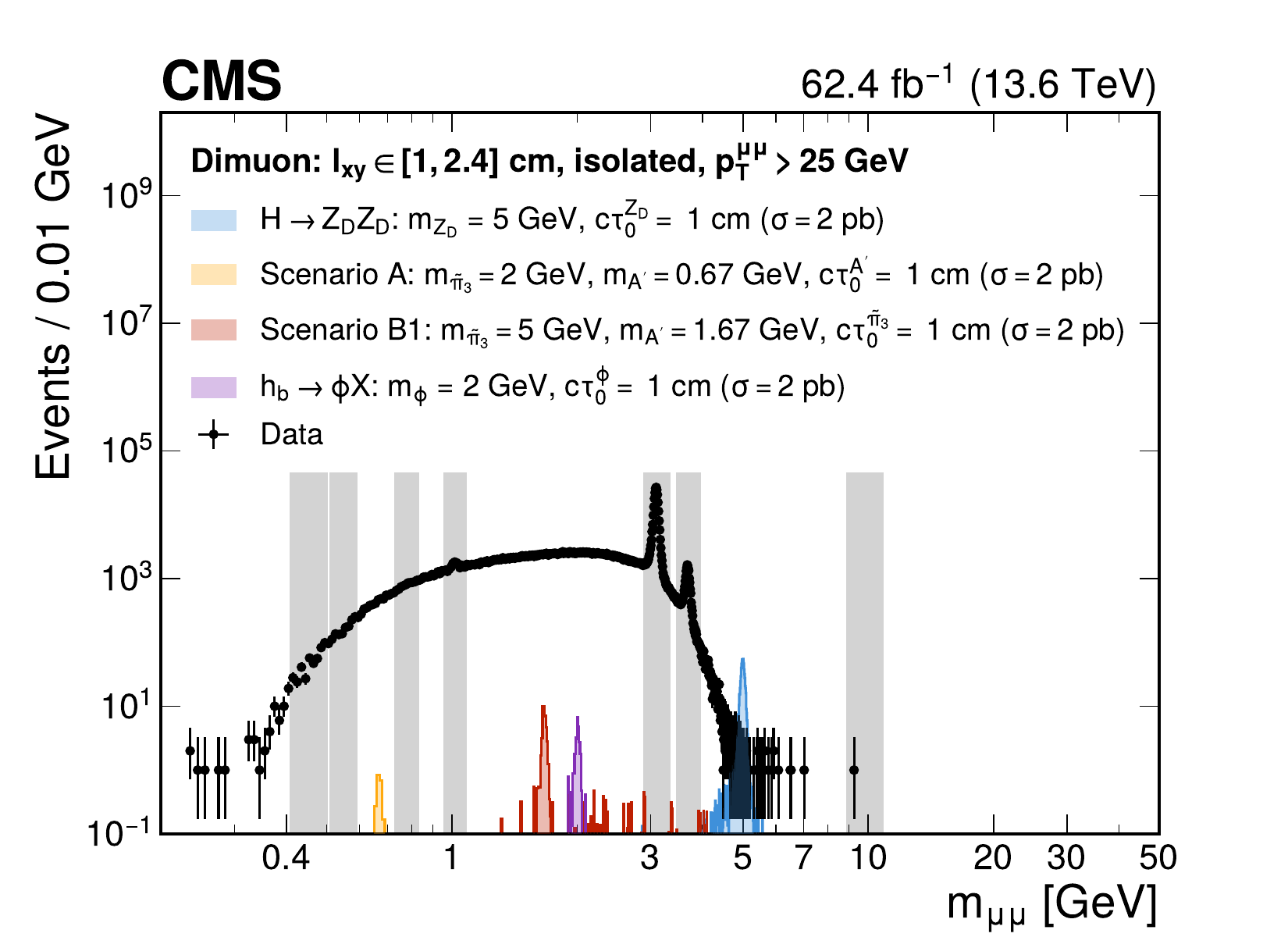}
  \includegraphics[width=.49\textwidth]{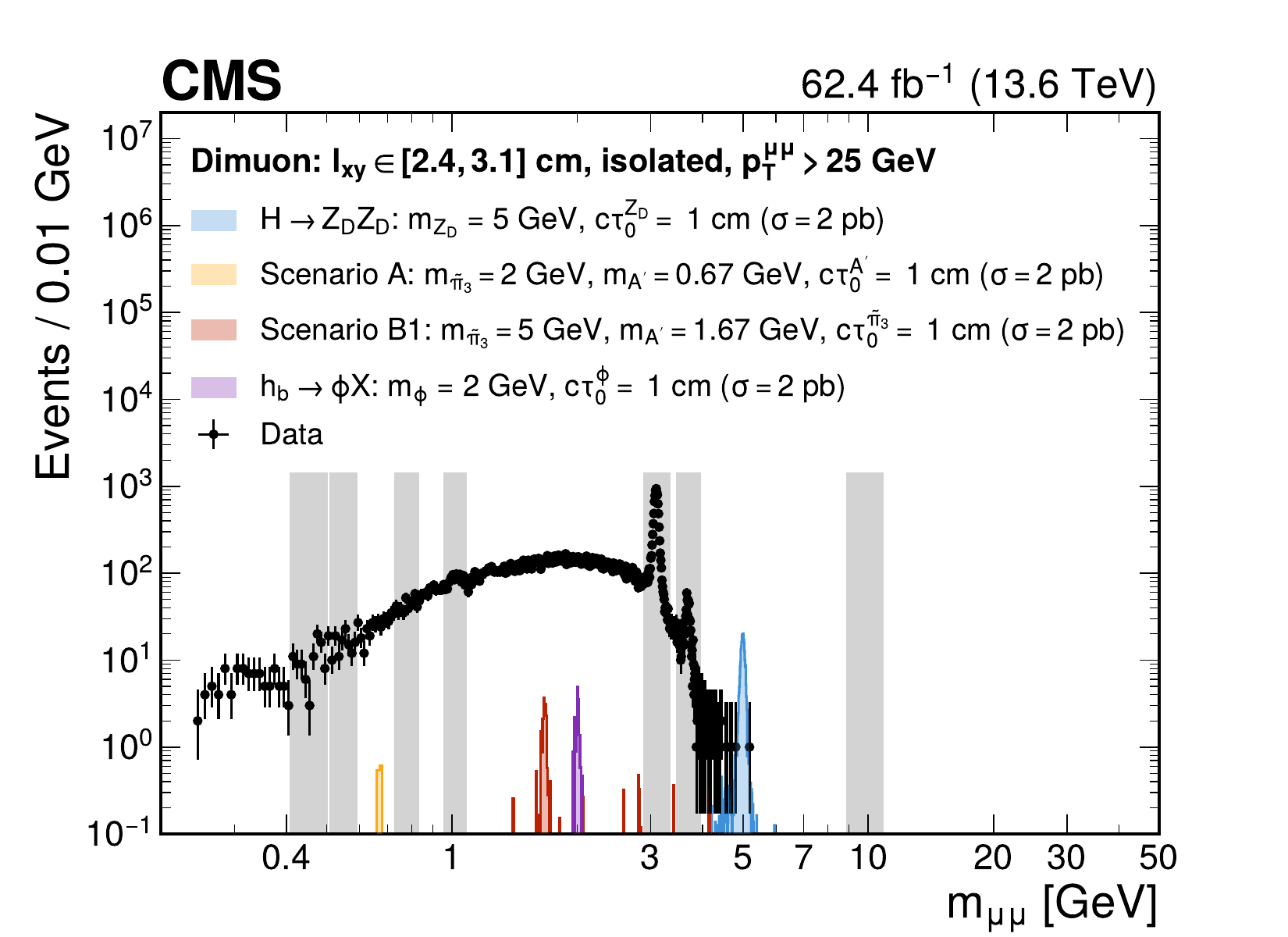}
  \includegraphics[width=.49\textwidth]{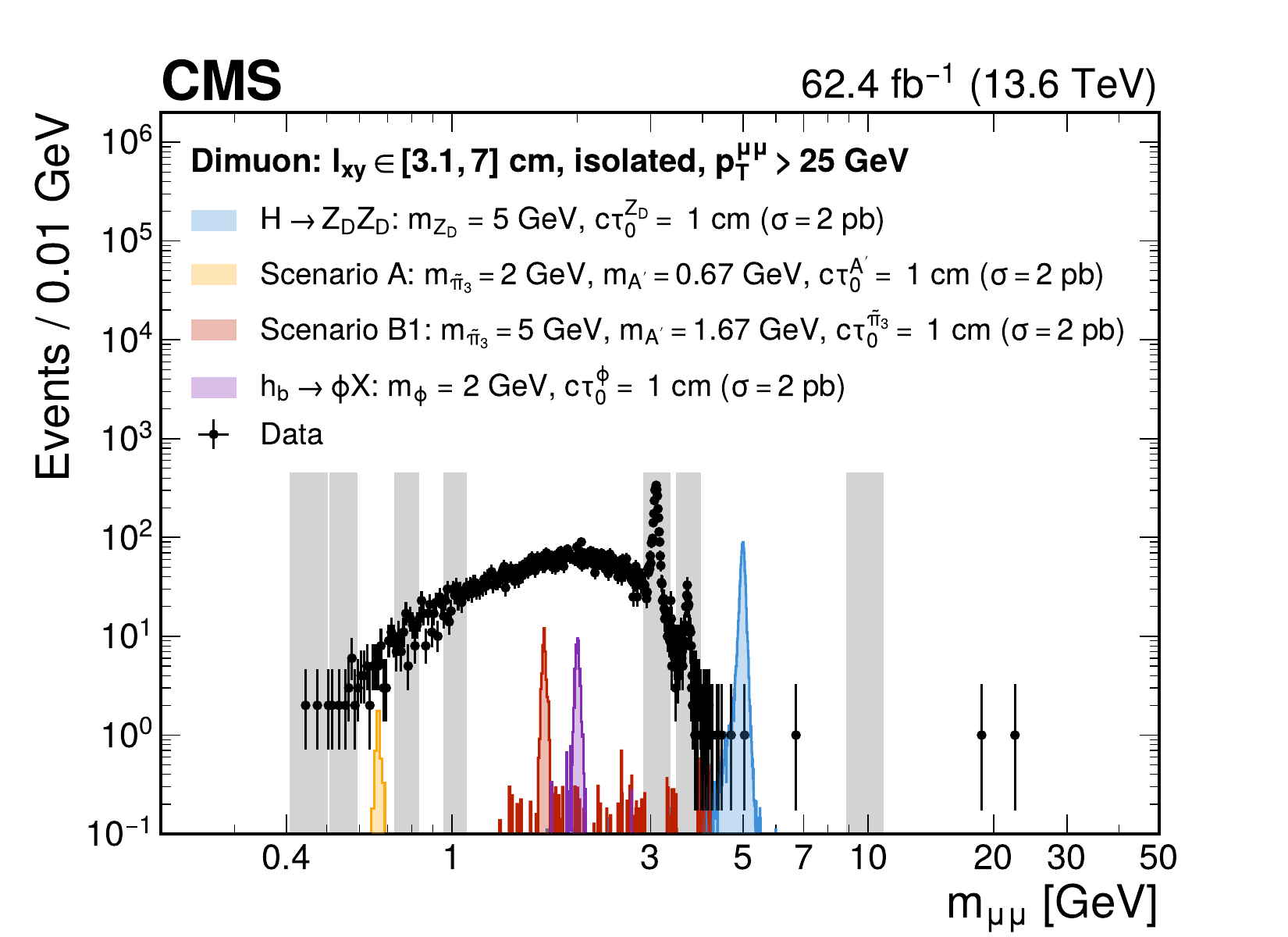}
  \includegraphics[width=.49\textwidth]{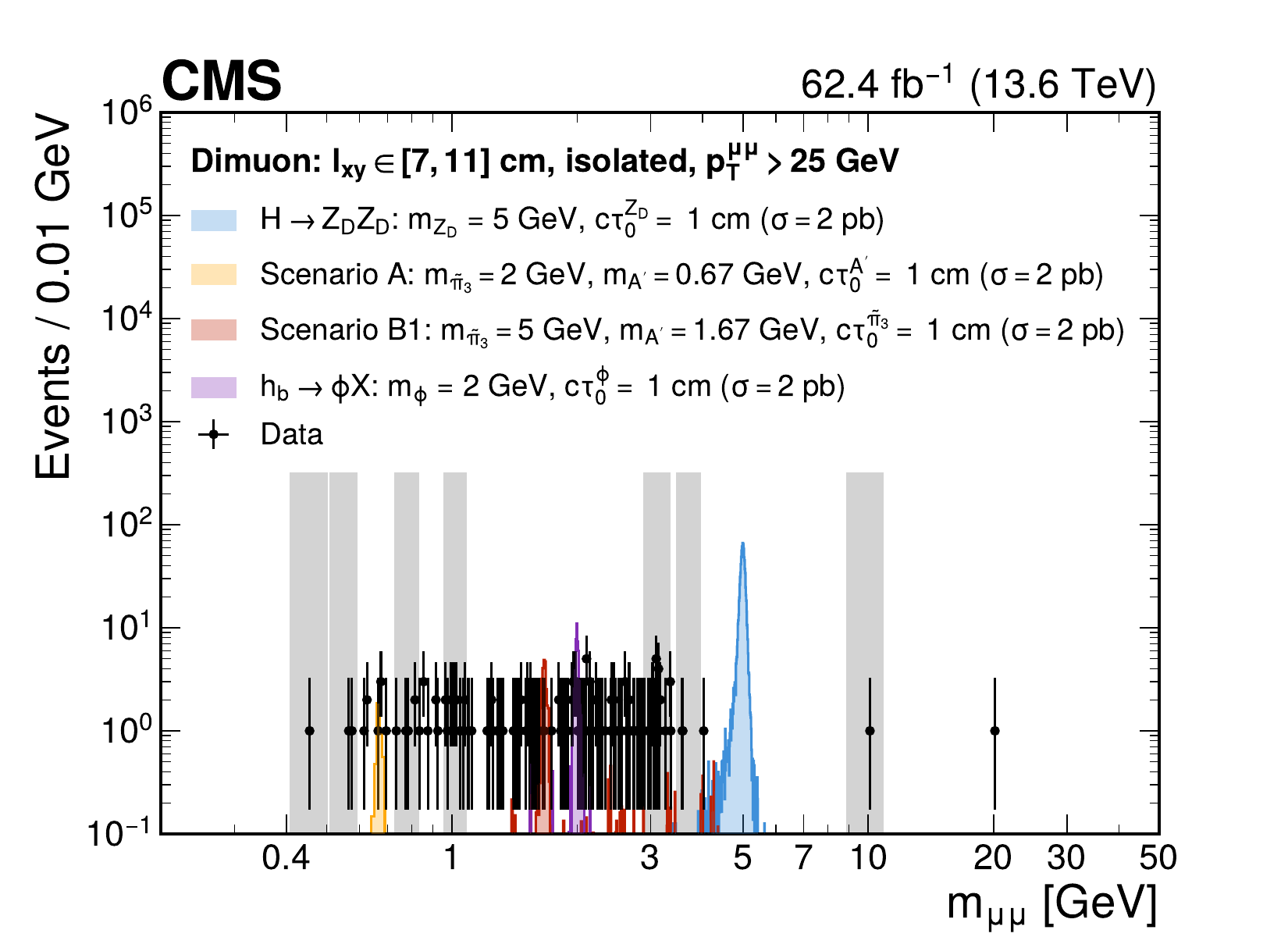}
  \caption{
    Dimuon invariant mass distributions of selected events with a single, isolated muon pair with $\ptmumu > 25\GeV$.
    Distributions are shown separately for each of the six $\lxy<11\cm$ bins.
    The vertical gray bands indicate mass ranges containing known SM resonances, which are masked for the purpose of this search.
    Distributions are shown as obtained in data (black points with uncertainties)
    and for representative signal models (colored histograms),
    assuming a cross section $\sigma = 2\unit{pb}$ for all models, for visualization purposes.
  }
  \label{fig:above}
\end{figure}

\begin{figure}[tbp]
  \centering
  \includegraphics[width=.49\textwidth]{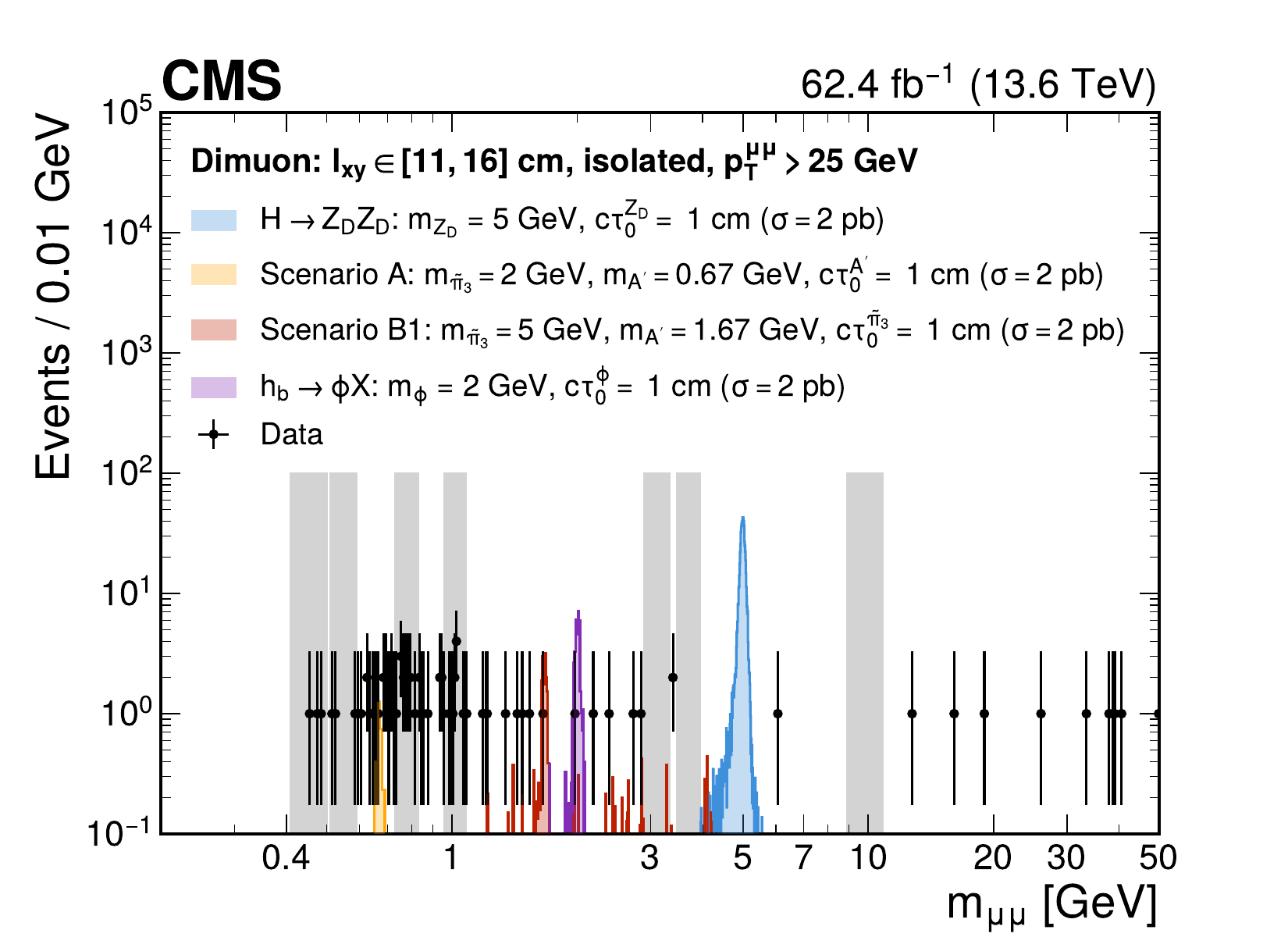}
  \includegraphics[width=.49\textwidth]{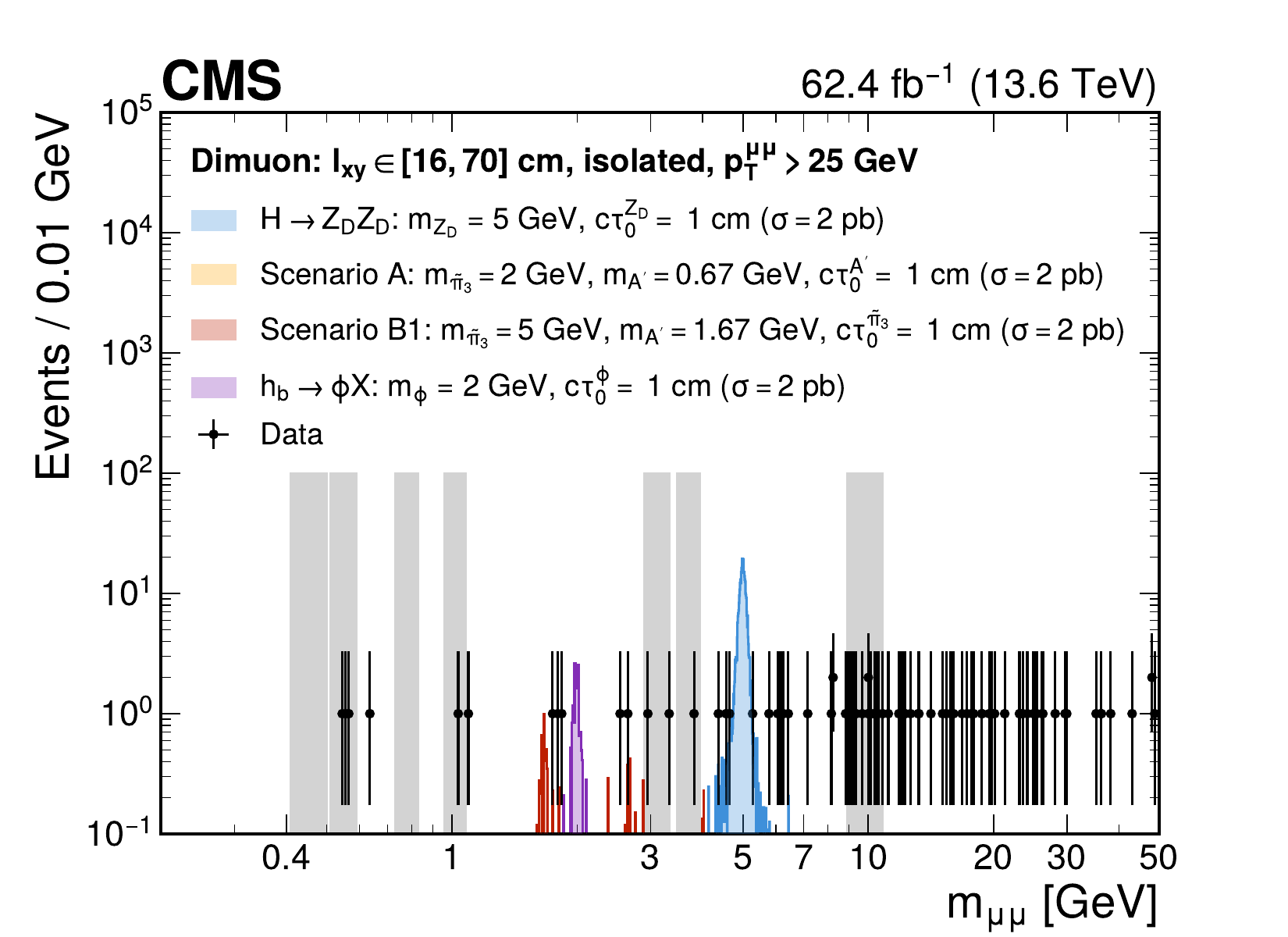}
  \caption{
    Dimuon invariant mass distributions of selected events with a single, isolated muon pair with $\ptmumu > 25\GeV$.
    Distributions are shown separately for each of the two $\lxy>11\cm$ bins
    (\ie, for the displacement range inaccessible to the previous results presented in Ref.~\cite{EXO-20-014}).
    The vertical gray bands indicate mass ranges containing known SM resonances, which are masked for the purpose of this search.
    Distributions are shown as obtained in data (black points with uncertainties)
    and for representative signal models (colored histograms),
    assuming a cross section $\sigma = 2\unit{pb}$ for all models, for visualization purposes.
  }
  \label{fig:aboveadd}
\end{figure}

\begin{figure}[tbp]
  \centering
  \includegraphics[width=.49\textwidth]{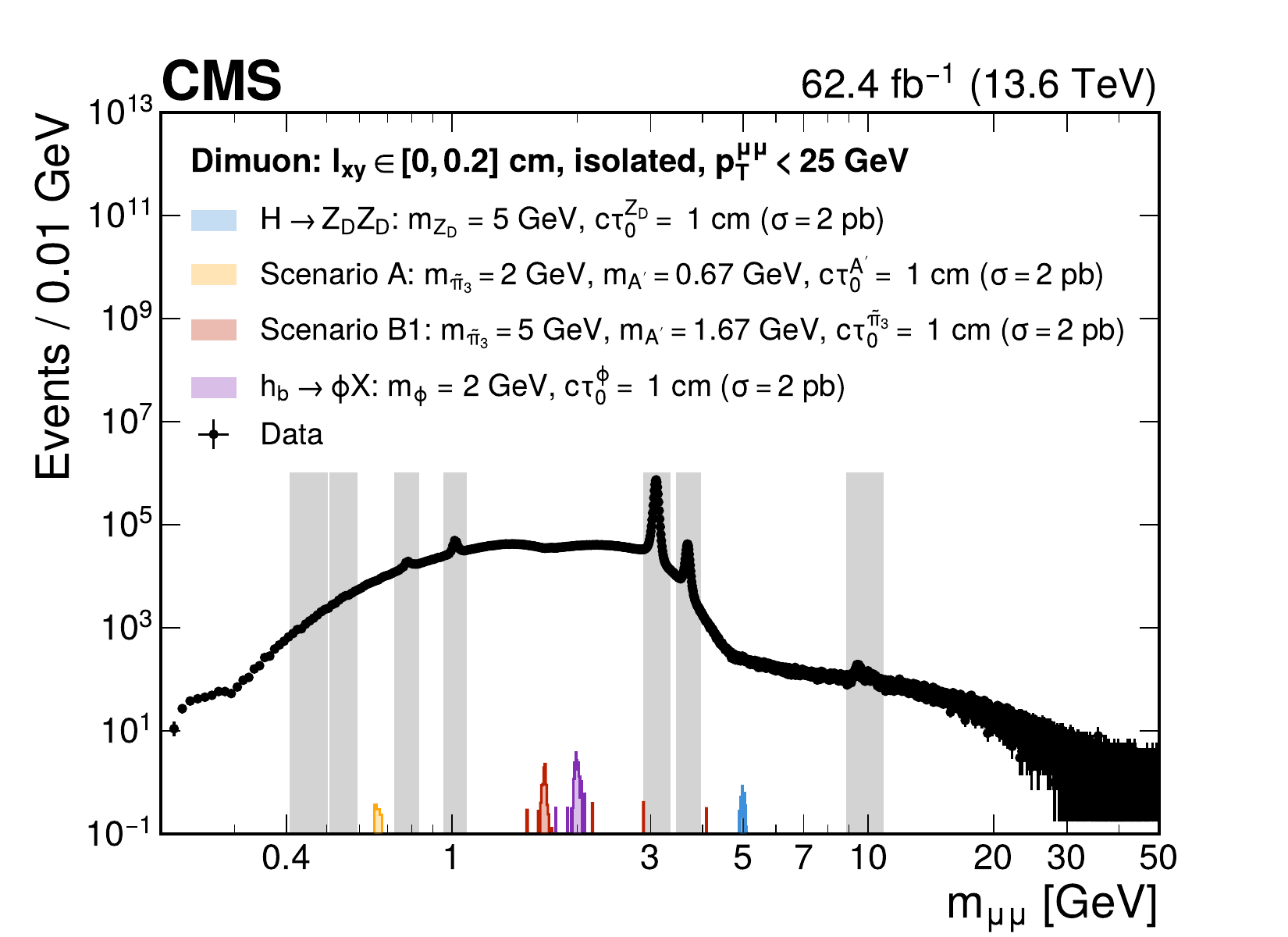}
  \includegraphics[width=.49\textwidth]{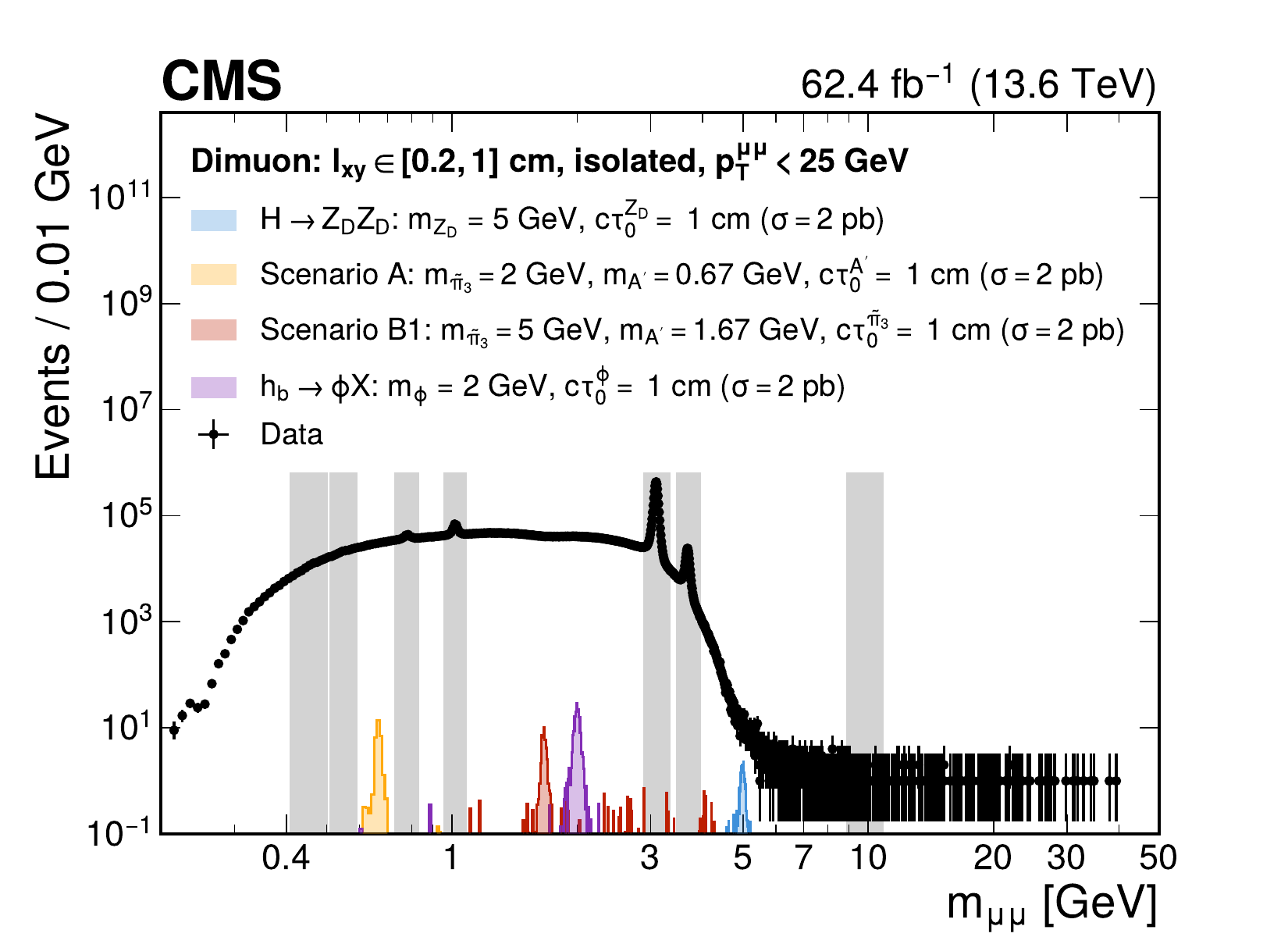}
  \includegraphics[width=.49\textwidth]{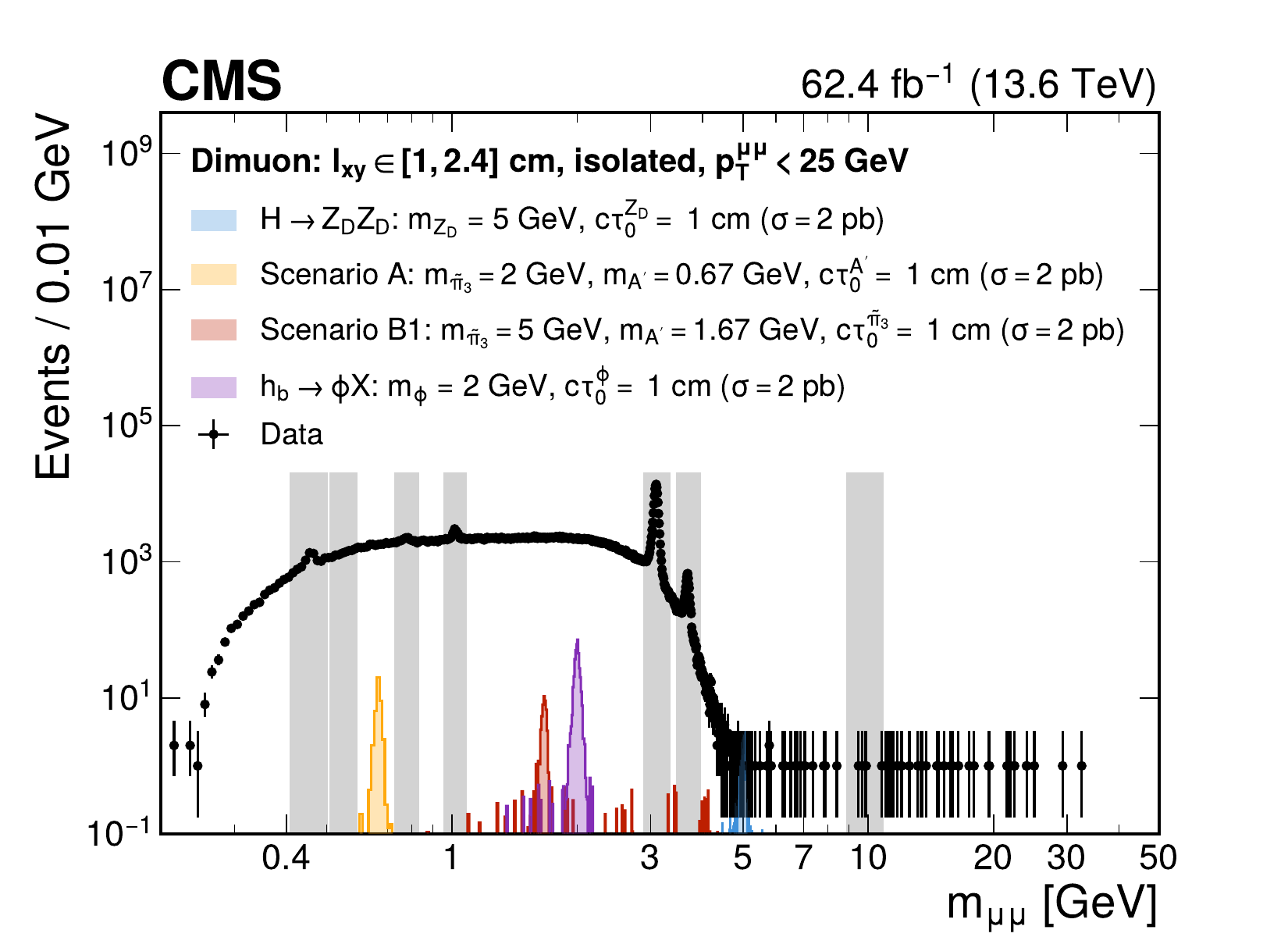}
  \includegraphics[width=.49\textwidth]{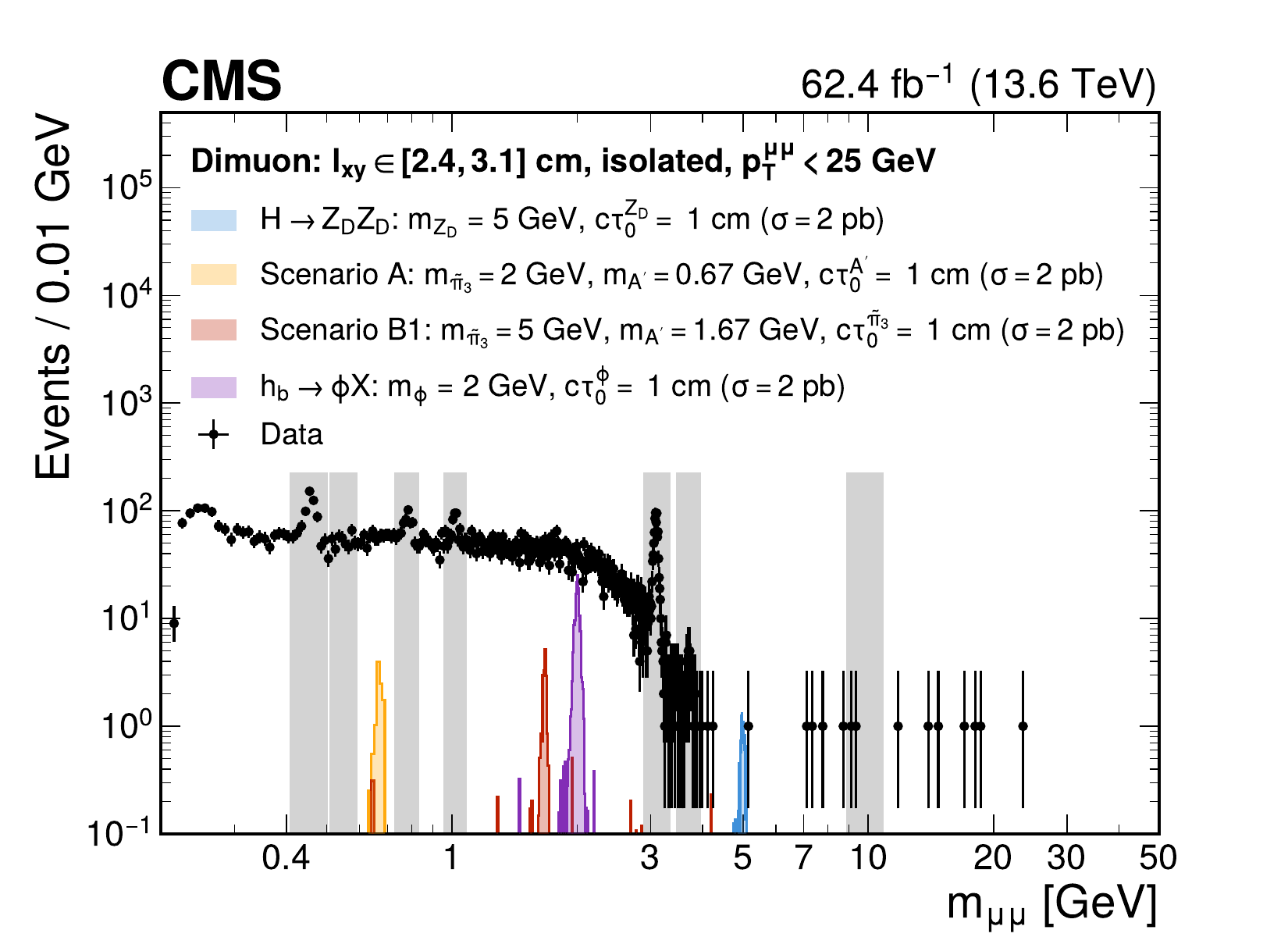}
  \includegraphics[width=.49\textwidth]{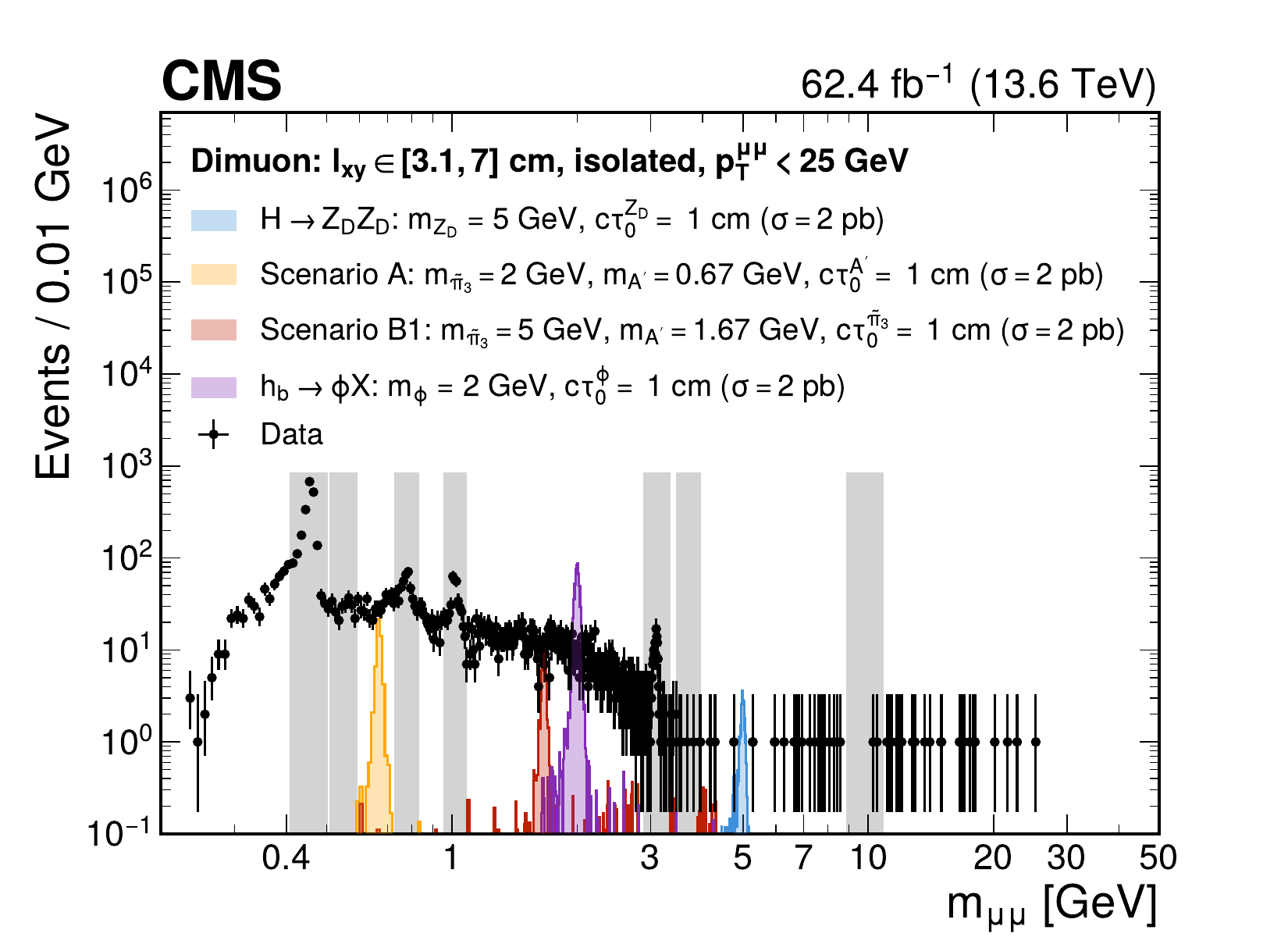}
  \includegraphics[width=.49\textwidth]{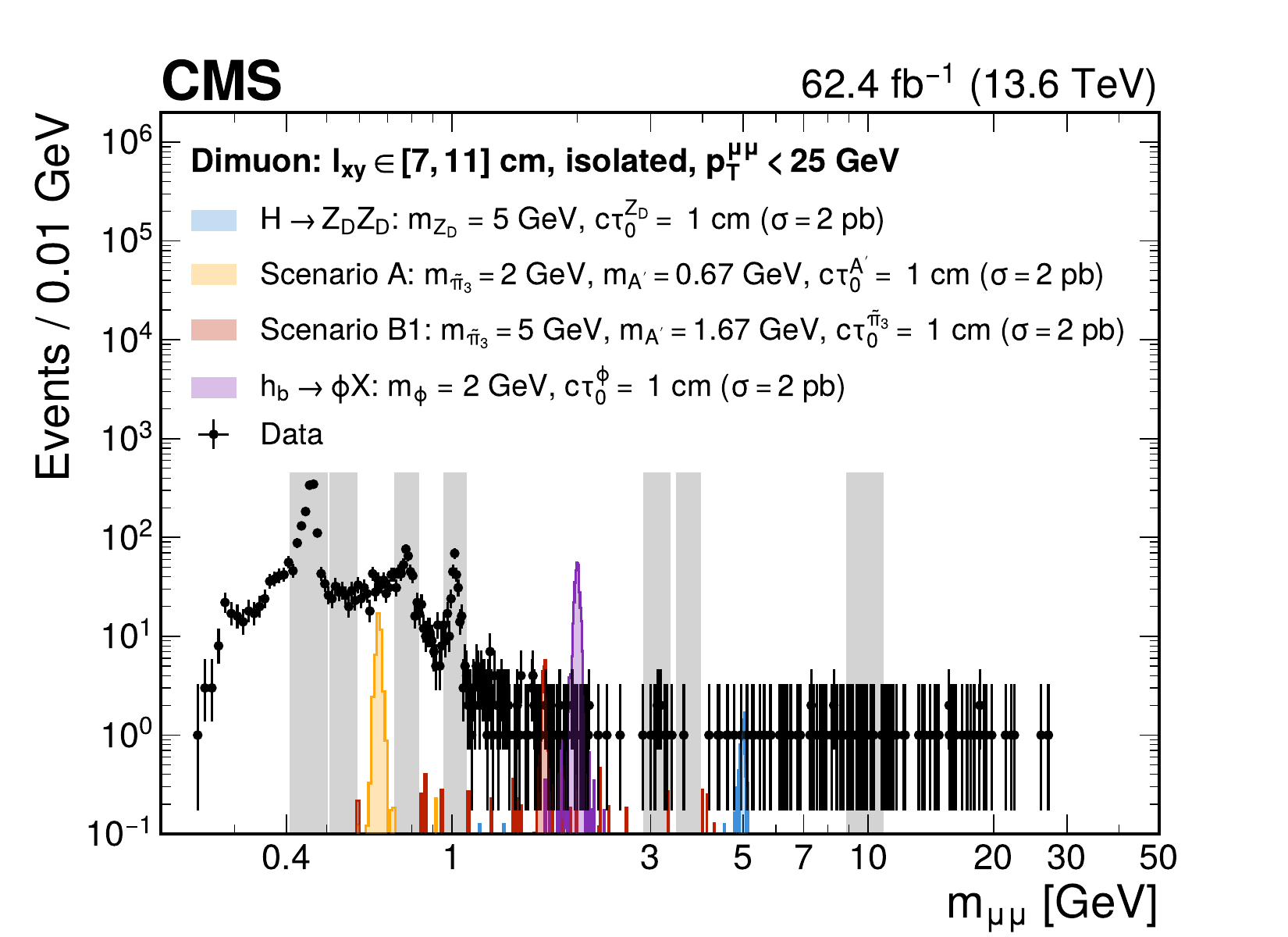}
  \caption{
    Dimuon invariant mass distributions of selected events with a single, isolated muon pair with $\ptmumu < 25\GeV$.
    Distributions are shown separately for each of the six $\lxy<11\cm$ bins.
    The vertical gray bands indicate mass ranges containing known SM resonances, which are masked for the purpose of this search.
    Distributions are shown as obtained in data (black points with uncertainties)
    and for representative signal models (colored histograms),
    assuming a cross section $\sigma = 2\unit{pb}$ for all models, for visualization purposes.
  }
  \label{fig:below}
\end{figure}

\begin{figure}[tbp]
  \centering
  \includegraphics[width=.49\textwidth]{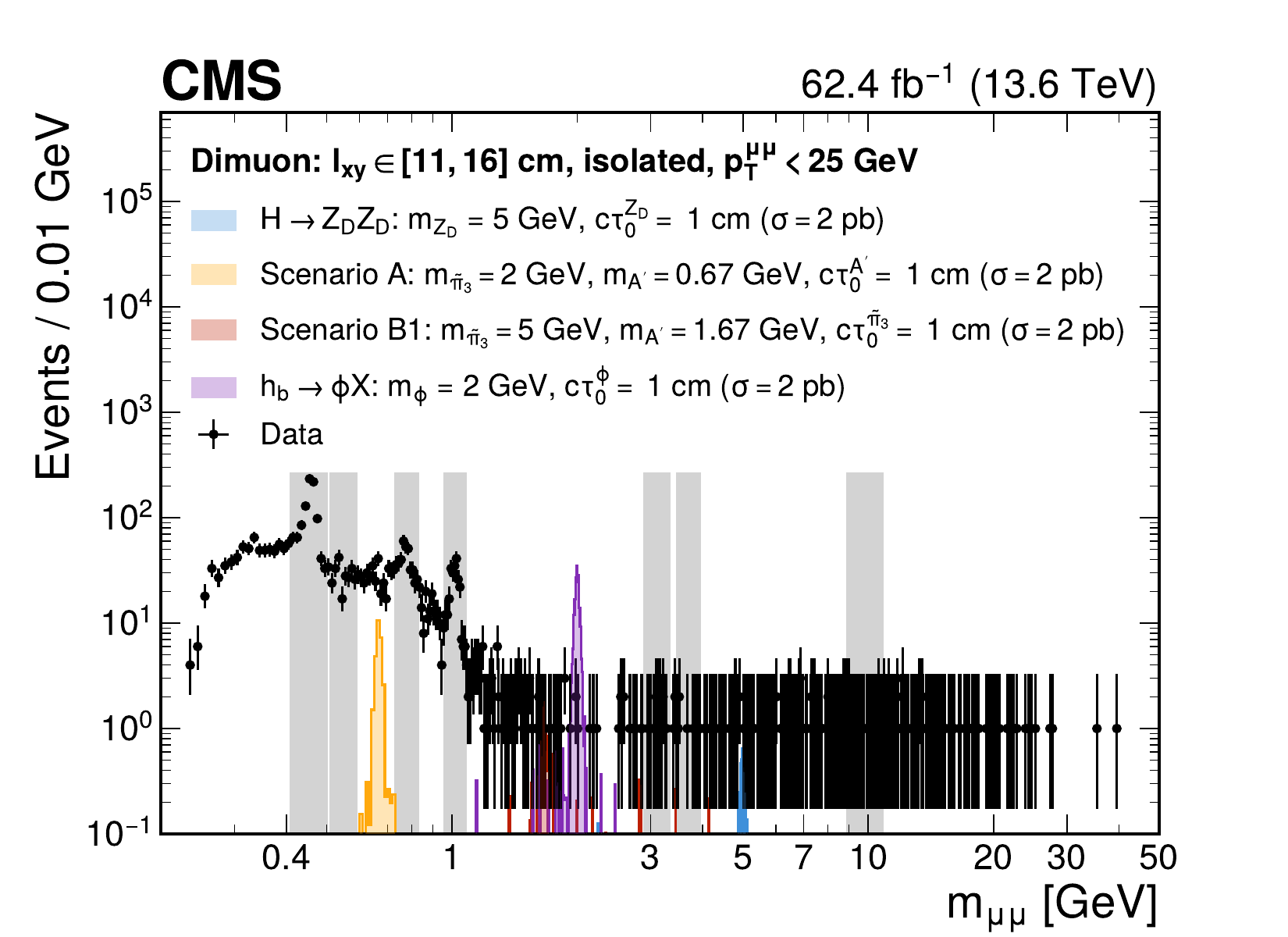}
  \includegraphics[width=.49\textwidth]{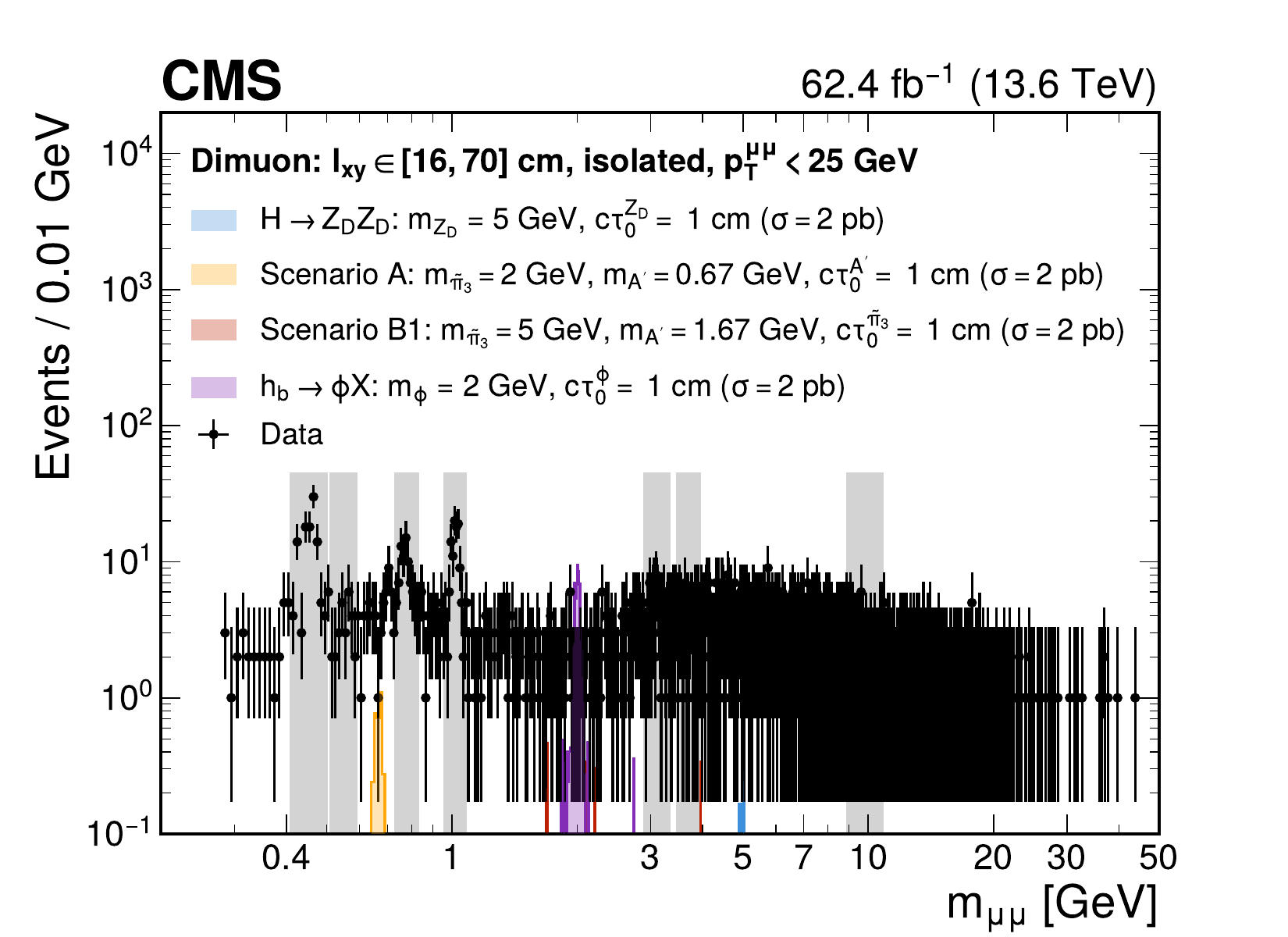}
  \caption{
    Dimuon invariant mass distributions of selected events with a single, isolated muon pair with $\ptmumu < 25\GeV$.
    Distributions are shown separately for each of the two $\lxy>11\cm$ bins
    (\ie, for the displacement range inaccessible to the previous results presented in Ref.~\cite{EXO-20-014}).
    The vertical gray bands indicate mass ranges containing known SM resonances, which are masked for the purpose of this search.
    Distributions are shown as obtained in data (black points with uncertainties)
    and for representative signal models (colored histograms),
    assuming a cross section $\sigma = 2\unit{pb}$ for all models, for visualization purposes.
  }
  \label{fig:belowadd}
\end{figure}

As an example, in Fig.~\ref{fig:4mu} we show the $m_{4\PGm}$ spectrum in the multivertex and overlapping vertex categories.
In Figs.~\ref{fig:above} and~\ref{fig:aboveadd} (\ref{fig:below} and~\ref{fig:belowadd}),
we show the \mmumu distribution in bins of \lxy, as obtained from selected dimuon events where both muons are isolated
and have \ptmumu above (below) 25\GeV.

In each mass window, the SM background is modeled by fitting the multimuon invariant mass distribution in data, 
as further described in Section~\ref{sec:fit}.
Finally, the mass windows in all search bins are used in combination
as input to a final likelihood fit for the interpretation of the results, as described in Section~\ref{sec:results}.

\section{Parameterization of the invariant mass distribution}
\label{sec:fit}

The search is performed by fitting the sum of signal and the SM background models to mass distributions in the various categories, 
which are separate for the 2022 and 2023 data-taking periods.
We target low-mass multimuon resonances whose intrinsic width is assumed to be much narrower than the detector resolution.
The signal multimuon invariant mass resolution
($\sigma_{\text{mass}}^{\PGm\PGm}/\mmumu \sim \sigma_{\text{mass}}^{4\PGm}/m_{4\PGm}$) is estimated from simulated events to be about 1.8\% of the mass, 
with an uncertainty of about 50\%, and both are independent of the signal mass and lifetime hypotheses.
We use the sum of a Gaussian function and a double-sided Crystal Ball function~\cite{dcb1,dcb2},
with common mean and width ($\sigma_{\text{mass}}$) parameters,
to model signal invariant mass distributions in dimuon and $4\PGm$ events.
The parameters are determined by fitting the corresponding signal MC simulation. 
Additionally, an interpolation of the signal model parameters is performed between different mass hypotheses 
to provide a continuous description of the signal shape across the probed mass range.

In each event category, 
the SM background is modeled in sliding windows of $\pm 5\sigma_{\text{mass}}$ around the considered signal mass hypothesis.
These windows are required not to overlap with the masked mass regions corresponding to known SM resonances
(as discussed in Section~\ref{sec:strategy}).
A number of functional forms are considered in order to model the background mass distribution.
These include Bernstein polynomials, power law, and exponential functions.  
For each family of functional forms,
the best order in each mass window and event category is selected 
by means of a Fisher test~\cite{ftest}.
First, the lowest order ($N$) function is used to fit the data.
Then, the next-order ($N+1$) is used, and the difference
$2\Delta\mathrm{NLL}_{N+1}=2(\mathrm{NLL}_{N}-\mathrm{NLL}_{N+1})$ 
(with NLL denoting the negative logarithm of the likelihood of the fit)
is evaluated to determine whether the data support the need for a higher-order function. 
This decision is based on the fact that 
$2\Delta\mathrm{NLL}_{N+1}$ is asymptotically described by a $\chi^{2}$ distribution with $M$ degrees of freedom, 
where $M$ is the difference in the number of free parameters in the $(N+1)$th and $N$th order functions (in this case $M=1$). 
A $p$-value \cite{Gross:2010qma} is calculated as $P_M(\chi^{2} > 2\Delta\mathrm{NLL}_{N+1})$ where $P_M(\chi_{\mathrm{min}}^2)$
is the $\chi^{2}$ tail probability for $M$ degrees of freedom.
If the $p$-value is less than 0.05, 
the higher-order function is retained, 
since it is determined to significantly improve the description of the data.
Once the best order $N$ for each family of functions has been determined, 
the corresponding functional forms are included in an envelope, which is then used 
in the discrete profiling method~\cite{discreteprofile}, 
where the choice of the background function is treated as a discrete nuisance parameter in the fit
to account for the uncertainty associated with the arbitrary choice of the function.
In addition, we use a goodness-of-fit test to remove background models that do not describe the data appropriately.
The test is based on a $\chi^{2}$ test statistic, which is converted into a $p$-value
through a set of pseudo-experiments. Models with $p < 0.01$ are not considered.

The potential presence of a bias in the measurement of a signal due to the choice of the functional forms used to model the SM background is also assessed
by means of pseudo-experiments. 
First, a varying amount of signal is injected on top of the background generated according to the selected functional form. 
Then, a background + signal fit is performed, allowing the signal yield to float freely. 
The case of a zero injected signal is also considered.
The bias, quantified as the difference between the measured and injected signal yields relative to the statistical uncertainty in the measured signal yield, 
is found to be negligible.

\begin{figure}[htb!]
  \centering
    \includegraphics[width=0.49\textwidth]{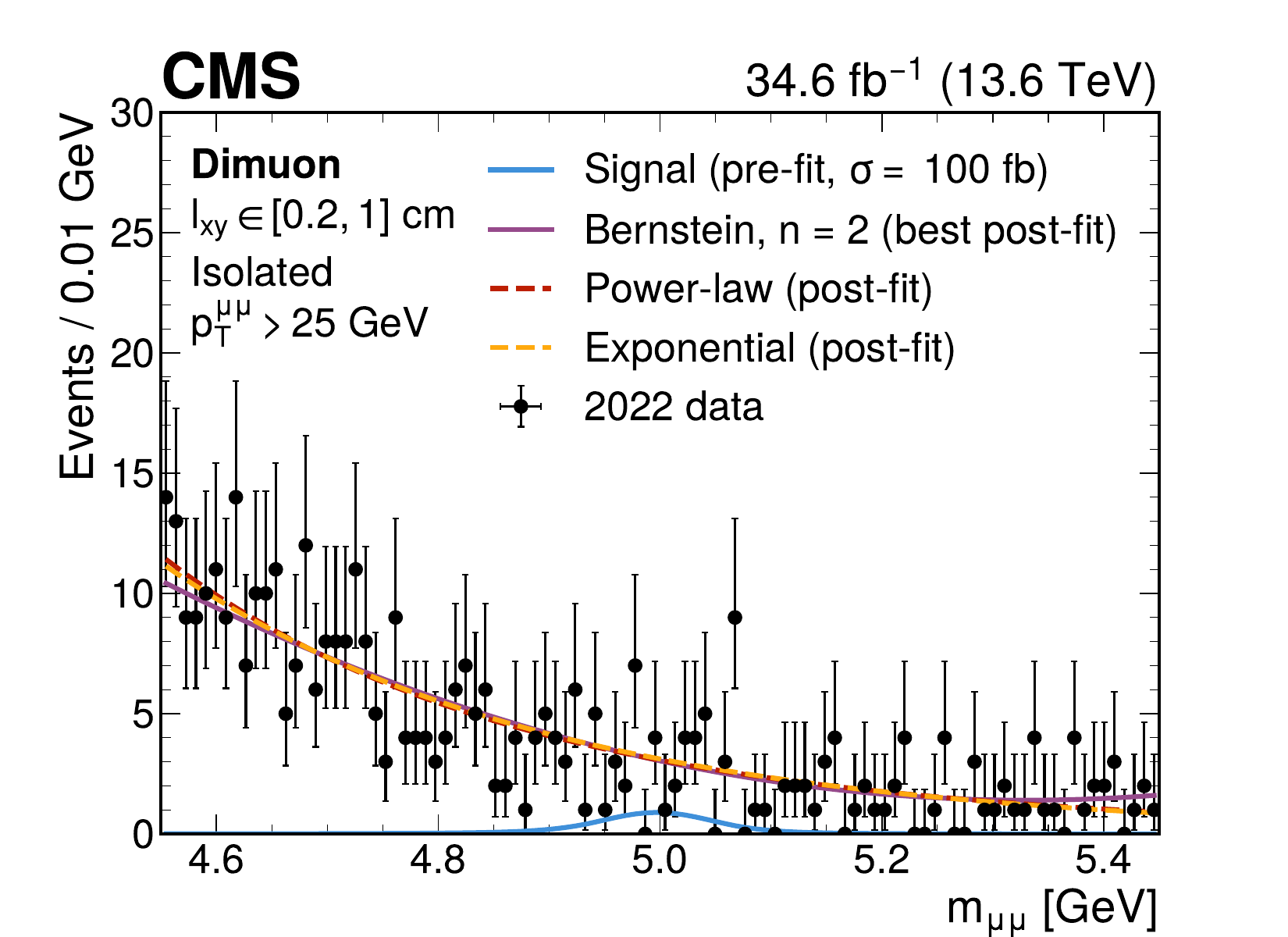}
    \includegraphics[width=0.49\textwidth]{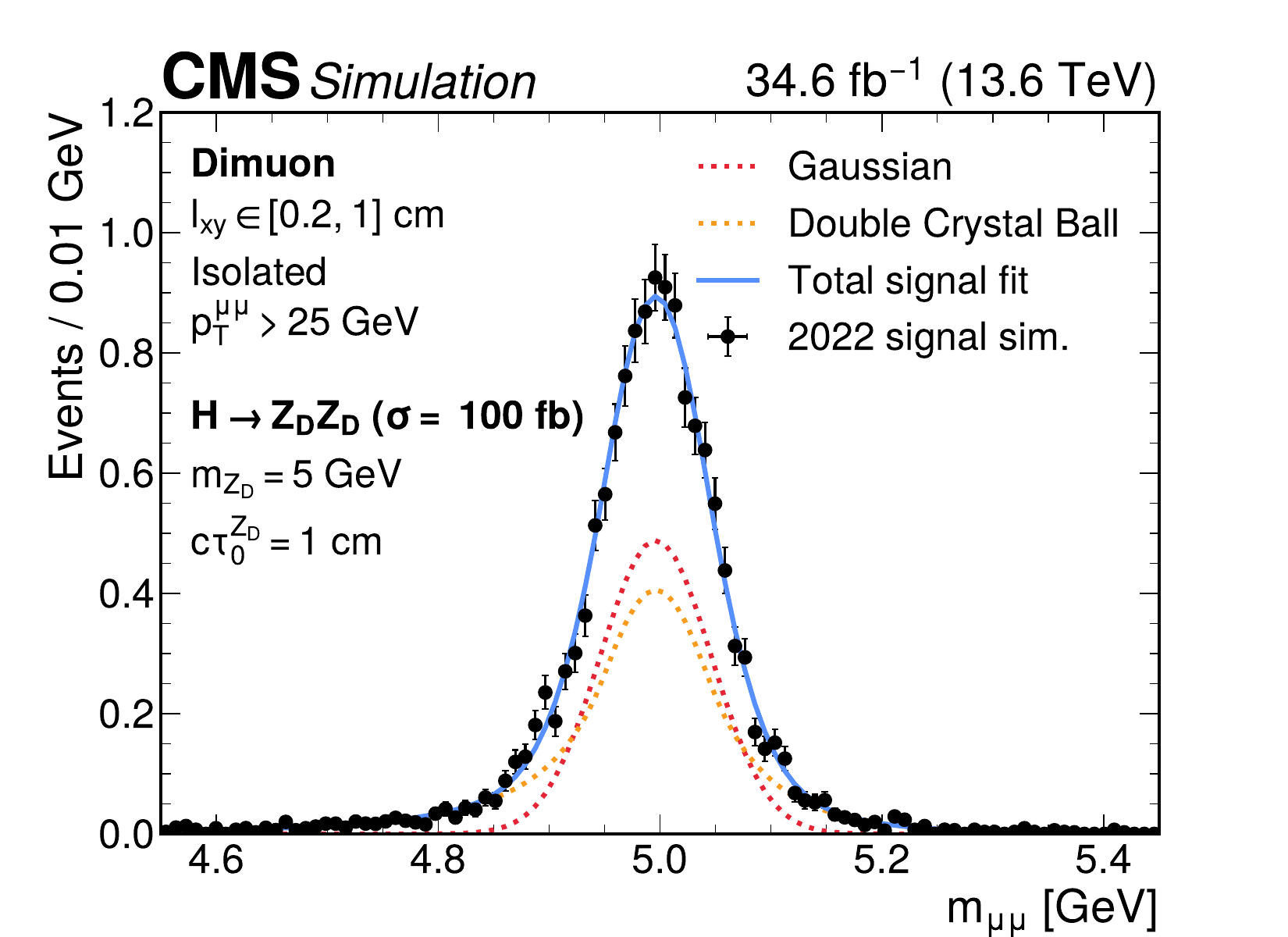}
    \caption{
      The dimuon invariant mass distribution (left) is shown for data collected in 2022 (black markers)
      in a mass window centered around 5\GeV, in one of the dimuon search bins ($0.2<\lxy<1\cm$, $\ptmumu>25\GeV$, with two isolated muons).
      The results of background+signal fits to the data of three selected functional forms are also shown (different colors, partially overlapping lines),
      highlighting the Bernstein polynomial as the best fit (solid, purple line) for this mass window.
      The corresponding pre-fit function (solid blue line) for a representative $\PH\to\PZ_{\mathrm{D}}\PZ_{\mathrm{D}}$ signal,
      with $m_{\PZ_{\mathrm{D}}}=5\GeV$ and $c\tau_{0}^{\PZ_{\mathrm{D}}}=1\cm$,
      assuming a cross section $\sigma(\PH\to\PZ_{\mathrm{D}}\PZ_{\mathrm{D}}) = 100\unit{fb}$, is shown for visualization purposes.
      The fit (solid blue line for the total fit function, and dashed red and orange lines for its individual components) to the dimuon invariant mass distribution (right)
      expected in the same dimuon search bin is shown for the same signal (black markers).
      }
      \label{fig:fit}
\end{figure}

The background-only fit results for the selected functional forms in one of the dimuon search bins using 2022 data
are shown in Fig.~\ref{fig:fit}, as an example,
together with the dimuon invariant mass distribution expected for a representative signal hypothesis.

\section{Results and interpretation}
\label{sec:results}

Unbinned maximum likelihood fits to the data
are performed simultaneously in all search bins, 
under either background-only or background+signal hypotheses, using background and signal
models and uncertainties described in Section~\ref{sec:fit}.
Additional log-normal constraint terms are used to account 
for the uncertainties in the signal yields, when considered (Section~\ref{sec:signalsysts}).

No significant peak-like structures are observed in data.
The background+signal fits are used to set 95\% confidence level (\CL) upper limits 
on the branching fractions for the signal models under consideration.  
Limits are set using a modified frequentist approach, employing the \CLs 
criterion~\cite{Junk:1999kv,Read:2002hq,ATLAS:2011tau}
within the CMS statistical analysis tool \textsc{Combine}~\cite{CMS:2024onh},
which is based on the \textsc{RooFit}~\cite{Verkerke:2003ir} and \textsc{RooStats}~\cite{Moneta:2010pm} frameworks.
These limits are then used, in conjunction
with the theoretical cross section calculations, 
to exclude ranges of masses and/or lifetimes 
for the BSM particles of the signal models.

\subsection{Systematic uncertainties in the signal yields}
\label{sec:signalsysts}

A systematic uncertainty in the trigger efficiency
(discussed in Section~\ref{sec:triggers}) is assessed 
to be in the range between 1 and 15\%, 
depending on the \pt and displacement of the muons.
A separate systematic uncertainty is considered, ranging from 1 to 30\% (largest for high-$\lxy$ bins), to account for the potential mismodeling of the signal simulation
that may introduce a difference in the event selection acceptance for the selection described in Section~\ref{sec:evtsel}.
Both uncertainties are derived from studies of $\PJGy \to \PGmp\PGmm$ data
and account for the differences between the measurements in data and simulation.
Additionally, we assess a systematic uncertainty in the expected signal yields
arising from the uncertainty in the integrated luminosity measurement,
of 1.4 (1.3)\% for 2022 (2023) data~\cite{LUM-17-003,LUM-22-001,CMS-DP-2024-068}.
The overall uncertainty in the luminosity measurement for the 2022--2023 period is 1.01\%.
Finally, we account for uncertainties resulting from the limited size of the simulated signal samples.
Uncertainties arising from the choice of the PDFs,
and of the renormalization and factorization scales used in the event generators are negligible compared to others.
Uncertainties in the trigger efficiency and integrated luminosity are treated as correlated across the search bins.
Other uncertainties are taken as uncorrelated.
Since we find no significant difference between 2022 and 2023 data and simulations,
all uncertainties are treated as correlated across the two data-taking periods.

\subsection{Constraints on models of BSM physics}

\begin{figure}[htb!]
  \centering
    \includegraphics[width=0.49\textwidth]{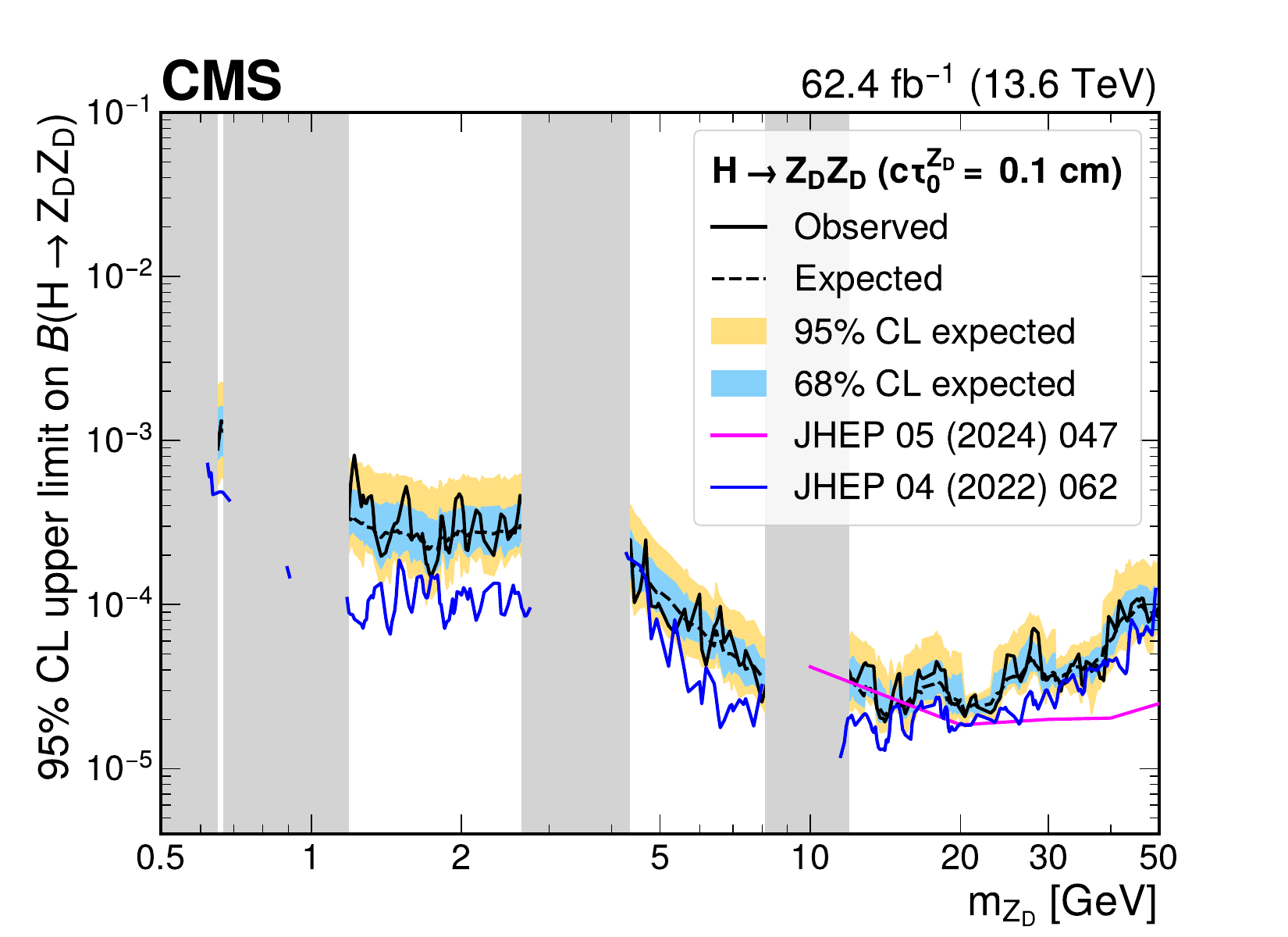}
    \includegraphics[width=0.49\textwidth]{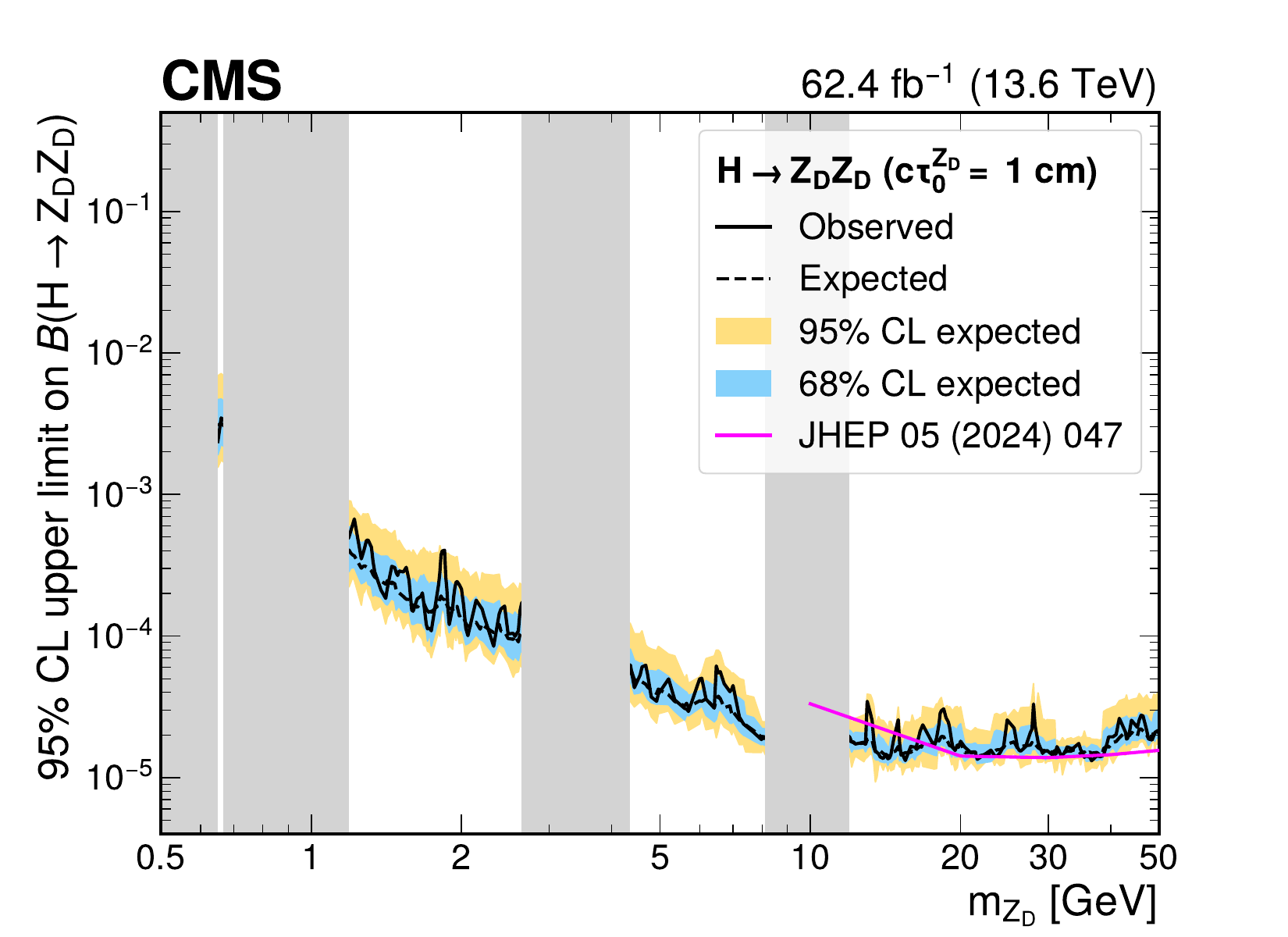}
    \includegraphics[width=0.49\textwidth]{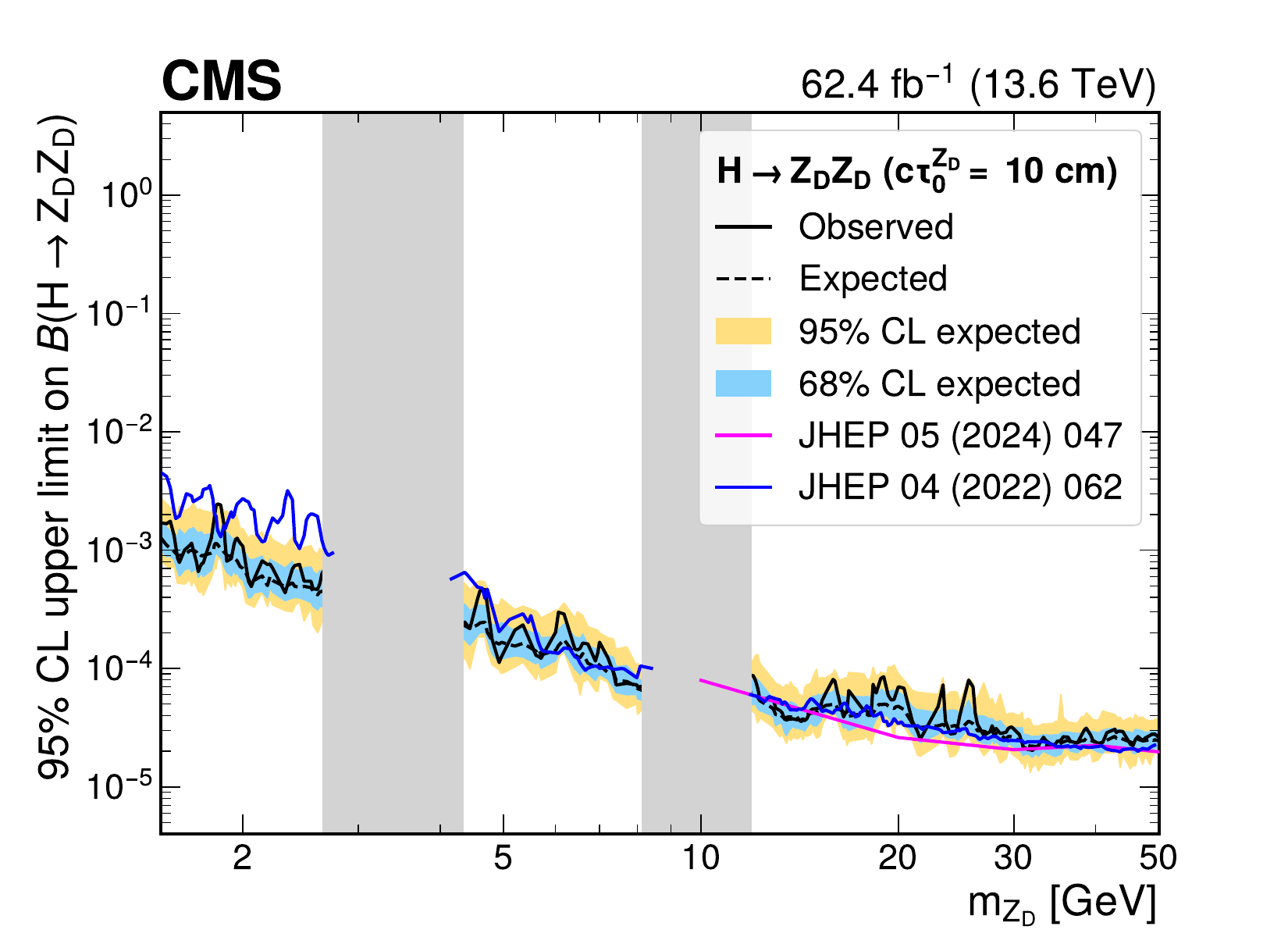}
    \includegraphics[width=0.49\textwidth]{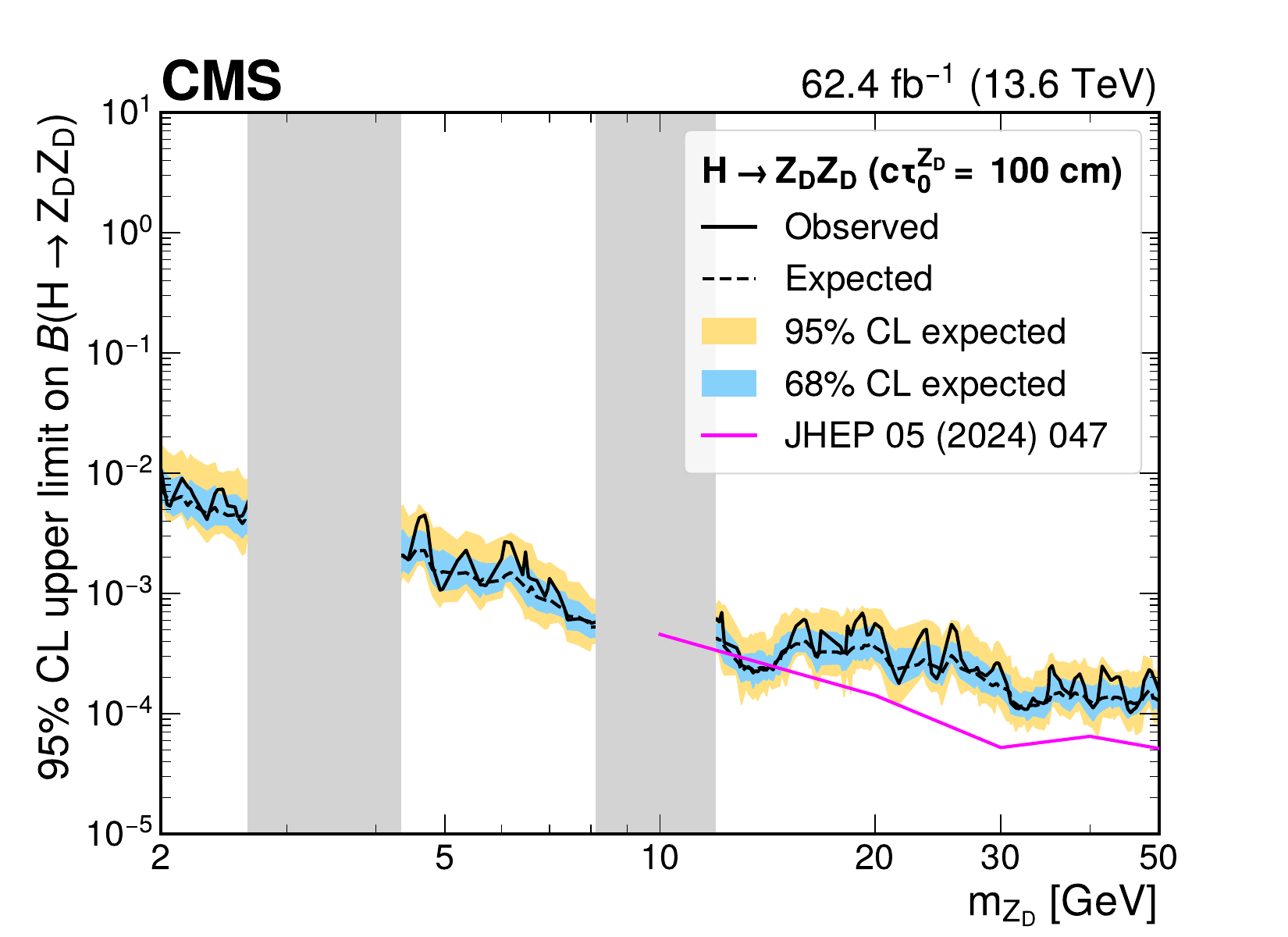}
    \caption{
      Upper limits at 95\%~\CL 
      on the branching fraction $\mathcal{B}(\PH\to\PZ_{\mathrm{D}}\PZ_{\mathrm{D}})$
      as functions of the dark-photon mass, $m_{\PZ_{\mathrm{D}}}$,
      for $c\tau_{0}^{\PZ_{\mathrm{D}}}=0.1$ (upper left), 1 (upper right),
      10 (lower left), and 100\cm (lower right).
      The solid black (dashed black) line represents the observed (median expected) exclusion. 
      The inner blue (outer yellow) band indicates the region containing 68~(95)\% 
      of the distribution of limits expected under the background-only hypothesis.
      The limits are obtained using the combination of all event categories.
      The dark blue (magenta) solid line represents the observed upper limits previously set in Ref.~\cite{EXO-20-014} (\cite{EXO-23-014}).
      The vertical gray bands indicate mass ranges containing known SM resonances,
      which are masked for the purpose of this search.
      The blue (magenta) line can overlap with the gray bands, owing to the different mass ranges covered by the results from Ref.~\cite{EXO-20-014} (\cite{EXO-23-014}).
      }
      \label{fig:hzdzd}
\end{figure}

\begin{figure}[htb!]
  \centering
    \includegraphics[width=0.49\textwidth]{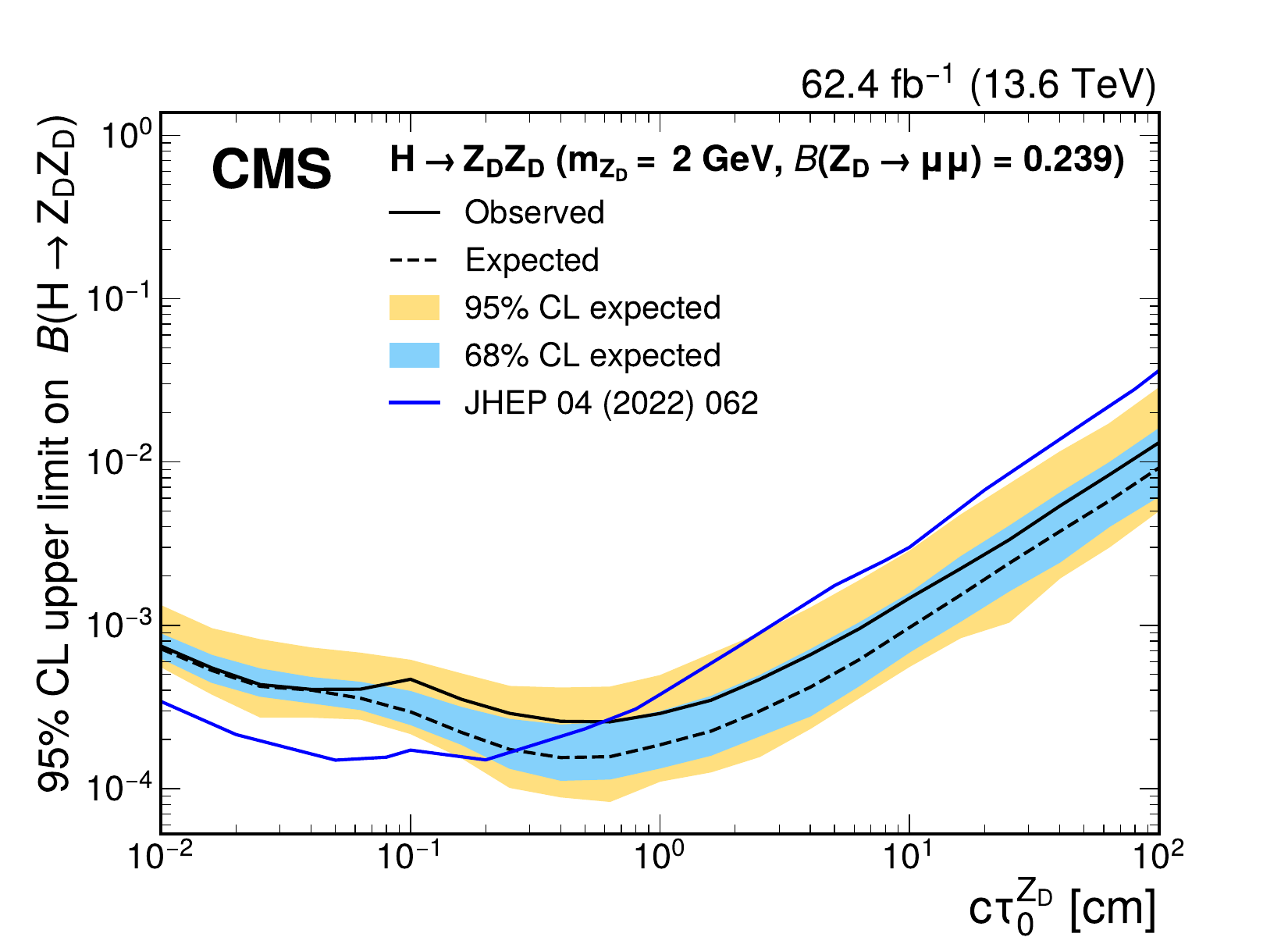}
    \includegraphics[width=0.49\textwidth]{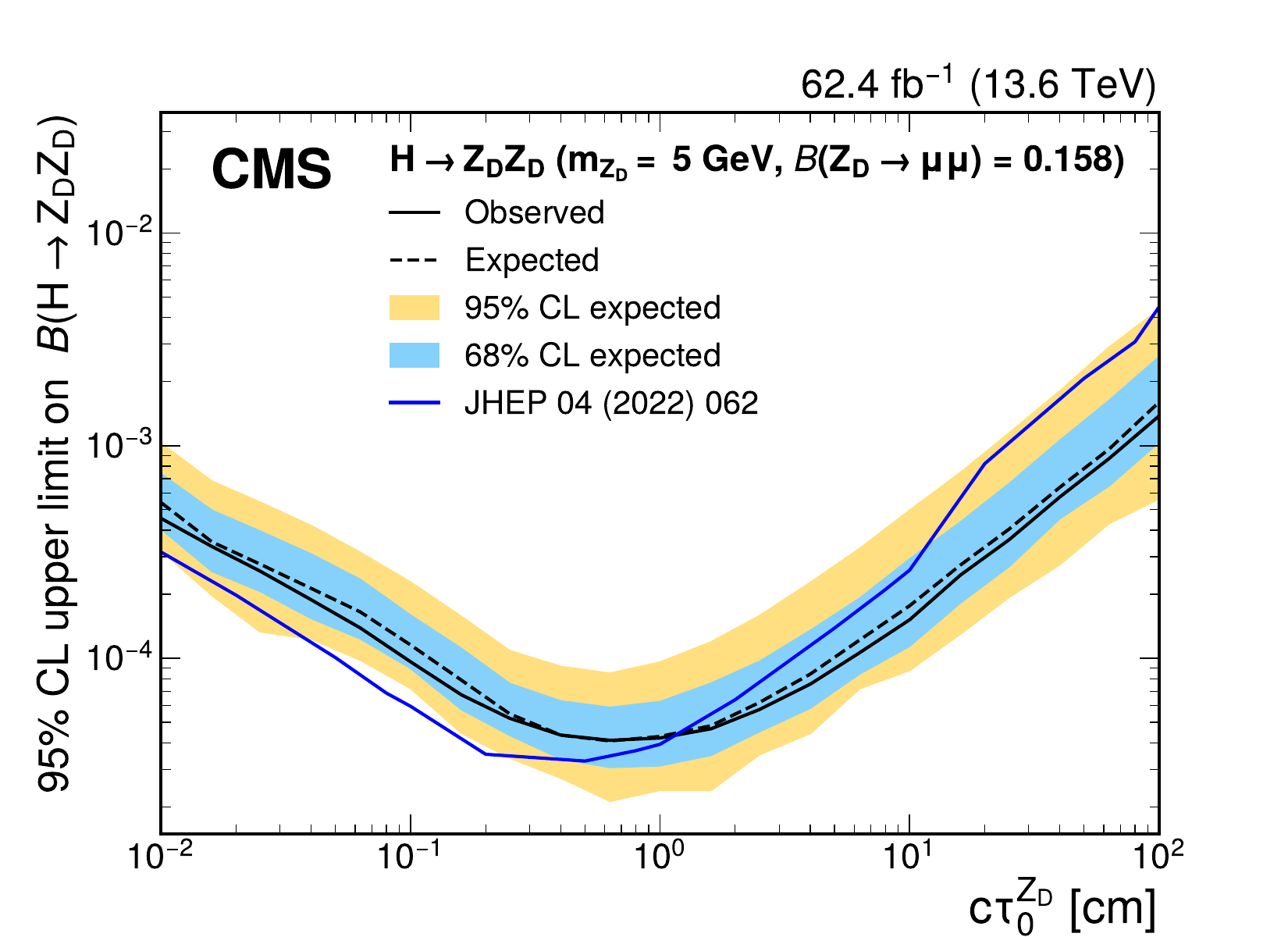}
    \includegraphics[width=0.49\textwidth]{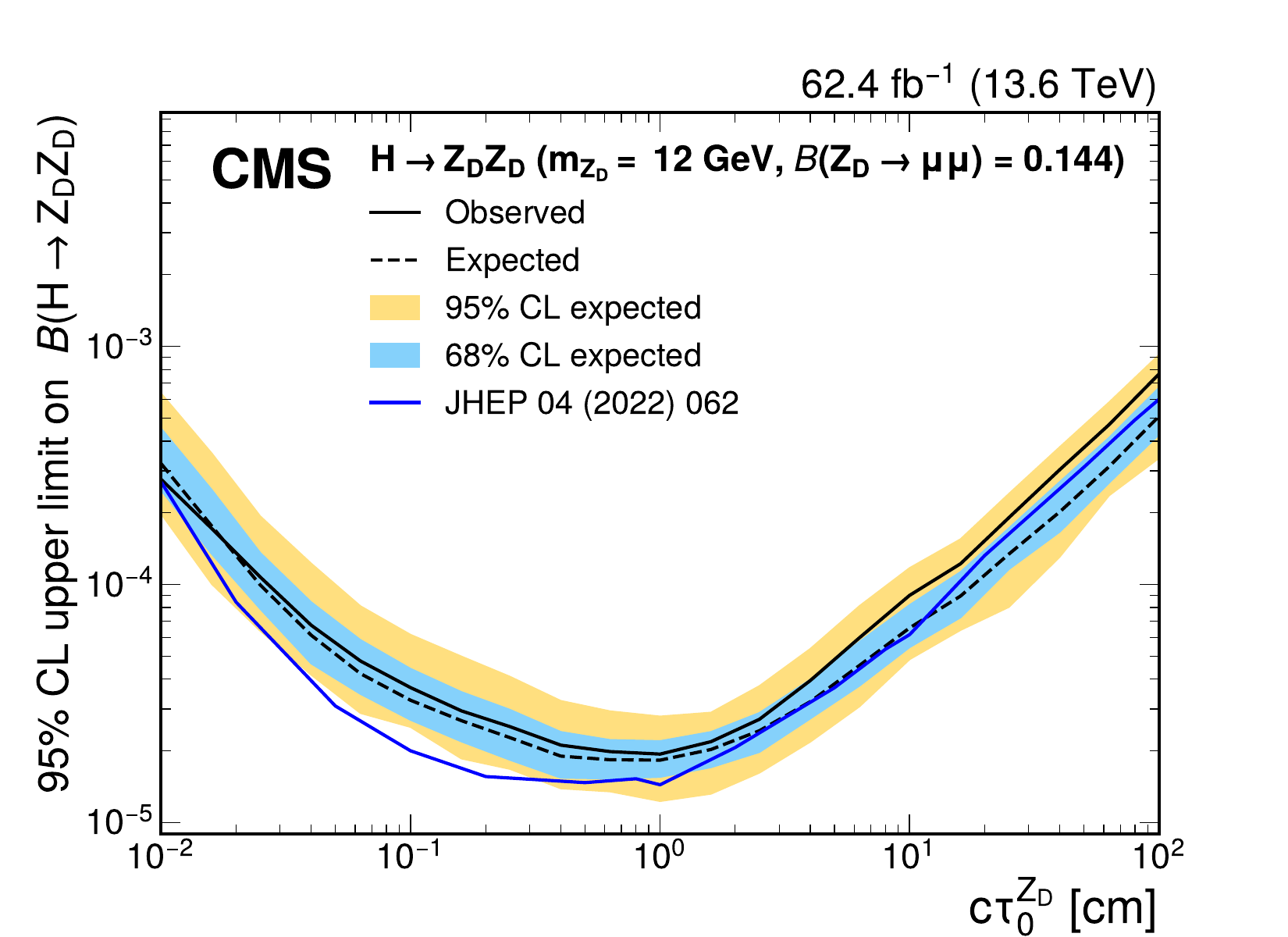}
    \includegraphics[width=0.49\textwidth]{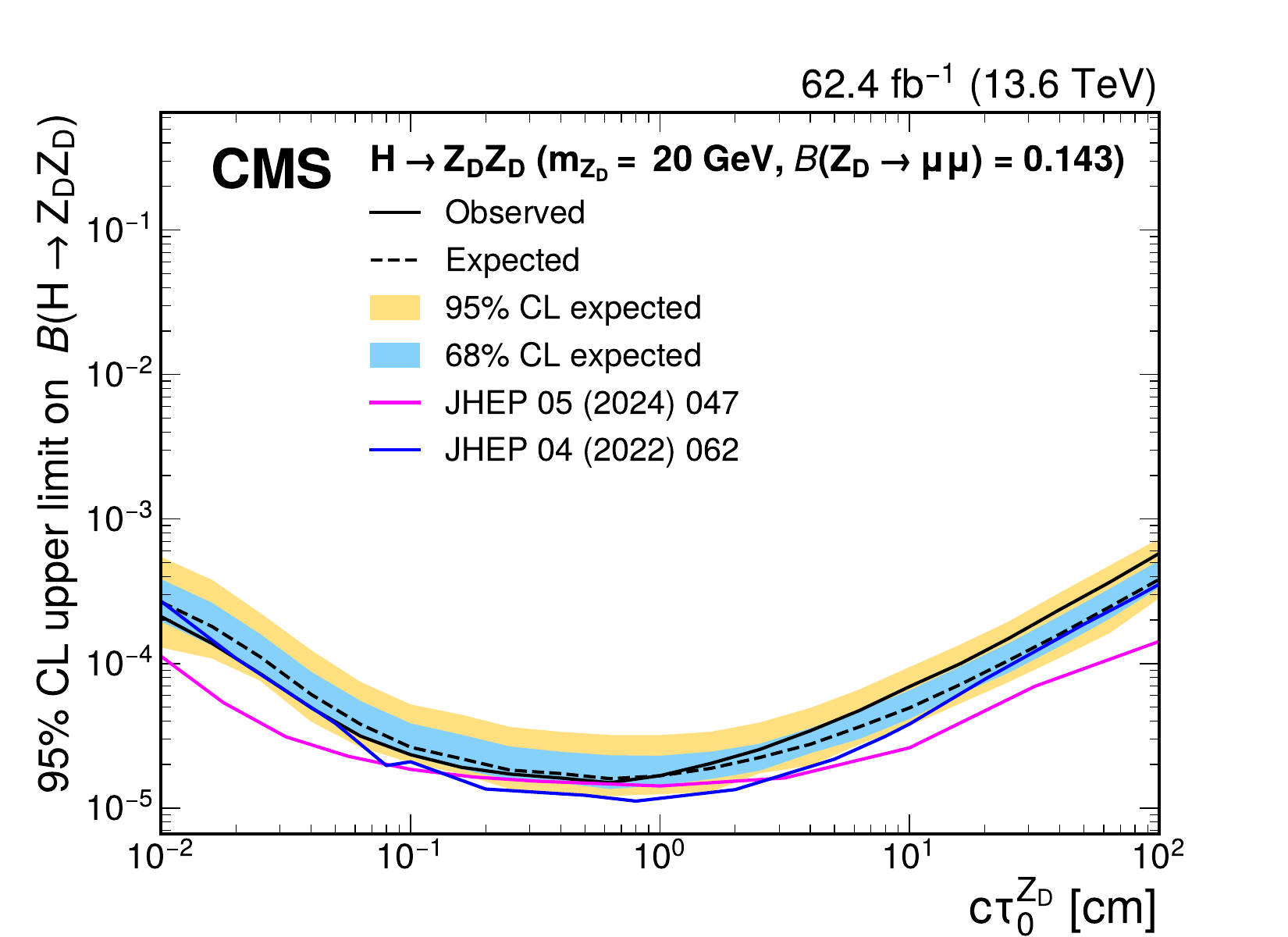}
    \caption{
      Upper limits at 95\%~\CL 
      on the branching fraction $\mathcal{B}(\PH\to\PZ_{\mathrm{D}}\PZ_{\mathrm{D}})$
      as functions of $c\tau_{0}^{\PZ_{\mathrm{D}}}$, for representative mass hypotheses, $m_{\PZ_{\mathrm{D}}} = 2$
      (upper left), 5 (upper right), 12 (lower left), and 20~(lower right)\GeV.
      The solid black (dashed black) line represents the observed (median expected) exclusion. 
      The inner blue (outer yellow) band indicates the region containing 68~(95)\% 
      of the distribution of limits expected under the background-only hypothesis.
      The limits are obtained using the combination of all event categories.
      The dark blue (magenta) solid line represents the observed upper limits previously set in Ref.~\cite{EXO-20-014} (\cite{EXO-23-014}).
      }
      \label{fig:hzdzdctau}
\end{figure}

We set upper limits at 95\%~\CL on the branching fraction 
$\mathcal{B}(\PH\to\PZ_{\mathrm{D}}\PZ_{\mathrm{D}})$,
using values of $\mathcal{B}(\PZ_{\mathrm{D}}\to\PGm\PGm)$ from the model of Ref.~\cite{hahm_2015}.
The constraints we obtain are the most stringent to date for a wide range of signal hypotheses.
Upper limits are shown in Fig.~\ref{fig:hzdzd} as functions of the mass of the dark photon, $m_{\PZ_{\mathrm{D}}}$,
for dark photon lifetimes $c\tau_{0}^{\PZ_{\mathrm{D}}}=0.1$, 1, 10, and 100\cm,
and in Fig.~\ref{fig:hzdzdctau} as functions of the dark photon lifetime for a representative set of $m_{\PZ_{\mathrm{D}}}$ hypotheses.
We improve and extend the constraints from previous results~\cite{EXO-20-014, EXO-23-014} at small $m_{\PZ_{\mathrm{D}}}$ and large $c\tau_{0}^{\PZ_{\mathrm{D}}}$.
Compared to Ref.~\cite{EXO-20-014}, this is achieved despite the smaller amount of available data, thanks to the improved muon scouting triggers,
which allow for the reconstruction of displaced muons with $\lxy>11\cm$, and to the refined event selection and categorization.
Compared to Ref.~\cite{EXO-23-014}, we are able to probe masses $m_{\PZ_{\mathrm{D}}}<10\GeV$, thanks to the lower muon \pt thresholds required in the scouting triggers used in this search.

\begin{figure}[htb!]
  \centering
    \includegraphics[width=0.49\textwidth]{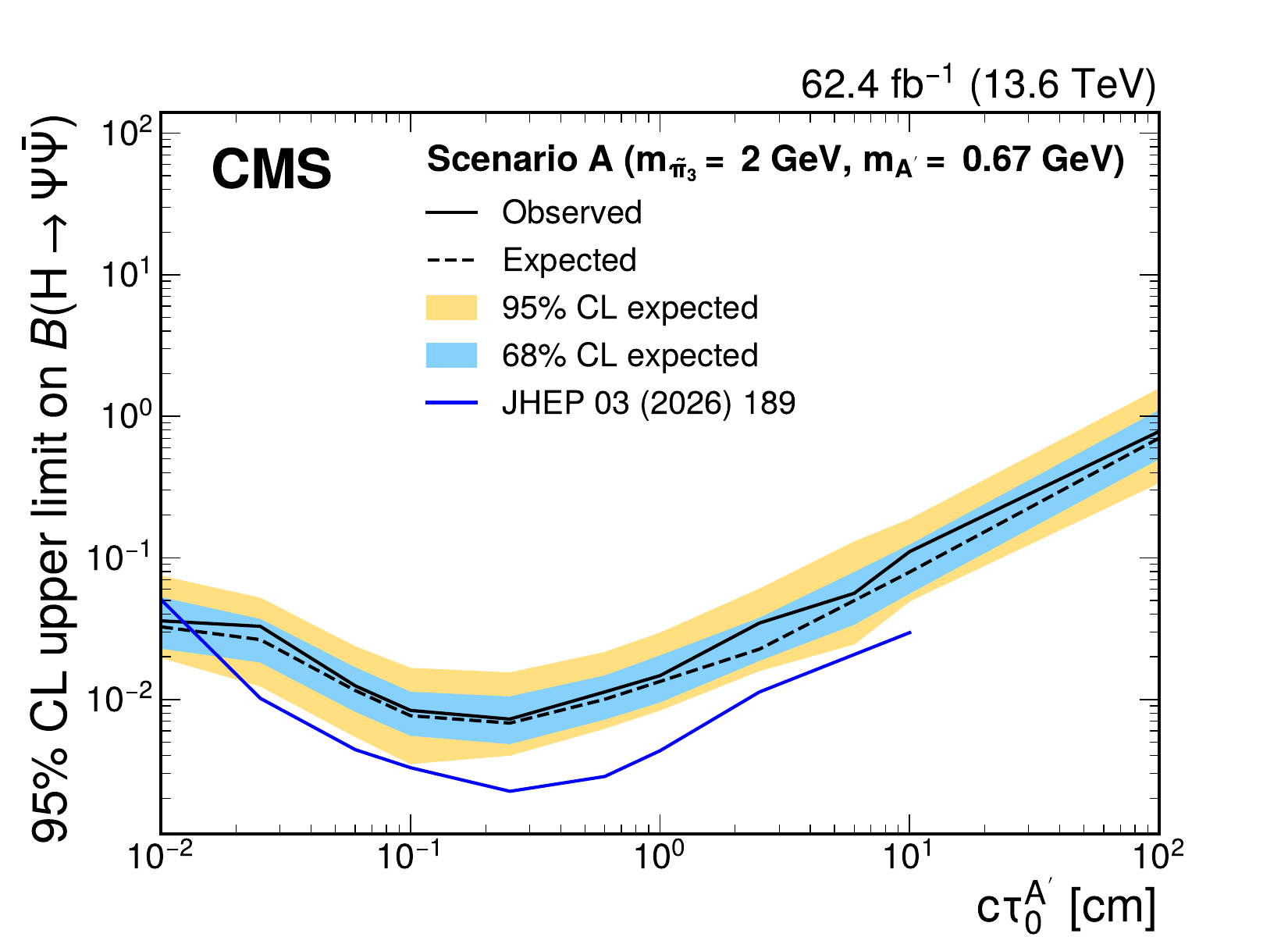}
    \includegraphics[width=0.49\textwidth]{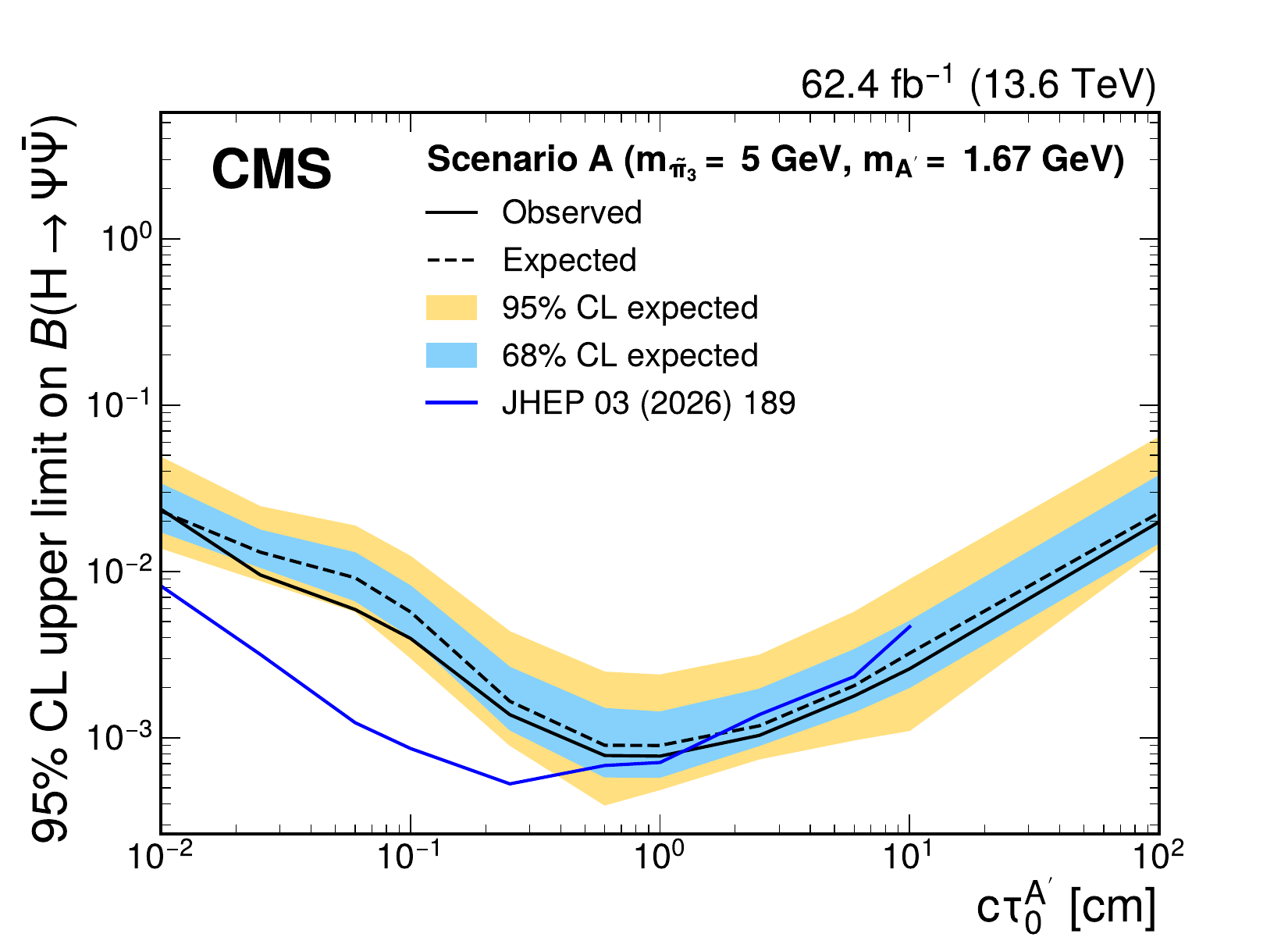}
    \includegraphics[width=0.49\textwidth]{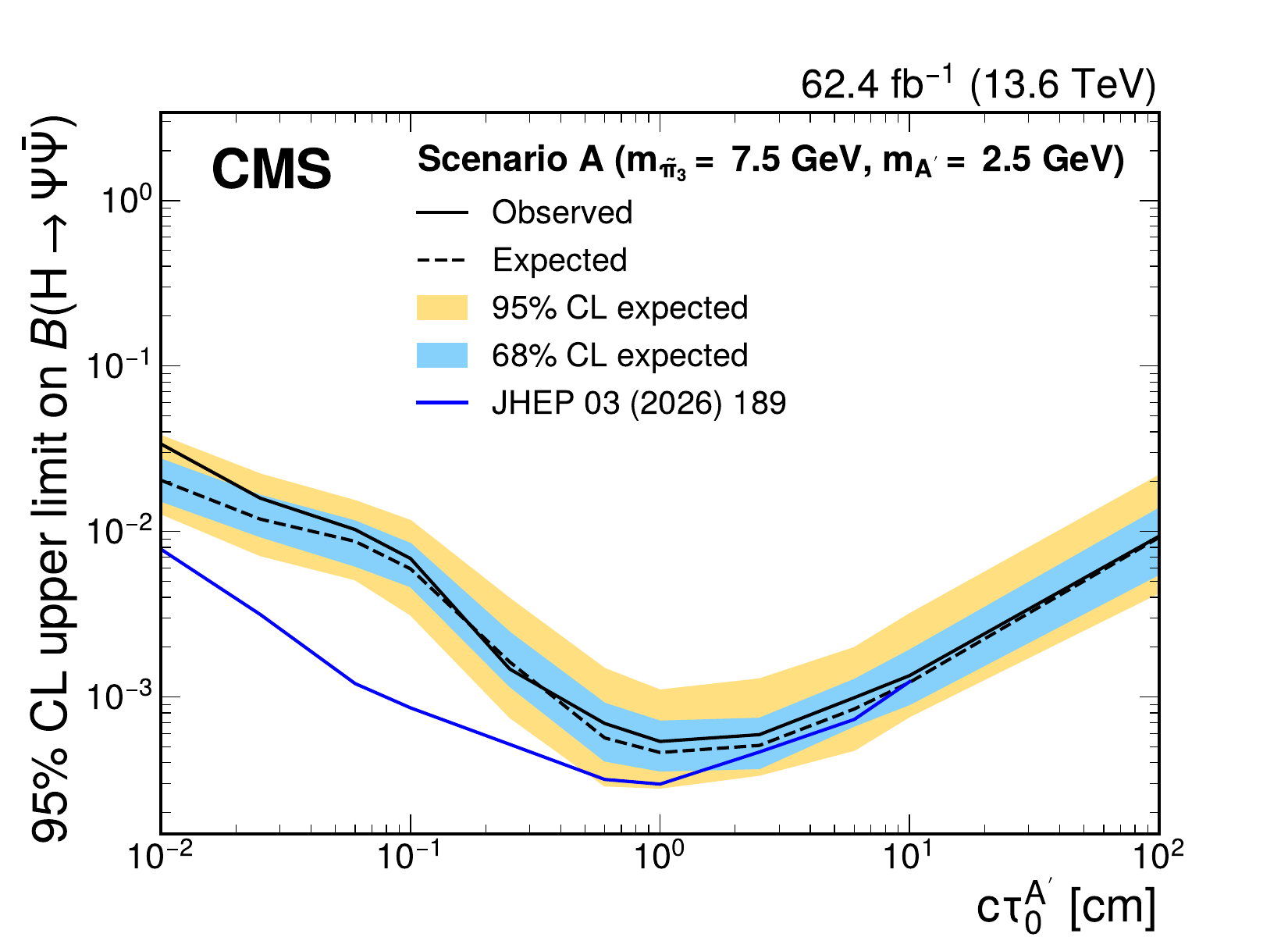}
    \caption{
      Upper limits at 95\%~\CL 
      on the branching fraction $\mathcal{B}(\PH\to\PGY\PAGY)$,
      for representative mass hypotheses of the Scenario A dark-shower model.
      Limits are shown as functions of $c\tau_{0}^{\PApr}$, for $m_{\PSGp} = 2\GeV$ and $m_{\PApr} = 0.67\GeV$ (upper left),
      $m_{\PSGp} = 5\GeV$ and $m_{\PApr} = 1.67\GeV$ (upper right),
      and $m_{\PSGp} = 7.5\GeV$ and $m_{\PApr} = 2.5\GeV$ (lower).
      For this model, it is assumed $\tilde{\Lambda} = m_{\PSGh} = 4m_{\PSGp}$, $\sin\theta = 0.1$
      and $\mathcal{B}(\PSGp_3 \to \PApr\PApr)=1$.
      The solid black (dashed black) line represents the observed (median expected) exclusion. 
      The inner blue (outer yellow) band indicates the region containing 68~(95)\% 
      of the distribution of limits expected under the background-only hypothesis.
      The limits are obtained using the combination of all event categories.
      The dark blue solid line represents the observed upper limits previously set in Ref.~\cite{CMS-EXO-24-008},
      rescaled to the same Higgs boson production cross section of $59.8\unit{pb}$~\cite{Cepeda:2019klc, LHCHWG136TeV}.
      }
      \label{fig:limdqcd_A}
\end{figure}

\begin{figure}[htb!]
  \centering
    \includegraphics[width=0.49\textwidth]{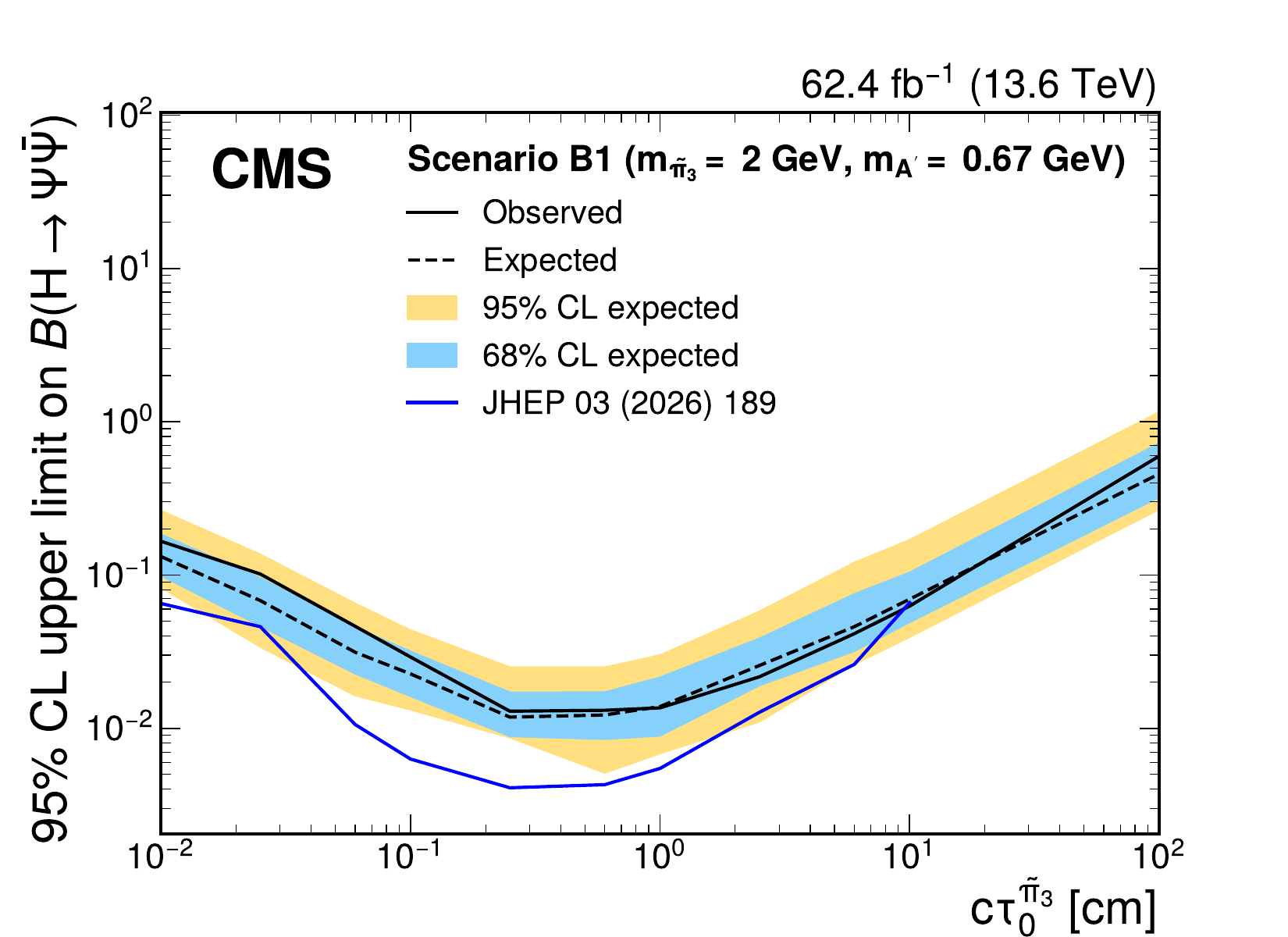}
    \includegraphics[width=0.49\textwidth]{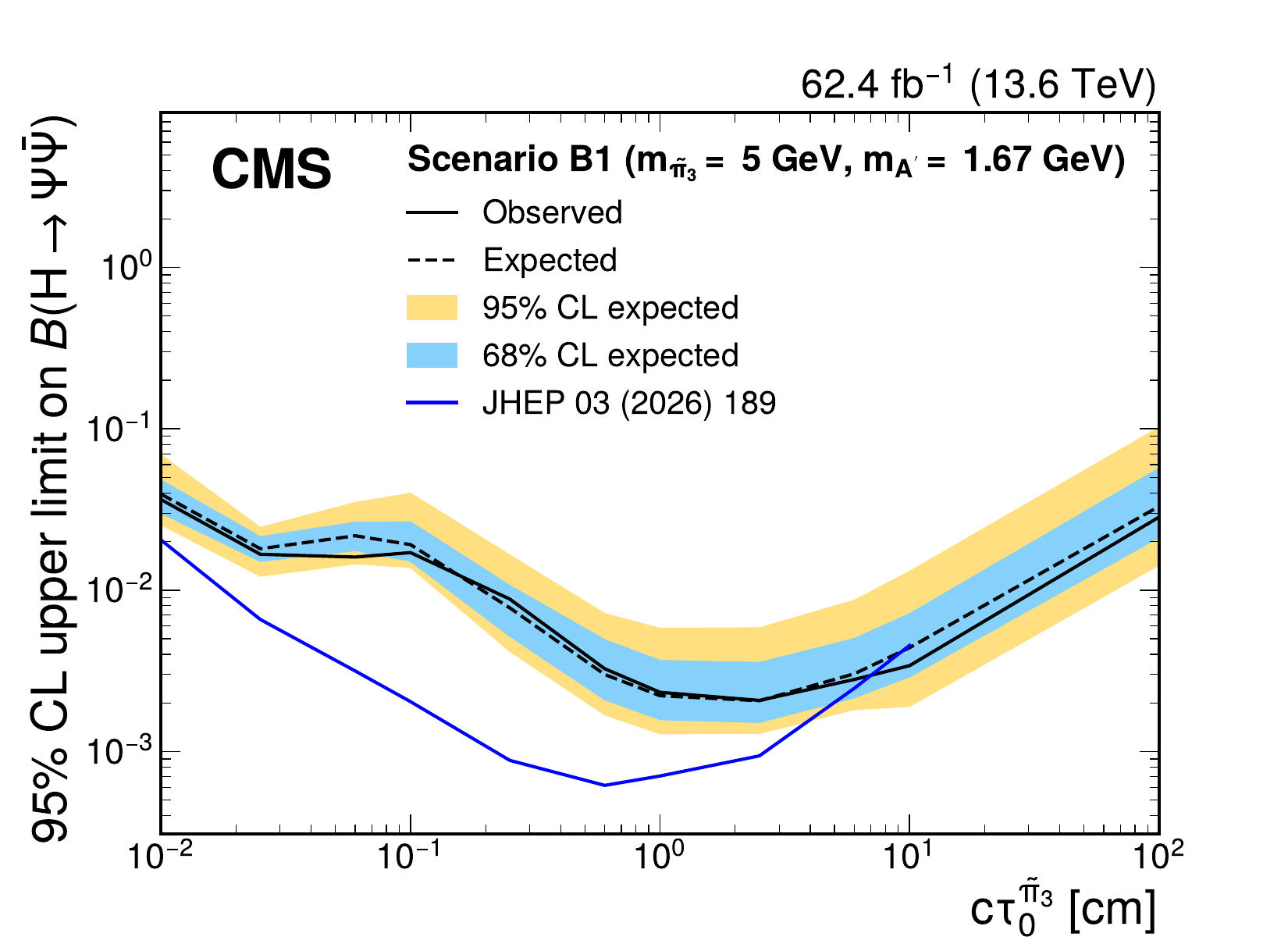}
    \includegraphics[width=0.49\textwidth]{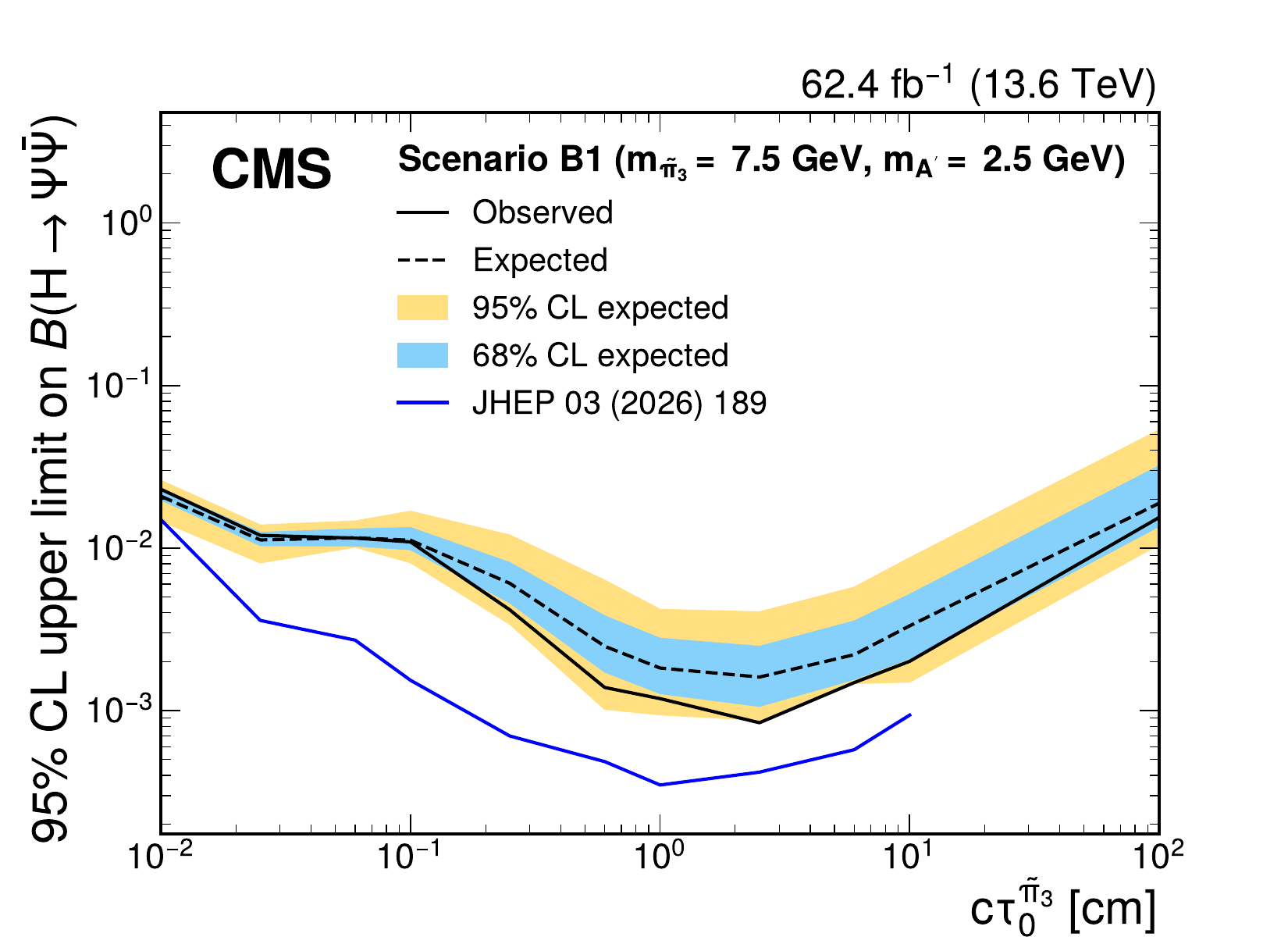}
    \caption{
      Upper limits at 95\%~\CL 
      on the branching fraction $\mathcal{B}(\PH\to\PGY\PAGY)$,
      for representative mass hypotheses of the Scenario B1 dark-shower model.
      Limits are shown as functions of $c\tau_{0}^{\PSGp_3}$, for $m_{\PSGp} = 2\GeV$ and $m_{\PApr} = 0.67\GeV$ (upper left),
      $m_{\PSGp} = 5\GeV$ and $m_{\PApr} = 1.67\GeV$ (upper right),
      and $m_{\PSGp} = 7.5\GeV$ and $m_{\PApr} = 2.5\GeV$ (lower).
      For this model, it is assumed $\tilde{\Lambda} = m_{\PSGh} = 4m_{\PSGp}$, $\sin\theta = 0.1$
      and $\mathcal{B}(\PSGp_3 \to \PApr\PApr)=1$.
      The solid black (dashed black) line represents the observed (median expected) exclusion. 
      The inner blue (outer yellow) band indicates the region containing 68~(95)\% 
      of the distribution of limits expected under the background-only hypothesis.
      The limits are obtained using the combination of all event categories.
      The dark blue solid line represents the observed upper limits previously set in Ref.~\cite{CMS-EXO-24-008},
      rescaled to the same Higgs boson production cross section of $59.8\unit{pb}$~\cite{Cepeda:2019klc, LHCHWG136TeV}.    
      }
      \label{fig:limdqcd_B1}
\end{figure}

We also set upper limits at 95\%~\CL on the branching fraction 
$\mathcal{B}(\PH\to\PGY\PAGY)$ considering two different dark shower scenarios.
Mean proper decay lengths of $\PApr$ and $\PSGp_3$ ranging from 0.01 to 100\cm are constrained for Scenarios A and B1, respectively.
Results are presented in Figs.~\ref{fig:limdqcd_A} and~\ref{fig:limdqcd_B1} for representative mass hypotheses
of the dark pion $\PSGp_3$ and of the dark photon $\PApr$ in Scenarios A and B1, respectively.
The results are compared with those previously obtained
in Ref.~\cite{CMS-EXO-24-008}, rescaled to the same Higgs boson production cross section of $59.8\unit{pb}$~\cite{Cepeda:2019klc, LHCHWG136TeV},
showing the complementarity of the two analysis strategies.
The constraints set in this paper extend the previous limits on the mean proper decay length of the $\PApr$ ($\PSGp_3$) in Scenario A (B1)
in the range from 10 to 100\cm,
thanks to the removal of the pixel hit multiplicity requirement in the scouting trigger streams used in this search,
extending the sensitivity to larger lifetime hypotheses.

\begin{figure}[htb!]
  \centering
    \includegraphics[width=0.49\textwidth]{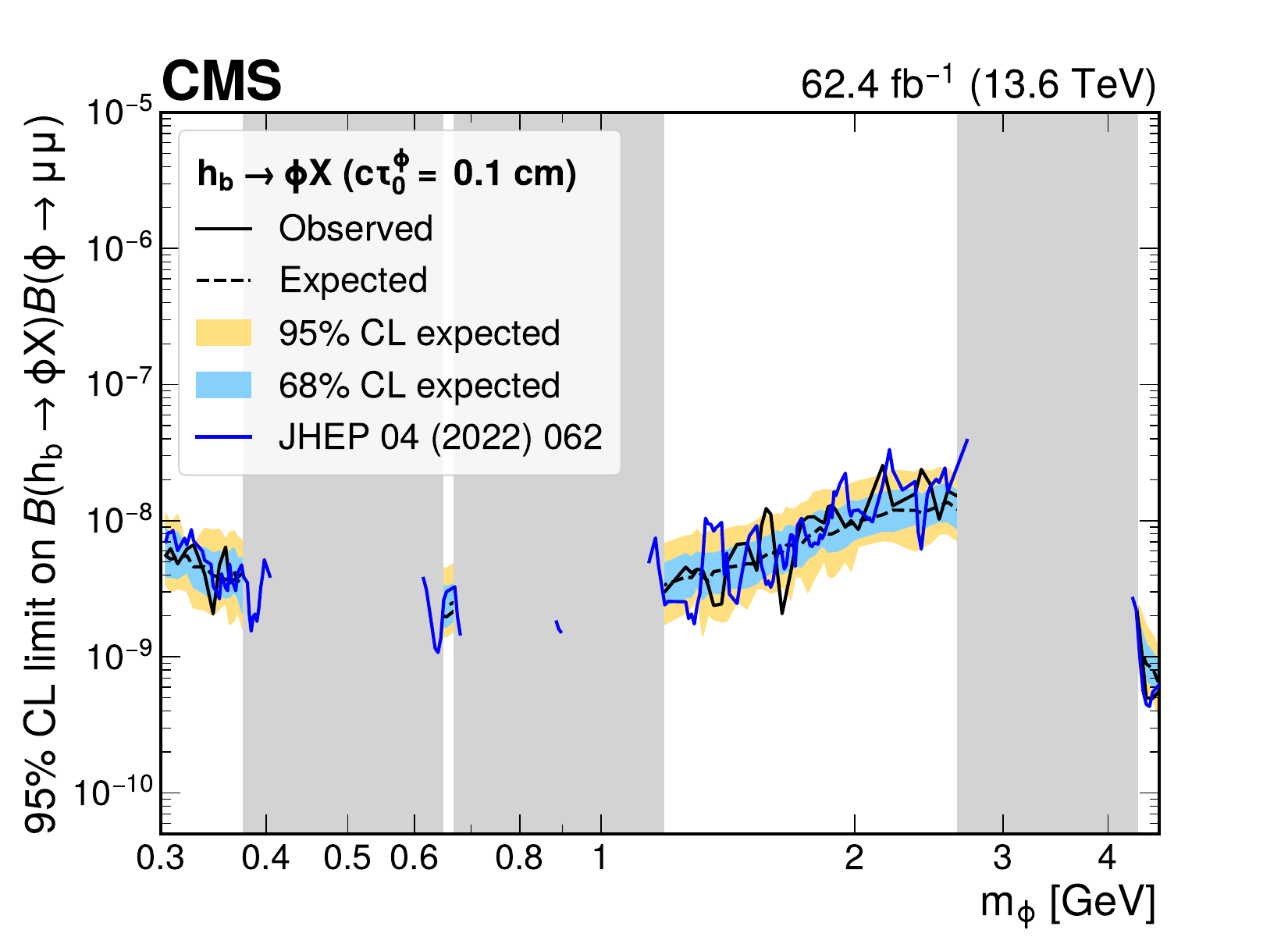}
    \includegraphics[width=0.49\textwidth]{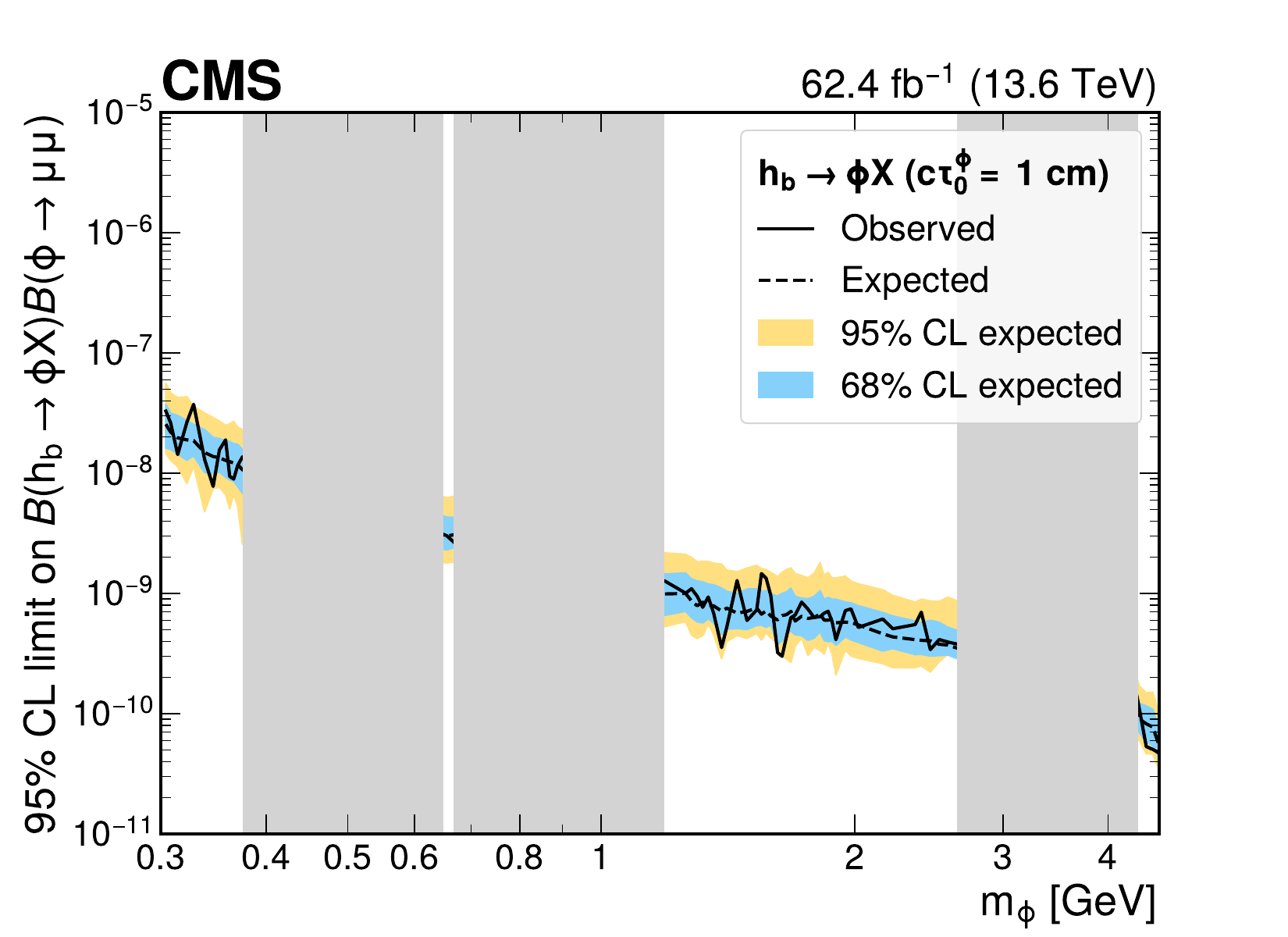}
    \includegraphics[width=0.49\textwidth]{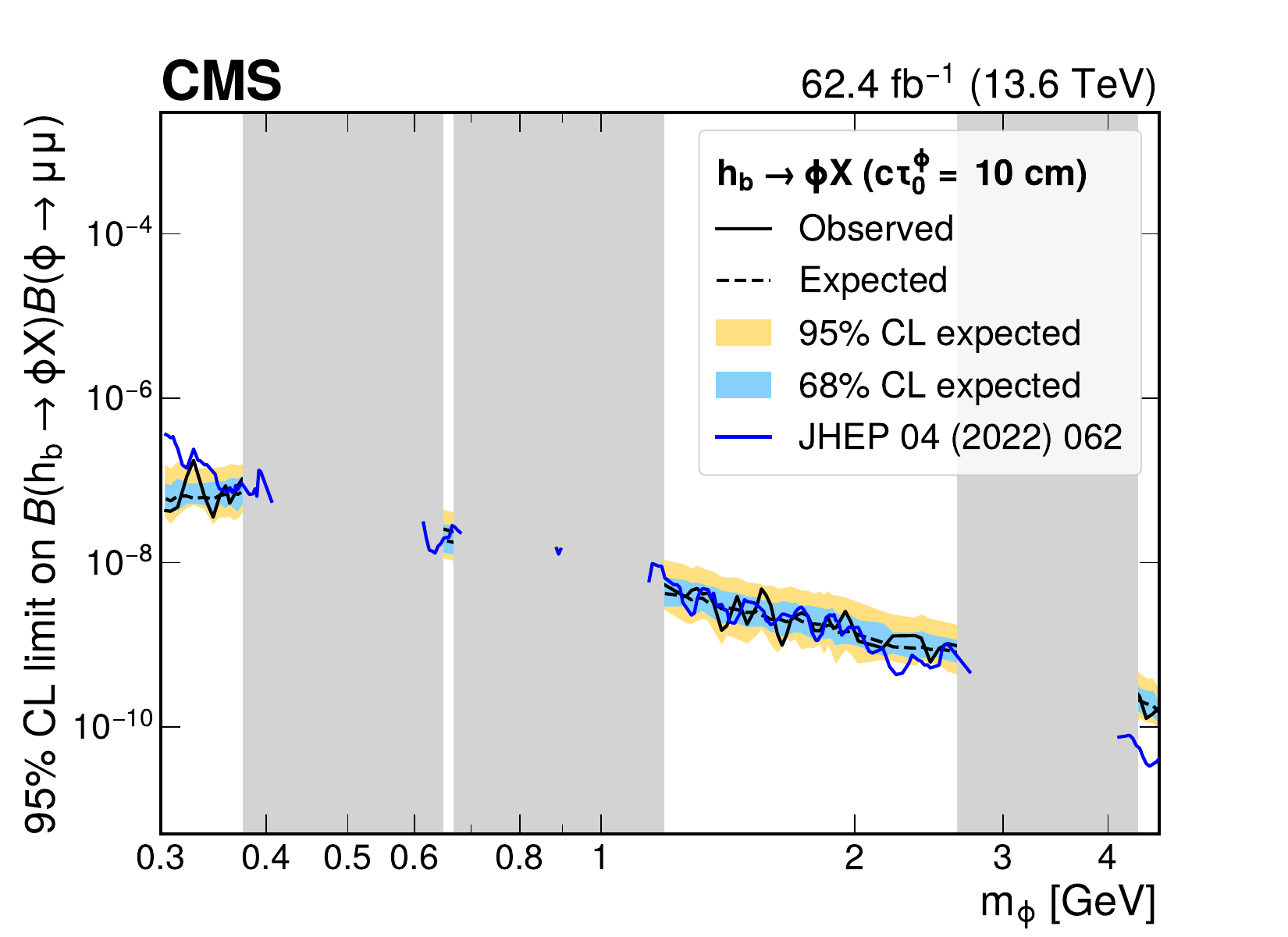}
    \caption{
      Upper limits at 95\%~\CL 
      on the branching fraction product $\mathcal{B}(\Ph_\PQb \to \PGf X)\mathcal{B}(\PGf\to\PGm\PGm)$
      as functions of the long-lived scalar resonance mass, $m_{\PGf}$,
      for $c\tau_{0}^{\PGf}=0.1$ (upper left), 1 (upper right), and
      10\cm (lower).
      The solid black (dashed black) line represents the observed (median expected) exclusion. 
      The inner blue (outer yellow) band indicates the region containing 68~(95)\% 
      of the distribution of limits expected under the background-only hypothesis.
      The limits are obtained using the combination of all dimuon event categories.
      The dark blue solid line represents the exclusion limits previously set in Ref.~\cite{EXO-20-014}.
      The vertical gray bands indicate mass ranges containing known SM resonances,
      which are masked for the purpose of this search.
      The blue line can overlap with the gray bands, owing to the different mass ranges covered by the results from Ref.~\cite{EXO-20-014}.
    }
    \label{fig:limB}
\end{figure}

Finally, we set 95\%~\CL upper limits on the branching fraction product $\mathcal{B}(\Ph_\PQb \to \PGf X)\mathcal{B}(\PGf\to\PGm\PGm)$
in a model where a long-lived scalar resonance $\PGf$ arises from the decay of a \PQb hadron.
Figure~\ref{fig:limB} shows the limits obtained as functions of $m_{\PGf}$, for $c\tau_0^{\PGf} = 0.1$, 1, and 10\cm.
They are found to be mostly consistent with the ones set in Ref.~\cite{EXO-20-014},
because for such signatures the smaller amount of available data is compensated
by the improvements in the scouting triggers and analysis strategy.

\clearpage

\section{Summary}
\label{sec:summary}

A search for displaced multimuon resonances has been performed using proton-proton collisions 
at a center-of-mass energy of 13.6\TeV, 
collected by the CMS experiment at the LHC in 2022 and 2023, corresponding to an integrated luminosity of \lumivalscouting.
The data were collected using a dedicated dimuon trigger stream with low transverse momentum thresholds, 
recorded at high rate by retaining a reduced amount of information at the high-level trigger, 
to explore otherwise inaccessible phase space at low multimuon invariant mass and nonzero displacement from the primary interaction vertex.
No significant excess above the standard model prediction is found.
The data are used to set constraints on a wide range of mass and lifetime hypotheses for various scenarios of physics beyond the standard model.
Three sets of models are considered:
the hidden Abelian Higgs model, where the Higgs boson decays into a pair of long-lived dark photons;
dark-shower models, where dark showers originate from the Higgs boson decays into dark quarks;
and models where a long-lived scalar resonance is produced from the decay of a \PQb hadron.
In the hidden Abelian Higgs model,
the most stringent limits to date are obtained for dark-photon proper decay lengths greater than 1\cm and
dark-photon masses less than 5\GeV.
In the dark-shower models, the most stringent limits to date are set for proper decay lengths larger than 10\cm.
In models with a \PQb hadron, the improved trigger and analysis strategy compensate for the smaller data set, 
leading to constraints comparable to the previous results.

\begin{acknowledgments}
We congratulate our colleagues in the CERN accelerator departments for the excellent performance of the LHC and thank the technical and administrative staffs at CERN and at other CMS institutes for their contributions to the success of the CMS effort. In addition, we gratefully acknowledge the computing centers and personnel of the Worldwide LHC Computing Grid and other centers for delivering so effectively the computing infrastructure essential to our analyses. Finally, we acknowledge the enduring support for the construction and operation of the LHC, the CMS detector, and the supporting computing infrastructure provided by the following funding agencies: SC (Armenia), BMBWF and FWF (Austria); FNRS and FWO (Belgium); CNPq, CAPES, FAPERJ, FAPERGS, and FAPESP (Brazil); MES and BNSF (Bulgaria); CERN; CAS, MoST, and NSFC (China); MINCIENCIAS (Colombia); MSES and CSF (Croatia); RIF (Cyprus); SENESCYT (Ecuador); ERC PRG and PSG, TARISTU24-TK10 and MoER TK202 (Estonia); Academy of Finland, MEC, and HIP (Finland); CEA and CNRS/IN2P3 (France); SRNSF (Georgia); BMFTR, DFG, and HGF (Germany); GSRI (Greece); MATE and NKFIH (Hungary); DAE and DST (India); IPM (Iran); SFI (Ireland); INFN (Italy); MSIT and NRF (Republic of Korea); MES (Latvia); LMTLT (Lithuania); MOE and UM (Malaysia); BUAP, CINVESTAV, CONACYT, LNS, SEP, and UASLP-FAI (Mexico); MOS (Montenegro); MBIE (New Zealand); PAEC (Pakistan); MSHE, NSC, and NAWA (Poland); FCT (Portugal); MESTD (Serbia); MICIU/AEI and PCTI (Spain); MOSTR (Sri Lanka); Swiss Funding Agencies (Switzerland); MST (Taipei); MHESI (Thailand); TUBITAK and TENMAK (T\"{u}rkiye); NASU (Ukraine); STFC (United Kingdom); DOE and NSF (USA).

\begin{sloppy}
\setlength\emergencystretch{\hsize}
\hyphenation{Rachada-pisek} Individuals have received support from the Marie-Curie program and the European Research Council and Horizon 2020 Grant, contract Nos.\ 675440, 724704, 752730, 758316, 765710, 824093, 101115353, 101002207, 101001205, and COST Action CA16108 (European Union); the Leventis Foundation; the Alfred P.\ Sloan Foundation; the Alexander von Humboldt Foundation; the Science Committee, project no. 22rl-037 (Armenia); the Fonds pour la Formation \`a la Recherche dans l'Industrie et dans l'Agriculture (FRIA) and Fonds voor Wetenschappelijk Onderzoek contract No. 1228724N (Belgium); the Beijing Municipal Science \& Technology Commission, No. Z191100007219010, the Fundamental Research Funds for the Central Universities, the Ministry of Science and Technology of China under Grant No. 2023YFA1605804, the Natural Science Foundation of China under Grant No. 12535004, and USTC Research Funds of the Double First-Class Initiative No.\ YD2030002017 (China); the Ministry of Education, Youth and Sports (MEYS) of the Czech Republic; the Shota Rustaveli National Science Foundation (Georgia); the Deutsche Forschungsgemeinschaft (DFG), among others, under Germany's Excellence Strategy -- EXC 2121 ``Quantum Universe" -- 390833306, and under project number 400140256 - GRK2497; the Hellenic Foundation for Research and Innovation (HFRI), Project Number 2288 (Greece); the Hungarian Academy of Sciences, the New National Excellence Program - \'UNKP, the NKFIH research grants K 131991, K 138136, K 143460, K 143477, K 147557, K 146913, K 146914, K 147048, TKP2021-NKTA-64, and 2025-1.1.5-NEMZ\_KI-2025-00004, and MATE KKP and KKPCs Research Excellence and Flagship Research Groups grants (Hungary); the Council of Science and Industrial Research, India; ICSC -- National Research Center for High Performance Computing, Big Data and Quantum Computing, FAIR -- Future Artificial Intelligence Research, and CUP I53D23001070006 (Mission 4 Component 1), funded by the NextGenerationEU program, the Italian Ministry of University and Research (MUR) under Bando PRIN 2022 -- CUP I53C24002390006, PRIN PRIMULA 2022RBYK7T (Italy); the Latvian Council of Science; the Ministry of Science and Higher Education, project no. 2022/WK/14, and the National Science Center, contracts Opus 2021/41/B/ST2/01369, 2021/43/B/ST2/01552, 2023/49/B/ST2/03273, and the NAWA contract BPN/PPO/2021/1/00011 (Poland); the Funda\c{c}\~ao para a Ci\^encia e a Tecnologia (Portugal); the National Priorities Research Program by Qatar National Research Fund;  MICIU/AEI/10.13039/501100011033, ERDF/EU, ``European Union NextGenerationEU/PRTR", projects PID2022-142604OB-C21, PID2022-139519OB-C21, PID2023-147706NB-I00, PID2023-148896NB-I00, PID2023-146983NB-I00, PID2023-147115NB-I00, PID2023-148418NB-C41, PID2023-148418NB-C42, PID2023-148418NB-C43, PID2023-148418NB-C44, PID2024-158190NB-C22, RYC2021-033305-I, RYC2024-048719-I, CNS2023-144781, CNS2024-154769 and Plan de Ciencia, Tecnolog{\'i}a e Innovaci{\'o}n de Asturias, Spain; the Chulalongkorn Academic into Its 2nd Century Project Advancement Project, the National Science, Research and Innovation Fund program IND\_FF\_68\_369\_2300\_097, and the Program Management Unit for Human Resources \& Institutional Development, Research and Innovation, grant B39G680009 (Thailand); the Eric \& Wendy Schmidt Fund for Strategic Innovation through the CERN Next Generation Triggers project under grant agreement number SIF-2023-004; the Kavli Foundation; the Nvidia Corporation; the SuperMicro Corporation; the Welch Foundation, contract C-1845; and the Weston Havens Foundation (USA). 
\end{sloppy}
\end{acknowledgments}

\bibliography{auto_generated}

\cleardoublepage \appendix\section{The CMS Collaboration \label{app:collab}}\begin{sloppypar}\hyphenpenalty=5000\widowpenalty=500\clubpenalty=5000\input{EXO-24-016-public-authorlist.tex}\end{sloppypar}
\end{document}

%% file: EXO-24-016-public-authorlist.tex
\cmsinstitute{Yerevan Physics Institute, Yerevan, Armenia}
{\tolerance=6000
A.~Gevorgyan\cmsorcid{0000-0003-2751-9489}, A.~Hayrapetyan, V.~Makarenko\cmsorcid{0000-0002-8406-8605}, A.~Tumasyan\cmsAuthorMark{1}\cmsorcid{0009-0000-0684-6742}
\par}
\cmsinstitute{Marietta Blau Institute for Particle Physics, Vienna, Austria}
{\tolerance=6000
W.~Adam\cmsorcid{0000-0001-9099-4341}, L.~Benato\cmsorcid{0000-0001-5135-7489}, T.~Bergauer\cmsorcid{0000-0002-5786-0293}, M.~Dragicevic\cmsorcid{0000-0003-1967-6783}, S.~Gundacker\cmsorcid{0000-0003-2087-3266}, A.K.~Guven\cmsorcid{0009-0004-5670-5138}, P.S.~Hussain\cmsorcid{0000-0002-4825-5278}, M.~Jeitler\cmsAuthorMark{2}\cmsorcid{0000-0002-5141-9560}, N.~Krammer\cmsorcid{0000-0002-0548-0985}, A.~Li\cmsorcid{0000-0002-4547-116X}, D.~Liko\cmsorcid{0000-0002-3380-473X}, M.~Matthewman, J.~Schieck\cmsAuthorMark{2}\cmsorcid{0000-0002-1058-8093}, R.~Sch\"{o}fbeck\cmsAuthorMark{2}\cmsorcid{0000-0002-2332-8784}, M.~Shooshtari\cmsorcid{0009-0004-8882-4887}, M.~Sonawane\cmsorcid{0000-0003-0510-7010}, N.~Van~Den~Bossche\cmsorcid{0000-0003-2973-4991}, W.~Waltenberger\cmsorcid{0000-0002-6215-7228}, C.E.~Wulz\cmsAuthorMark{2}\cmsorcid{0000-0001-9226-5812}
% ...
\par}
$^{1}$Also at Yerevan State University, Yerevan, Armenia\\
$^{2}$Also at Technische Universit\"{a}t Wien, Vienna, Austria\\
% ...
